\documentclass[12pt,a4paper,fleqn]{book}
\usepackage{epsfig}
\usepackage[tbtags]{amsmath}
\usepackage{amssymb}
\usepackage{mathrsfs}
\usepackage{appendix}
\usepackage[pdftex,hyperindex,bookmarks=true,colorlinks=true,citecolor=blue,linkcolor=blue]{hyperref}
\usepackage{fancyhdr}
\makeatletter
\def\cleardoublepage{\clearpage\if@twoside \ifodd\c@page\else%
    \hbox{}%
    \thispagestyle{empty}%              % Empty header styles
    \newpage%
    \if@twocolumn\hbox{}\newpage\fi\fi\fi}
\makeatother

\begin{document}

%%%%%%%%%%%%%%%%%%%%%%%%%%%%%%%%%%%%%%%%%%%%%%%%%%%%%%%%%%%%%%%%%%

\setlength{\parskip}{1.0ex plus 0.5ex minus 0.5ex}
%\frontmatter
%\include{titlepage}
%\include{cert_from_supervisor}
%\include{to_my_parents}
%\include{acknowledgements}
%\include{list_of_publications}
\frontmatter 
%%%%%%%%%%%%%%%%%%%%%%%%%%%%%%%%%%%%%%%%%%%%%%%%%%%%%%%%%%%%%%%%%%

\begin{titlepage}
\begin{center}

%%%%%%%%%%%%%%%%%%%%%%%%%%%%%%%%%%%%%%%%%%%%%%%%%%%%%%%%%%%%%%%%%%%%%%%%%%%%%%%%%%%%%%%%%%%
 {\bf \uppercase{\huge Q\Large uantum \huge T\Large unneling\\ in \\
\vspace{0.4 cm}\huge B\Large lack \huge H\Large oles\\
%\vspace{0.4 cm} \huge Q\Large uantum \huge T\Large unneling
}}
%(Corrected Copy)
%%%%%%%%%%%%%%%%%%%%%%%%%%%%%%%%%%%%%%%%%%%%%%%%%%%%%%%%%%%%%%%%%%%%%%%%%%%%%%%%%%%%%%%%%%
%\Large\textsc{Field theory aspects of  Cosmology \\and \\Black Holes}
%%%%%%%%%%%%%%%%%%%%%%%%%%%%%%%%%%%%%%%%%%%%%%%%%%%%%%%%%%%%%%%%%%%%%%%%%%%%%%%%%%%%%%%5
\vfill

\normalsize
{\Large Thesis submitted for the degree of}\\[2.2ex]
\textbf{\Large Doctor of Philosophy (Science)}\\[2ex]
{\Large of}\\[2ex]
\textbf{\Large University of Calcutta, India}

\vfill

{\Large December, 2010}

\vfill

\textbf{{\Large Bibhas Ranjan Majhi}}\\[2ex]
%{\large \mbox{Satyendra Nath Bose National Centre for Basic Sciences}}\\
%{\large JD Block, Sector 3, Salt Lake, Kolkata 700098, India}
{Department of Theoretical Sciences}

\end{center}
\end{titlepage}

%%%%%%%%%%%%%%%%%%%%%%%%%%%%%%%%%%%%%%%%%%%%%%%%%%%%%%%%%%%%%%%%%%

%\include{bibhas-cert-from-supervisor}
%%%%%%%%%%%%%%%%%%%%%%%%%%%%%%%%%%%%%%%%%%%%%%%%%%%%%%%%%%%%%%%%%%

%\chapter*{}
\thispagestyle{empty}
\vspace*{6.5 cm}
%%%%%%%%%%%%%%%%%%%%%%%%%%%%%%%%%%%%%%%%%%%%%%%%%%%%%%%%%%%%%%%%%%
%\begin{flushright}
%{\Large \bf To\hspace*{2 cm}\\my\hspace*{1.5 cm}\\mother}
%\end{flushright}
%%%%%%%%%%%%%%%%%%%%%%%%%%%%%%%%%%%%%%%%%%%%%%%%%%%%%%%%%%%%%%%%%%
\begin{figure}[th] 
\centering
\includegraphics[width=0.5\textwidth]{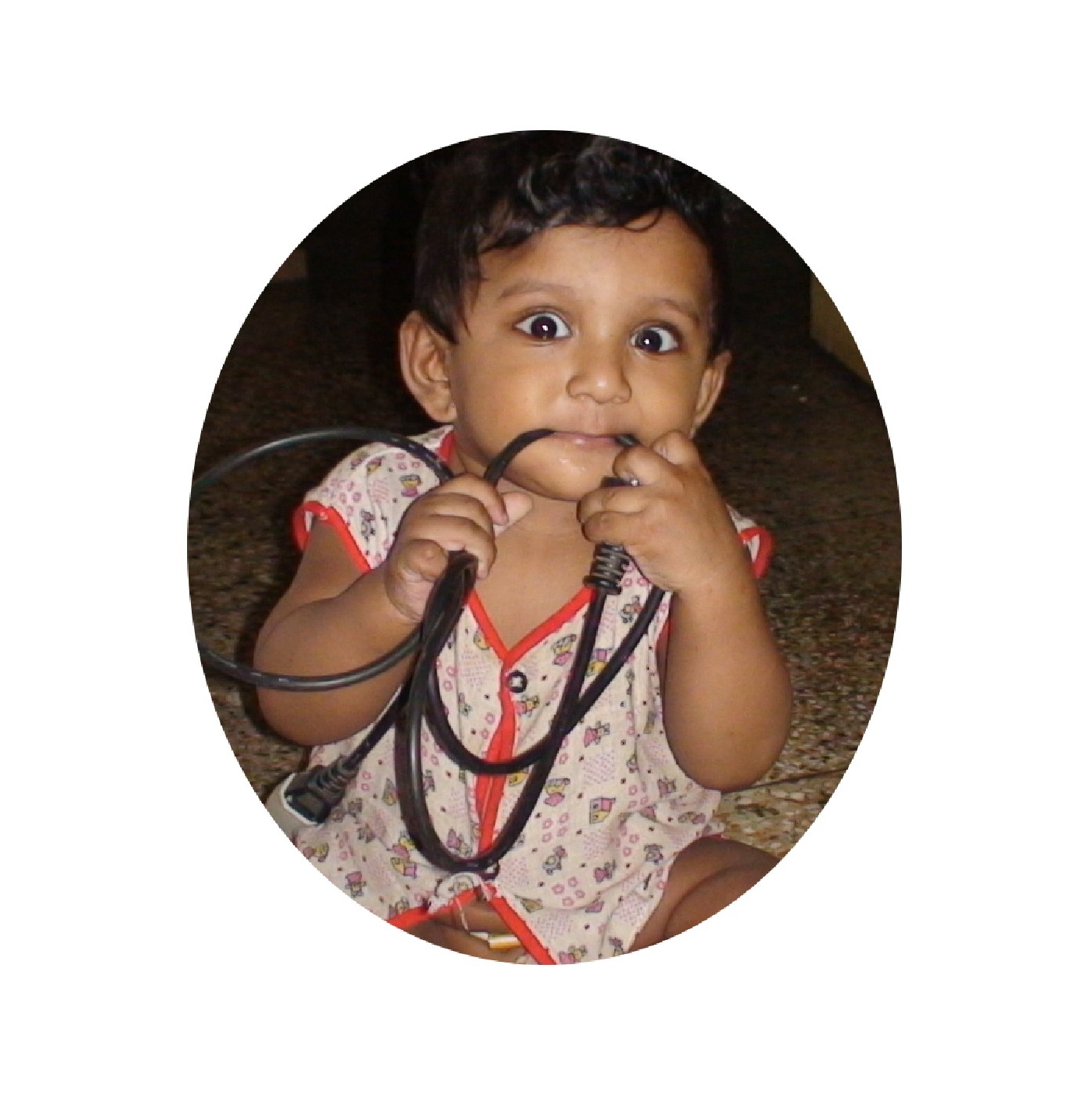}
\end{figure}
%%%%%%%%%%%%%%%%%%%%%%%%%%%%%%%%%%%%%%%%%%%%%%%%%%%%%%%%%%%%%%%
%\Large \bf \hspace*{8.2 cm}To\\ \hspace*{9.6 cm}my\\ \hspace*{10.5 cm}mother\\
\hspace*{8.6 cm} $\mathcal{TO}$\\ \hspace*{10 cm}$\mathcal{MY \ NEPHEW}$

%%%%%%%%%%%%%%%%%%%%%%%%%%%%%%%%%%%%%%%%%%%%%%%%%%%%%%%%%%%%%%%%%%

\chapter*{Acknowledgements}
\thispagestyle{empty}

%%%%%%%%%%%%%%%%%%%%%%%%%%%%%%%%%%%%%%%%%%%%%%%%%%%%%%%%%%%%%%%%%%
This thesis is the culminative outcome of four years work, which has been
 made possible by the blessings and support of many individuals. I take this
 opportunity to express my sincere gratitude to all of them.

First and foremost, I would like to thank Prof. Rabin Banerjee, my thesis supervisor.
 His uncanny ability to select a particular problem, a keen and strategic analysis of it
 and deep involvement among the students makes him something special.
 Thank you Sir for giving me all that could last my entire life.

  In addition to this, I like to express my sincere thanks to Prof. Claus Kiefer, Universit\"at zu K\"oln, Germany and Dr. Elias C. Vagenas, Academy of Athens, Greece for their constant academic help and stimulating collaboration throughout my research activity. I thank Prof. Subir Ghosh, ISI Kolkata. It was a quite nice experience to work with them.

     I would also like to express my gratitude Prof. Sandip K. Chakrabarti, Dr. Debashis Gangopadhyay, Dr. Archan S. Majumdar, Dr. Biswajit Chakraborty and Prof. Subhrangshu Sekhar Manna for helping me in my academics at Satyendra Nath National Center for Basic Sciences (SNBNCBS). 

  Prof. Jayanta Kumar Bhattacharjee and Prof. Binayak Dutta Roy have 
 always been available to clarify very elementary but at same time subtle concepts
 in physics. I sincerely thank them for giving me their valuable attention. 

%  I would like to thank Prof. Sushanta Dattagupta, ex. Director of SNBNCBS for allowing me
%  to carry out my research here. 

  I thank Prof. Arup K. Raychaudhuri,  Director of SNBNCBS, for providing an excellent academic atmosphere.

 I am also thankful to the Library staff of SNBNCBS for their support.

 I am indebted to Prof. Manoranjan Saha, Prof. Amitava Raychaudhuri, Prof. Anirban Kundu, Dr. Debnarayana Jana, University of Calcutta, for giving me 
 crucial suggestions and guidance, whenever I required. I would also like to thank Prof. Tapan Kumar Das , University of Calcutta and my college teachers Dr. Debabrata Das, Prof. Maynak Gupta.

  My life during this voyage has been made cherishable and interesting by some
 colourful and memorable personalities to whom I would like to express my heartiest feelings.
 I am indebted to my seniors - Chandrashekhar da and  Mitali di.
 Things, which I gained from them, turn out to be most important in my thesis work.
 I would like to thank all of them for their brotherly support. Saurav Samanta and Shailesh Kulkarni, seniors and friends, deserve special mention for
 those strategic discussions, on both the academic as well as non-academic fronts. 
  Then come my brilliant and helpful group mates,
 ranging from enthusiastic Sujoy, Debraj, Sumit, Dibakar to Biswajit. I am grateful to Himadri for his {\it sometimes} meaningless but still delightful gossips and also to his group for sharing precious moments. 
 I would also like to thank all the cricket and football players of SNBNCBS for making my
 stay enjoyable.    
% I would like to mention my {\it Sahyadri group} friends who have made
% my each trip to Pune, memorable - Bhalchandra, Manish, Chaitanya, Sandeep, Hemant, Dhananjay,
% Pramod and Sukumaran.  All of them  have something special, which is only visible 
% from the {\it cliffs} of {\it western ghats}. I am also thankful to Hrishikesh and Smita,
% my school friends, for their friendly attitude on many occasions.          
%

   I owe my deepest gratitude to each and every member of my of family. Especially, I would like to express my heartiest love and respect to my parents for their constant care, encouragement and blessings. Another two pairs of my family, who stands in the same footing as my parents, are my uncle (Jethu) and aunt (Jethima) and my would be father-in-law and mother-in-law. I give {\it paranams} to all. I express my love to my elder brother Pinku and elder sisters Tinku, Rinku and my brother-in-law, Subikash and sister-in-law, {\it Dona} for understanding and helping me, on many occasions.

    I am indebted to my better halves: my grandmothers ({\it Didima} and {\it Thakurma}) and my beloved nephew, Oishik ({\it Sona}) for making my life fill of joy and greetings. Finally,  I express my heartiest love to Priyanka ({\it Mousam}, my would be wife) with whom I shared all my moments ({\it sweet or bitter}) for making everything possible from my post-graduate life.   
% Archana for their help during my stay in Pune. I thank Kedar (my brother and friend),
% with whom I shared some {\it sweet} moments of my graduate life and uncountable
% visits to that {\it stop}.

%%%%%%%%%%%%%%%%%%%%%%%%%%%%%%%%%%%%%%%%%%%%%%%%%%%%%%%%%%%%%%%%%%

\chapter*{List of publications}
\thispagestyle{empty}

%%%%%%%%%%%%%%%%%%%%%%%%%%%%%%%%%%%%%%%%%%%%%%%%%%%%%%%%%%%%%%%%%%

%\begin{minipage}{13cm}
\begin{enumerate} \raggedright
%\vskip -0.8cm
\item \textbf{Gauge Theories on A(dS) space and Killing Vectors.}\\
    Rabin Banerjee and Bibhas Ranjan Majhi\\
    {\em Annals Phys.} {\bf 323}, 705 (2008) [arXiv:hep-th/0703207].

\item \textbf{ Crypto-Harmonic Oscillator in Higher Dimensions: Classical and Quantum Aspects.}\\
    Subir Ghosh and Bibhas Ranjan Majhi\\
    {\em J. Phys.} {\bf A41}, 065306 (2008) [arXiv:0709.4325].
     
\item \textbf{Quantum Tunneling and Back Reaction.}\\
    Rabin Banerjee and Bibhas Ranjan Majhi\\
    {\em Phys. Lett.} {\bf B662}, 62 (2008) [arXiv:0801.0200].
    
\item \textbf{Noncommutative Black Hole Thermodynamics.}\\
  Rabin Banerjee, Bibhas Ranjan Majhi and Saurav Samanta\\
  {\em Phys. Rev.} {\bf D77}, 124035 (2008) [arXiv:0801.3583].

\item \textbf{Noncommutative Schwarzschild Black Hole and Area Law.}\\
  Rabin Banerjee, Bibhas Ranjan Majhi and Sujoy Kumar Modak\\
  {\em Class. Quant. Grav.} {\bf 26}, 085010 (2009) [arXiv:0802.2176].

\item \textbf{Quantum Tunneling Beyond Semiclassical Approximation.}\\
  Rabin Banerjee and Bibhas Ranjan Majhi\\
  {\em JHEP} {\bf 0806}, 095 (2008) [arXiv:0805.2220].

\newpage

\item \textbf{Quantum Tunneling and Trace Anomaly.}\\
  Rabin Banerjee and Bibhas Ranjan Majhi\\
  {\em Phys. Lett.} {\bf B674}, 218 (2009) [arXiv:0808.3688].

\item \textbf{Fermion Tunneling Beyond Semiclassical Approximation.}\\
  Bibhas Ranjan Majhi\\
  {\em Phys. Rev.} {\bf D79}, 044005 (2009) [arXiv:0809.1508].

\item \textbf{Connecting anomaly and tunneling methods for Hawking effect through chirality.}\\
   Rabin Banerjee and Bibhas Ranjan Majhi\\
   {\em Phys. Rev.} {\bf D79}, 064024 (2009) [arXiv:0812.0497].

\item \textbf{Hawking Radiation due to Photon and Gravitino Tunneling.}\\
  Bibhas Ranjan Majhi and Saurav Samanta\\
  {\em Annals Phys.} {\bf 325}, 2410 (2010) [arXiv:0901.2258]. 

\item \textbf{Hawking black body spectrum from tunneling mechanism.}\\
   Rabin Banerjee and Bibhas Ranjan Majhi\\
   {\em Phys. Lett.} {\bf B675}, 243 (2009) [arXiv:0903.0250]. 

\item \textbf{Quantum tunneling and black hole spectroscopy.}\\
   Rabin Banerjee, Bibhas Ranjan Majhi and Elias C. Vagenas\\
   {\em Phys. Lett.} {\bf B686}, 279 (2010) [arXiv:0907.4271].

\item \textbf{Hawking radiation and black hole spectroscopy in Horava-Lifshitz gravity.}\\
  Bibhas Ranjan Majhi\\
  {\em Phys. Lett.} {\bf B686}, 49 (2010) [arXiv:0911.3239] .

\item \textbf{New Global Embedding Approach to Study Hawking and Unruh Effects.}\\
  Rabin Banerjee and Bibhas Ranjan Majhi\\
  {\em Phys. Lett.} {\bf B690}, 83 (2010) [arXiv:1002.0985].

\pagebreak

\item \textbf{Statistical Origin of Gravity.}\\
  Rabin Banerjee and Bibhas Ranjan Majhi\\
  {\em Phys. Rev.} {\bf D81}, 124006 (2010) [arXiv:1003.2312].

\item \textbf{A Note on the Lower Bound of Black Hole Area Change in Tunneling Formalism.}\\
  Rabin Banerjee, Bibhas Ranjan Majhi and Elias C. Vagenas\\
 {\em Europhys. Lett.} {\bf 92}, 20001 (2010) [arXiv:1005.1499].

\item \textbf{Quantum gravitational correction to the Hawking temperature from the Lemaitre-Tolman-Bondi model.}\\
  Rabin Banerjee, Claus Kiefer and Bibhas Ranjan Majhi\\
  {\em Phys. Rev.} {\bf D82}, 044013 (2010) [arXiv:1005.2264].

\item \textbf{Killing Symmetries and Smarr Formula for Black Holes in Arbitrary Dimensions.}\\
  Rabin Banerjee, Bibhas Ranjan Majhi, Sujoy Kumar Modak and Saurav Samanta\\
  {\em Phys. Rev.} {\bf D82}, 124002 (2010) [arXiv:1007.5204].

\end{enumerate}
This thesis is based on the papers numbered by [3,4,6,9,11,12,14,15] whose reprints are attached at
 the end of the thesis.

%%%%%%%%%%%%%%%%%%%%%%%%%%%%%%%%%%%%%%%%%%%%%%%%%%%%%%%%%%%%%%%%%%

%%%%%%%%%%%%%%%%%%%%%%%%%%%%%%%%%%%%%%%%%%%%%%%%%%%%%%%%%%%%%%%%%%

\chapter*{}
\pagenumbering{roman}
\thispagestyle{empty}

\begin{center}
\uppercase{Quantum Tunneling\\ in \\ Black Holes}
\end{center}

%%%%%%%%%%%%%%%%%%%%%%%%%%%%%%%%%%%%%%%%%%%%%%%%%%%%%%%%%%%%%%%%%%

%%%%%%%%%%%%%%%%%%%%%%%%%%%%%%%%%%%%%%%%%%%%%%%%%%%%%%%%%%%%%%%%%%

\tableofcontents

%%%%%%%%%%%%%%%%%%%%%%%%%%%%%%%%%%%%%%%%%%%%%%%%%%%%%%%%%%%%%%%%%%

%\include{misc}
%\setlength{\parskip}{1.0ex plus 0.4ex minus 0.4ex}

\mainmatter
\chapter{\label{chap:introduction} Introduction}
%%%%%%%%%%%%%%%%%%%%%%%%%%%%%%%%%%%%%%%%%%%%%%%%%%%%%%%%%%%%%%%%%%%%%%%%%%%%%%%%%%%
\section{\label{Overview}Overview}
This is a short overview of the vast subject of black holes. It specifically highlights those issues which are relevant for the present thesis.
          The search for a theory of quantum gravity drives a great deal of research in theoretical physics today, and much has been learned along the way, but convincing success remains elusive. There are two parts of general relativity: the framework of space-time curvature and its influence on matter, and the dynamics of the metric in response to energy-momentum (as described by Einstein's equation). Lacking the true theory of quantum gravity, we may still take the first part of GR - the idea that matter fields propagate on a curved space-time background - and consider the case where those matter fields are quantum mechanical. In other words, we take the metric to be fixed, rather than obeying some dynamical equations, and study quantum field theory in the curved space-time.

        Classical solutions of Einstein's equation gives several metrics of space-time in absence (Schwarzschild metric) or in presence (e.g. Reissner-Nordstrom metric) of matter fields. Both of these solutions show there exists a region of space-time in which information can enter, but nothing can come out from it. The partition that separates this region (known as {\it black hole}) is usually called the {\it{event horizon}}. The black holes are usually formed from the collapse of star etc. According to the No Hair theorem, {\it{collapse leads to a black hole endowed with
small number of macroscopic parameters (mass, charge,
angular momentum) with no other free parameters}}. All these are classical pictures.

         Hawking showed that {\it the area of a black hole never decreases} - known as {\it area theorem} \cite{Hawking:1971tu}. This fact attracted Bekenstein a lot. A simple thought experiment led him to associate entropy with the black hole. Then he \cite{Bekenstein:1973ur} proposed that a black hole has an entropy $S_{bh}$ which is some finite multiple $\eta$ of its area of the event horizon $A$. He was not able to determine the exact value of $\eta$, but gave heuristic arguments for conjecturing that it was $\frac{ln 2}{8\pi}$. Also, several investigations reveled that
classical black hole mechanics can be summarized by the following three basic
 laws \cite{Bardeen:1973gs},
\begin{enumerate}
 \item  Zeroth law : The surface gravity $\kappa$ of a black hole is constant
on the horizon.
\item First law :  The variations in the black hole parameters, i.e  mass $M$, area $A$, angular momentum
$L$, and charge $Q$, obey 
\begin{equation}
\delta M = \frac{\kappa}{8\pi} \delta A + \Omega \delta L - V \delta Q 
\label{1.02}
\end{equation}
where $\Omega$ and $V$ are the angular velocity and the electrostatic potential, respectively. 
\item Second law : The area of a black hole horizon $A$ is nondecreasing in time \cite{Hawking:1971tu},
\begin{equation}
\delta A \geq 0.
\label{1.03}
\end{equation}
\end{enumerate} 
These laws have a close resemblance to the corresponding
 laws of thermodynamics. The zeroth law of thermodynamics
says that the temperature is constant throughout a system in thermal
equilibrium. The first law states that in small variations between equilibrium
configurations of a system, the changes in the energy and entropy
of the system obey equation (\ref{1.02}), if the surface gravity $\kappa$
 is replaced by a term proportional to temperature of the system (other terms on the right hand side are interpreted as work terms).
 The second law of thermodynamics states that, for a closed system, entropy always increases
in any (irreversible or reversible) process.
Therefore from Bekenstein's argument and the first law of black hole mechanics one might say $T_H=\epsilon\kappa$ and $S_{bh}=\eta A$ with $8\pi\eta\epsilon=1$. Bekenstein proposed that $\eta$ is finite and it is equal to $\frac{ln 2}{8\pi}$. Then one would get $\epsilon=\frac{1}{ln 2}$ and so $T_H=\frac{\kappa}{ln 2}$.

      Later on, the study of QFT in curved space-time by Hawking in 1974-75 \cite{Hawking:1974rv,Hawking:1974sw} showed that black holes are not really black, instead emit thermal radiation at temperature ($T_H$) proportional to surface gravity ($\kappa$) of black hole - popularly known as {\it{Hawking effect}}. The exact expression was found to be
\cite{Hawking:1974sw}: 
\begin{eqnarray}
T_H=\frac{\hbar c \kappa}{2\pi k_B},
\label{1.01}
\end{eqnarray}
where $c$, $\hbar$ and $k_B$ are respectively the velocity of light, plank constant and Boltzmann constant. This is known as {\it Hawking temperature} {\footnote{Although in (\ref{1.01}) we keep all the fundamental constants explicitly, for later analysis, whenever any particular unit will be chosen, that will be mentioned there.}}. 
For the Schwarzschild black hole $\kappa= \frac{c^2}{4GM}$ where $M$ is the mass of the black hole and $G$ is the gravitational constant. All these reflects the fact that Hawking effect incorporates quantum mechanics, gravity as well as thermodynamics.  
The key idea behind quantum particle production in curved space-time
is that the definition of a particle is vacuum dependent. It depends on the
choice of reference frame. Since the theory is generally
covariant, any time coordinate, possibly defined only locally within
a patch, is a legitimate choice with which to define positive and negative
 frequency modes. Hawking considered a massless quantum scalar field moving in the background of a collapsing
 star. If the quantum field was initially in the vacuum state (no particle state) defined in the asymptotic past, then at late times it will appear as if particles are present in that state. Hawking showed \cite{Hawking:1974sw}, by explicit computation of the Bogoliubov coefficients (see also  \cite{birrell,ford} for detailed calculation of Bogoliubov coefficients) between the two sets of vacuum states defined at asymptotic past and future respectively, that the spectrum of the emitted particles is identical to that of black body with the temperature (\ref{1.01}).
This remarkable discovery indeed helps us to get various physical information about the classically forbidden region inside the horizon. Since then people thought that the black holes may play a major role in the attempts to shed some light on the nature of quantum theory of gravity as the role played by atoms in early development of quantum mechanics. Hence QFT on curved space-time and Hawking effect attracted the physicists for their beauty and usefulness in various aspects.

       Hawking then realised that Bekenstein's idea was consistent. In fact, since the black hole temperature is given by (\ref{1.01}), $\epsilon=\frac{1}{2\pi}$ and hence $\eta=\frac{1}{4}$. This leads to the famous Bekenstein-Hawking area law for entropy of black hole
\begin{eqnarray}
S_{bh}=\frac{A}{4},
\label{1.04}
\end{eqnarray} 
where $A$ is the area of the event horizon {\footnote{Here all the fundamental constants are chosen to be unity.}}.
This astonishing result is obtained using the approximation that the matter field behaves quantum mechanically but the gravitational field (metric) satisfy the classical Einstein equation. This semi-classical approximation holds good for energies below the Planck scale \cite{Hawking:1974sw}. Although it is a semi-classical result, Hawking's computation is considered an important clue in the search for a theory of quantum gravity. Any theory of quantum gravity that is proposed must predict black hole evaporation.

     Apart from Hawking's original calculation there are other semi-classical approaches. We summarise these briefly.
S. Hawking and G. Gibbons, in 1977 \cite{Gibbons:1976ue} developed an approach
 based on the Euclidean quantum gravity. In this approach they computed
 an action for gravitational field, including the boundary term, on the
 complexified space-time. The purely imaginary values of this
 action gives a contribution of the metrics to the partition function for a grand
 canonical ensemble at Hawking temperature (\ref{1.01}). Using this, they were able to 
 show that the entropy associated with these metrics is always equal to (\ref{1.04}). Almost at the same time, Christensen and Fulling \cite{Christensen:1977jc}, by 
 exploiting the structure of trace anomaly, were able to obtain the expectation value 
 for each component of the stress tensor $\langle T_{\mu\nu} \rangle$, which eventually
 lead to the Hawking flux. This approach is exact in $(1+1)$ dimensions, however in $3+1$ dimensions, the requirements
 of spherical symmetry, time independence and covariant conservation are not sufficient
 to fix completely the flux of Hawking radiation in terms of the trace anomaly \cite{birrell,Christensen:1977jc}.
 There is an additional arbitrariness in the expectation values of the angular components of 
 the stress tensor.

        Later on, S. Robinson and F. Wilczek \cite{Robinson:2005pd,Iso:2006wa,Iso:2006ut} gave a new approach to compute the Hawking flux
 from a black hole. This approach is based on gauge and gravitational or diffeomorphism anomalies. Basic and  
 essential fact used in their analysis is that the theory of matter fields
 (scalar or fermionic) in the $3+1$ dimensional static black hole background 
 can effectively be represented, in the vicinity of event horizon, by an
 infinite collection of free massless $1+1$ dimensional fields, each propagating
 in the background of an effective metric given by the $r-t$ sector of full $3+1$ dimensional
 metric \footnote{Such a dimensional reduction of matter fields 
 has been already used in the analysis of \cite{Carlip:2006fm,Solodukhin:1998tc} to compute the entropy of
 $2+1$ dimensional $BTZ$ black hole.}. By definition the horizon is a null surface and hence
 the region inside it is causally disconnected from the exterior. Thus, in the region near
 to the horizon the modes which are going into the black hole do not affect the 
 physics outside the horizon. In other words, the theory near the event horizon acquires 
   a definite chirality. Any two dimensional chiral theory in general curved background
 possesses both gauge and gravitational anomaly \cite{AlvarezGaume:1983ig}. This anomaly is manifested in the nonconservation
 of the current or the stress tensor. The theory far away from the event horizon
 is $3+1$ dimensional and anomaly free and the stress tensor in this region satisfies
 the usual conservation law. Consequently, the total energy-momentum tensor, which is a sum of two contribution
 from the two different regions, is also anomalous. However, it becomes anomaly free once we 
 take into account the contribution from classically irrelevant ingoing modes. This imposes 
 restrictions on the structure of the energy-momentum tensor and is ultimately responsible
 for the Hawking radiation \cite{Robinson:2005pd}. The expression for energy-momentum flux
 obtained by this anomaly cancellation approach is in exact agreement with the 
 flux from the perfectly black body kept at Hawking temperature \cite{Robinson:2005pd}.
%Soon this analysis was extended to compute Hawking fluxes 
% from the Reissner-Nordstrom (charged) black hole \cite{Iso:2006wa} and Kerr (rotating)
% black hole \cite{Iso:2006ut}. 
In this approach they used consistent expression for anomaly (satisfying Wess-Zumino consistency condition) 
%and not in covariant
%form) with 
but used a covariant boundary condition.

     Recently, a technically simple (only one Ward identity) and conceptually cleaner (covariant expression for anomaly with
covariant BC) derivation of Hawking flux was introduced by
Banerjee and Kulkarni \cite{Banerjee:2007qs,Banerjee:2008az}. In addition to this, a new method \cite{Banerjee:2007uc}, to obtain the Hawking flux using chiral effective action, was put forwarded by them. In all these approaches, the covariant boundary condition is applied by hand. Later on, it was shown again by them that such a boundary condition is compatible with the choice of Unruh vacuum \cite{Banerjee:2008wq}. 
%  It is worth to note that there are certain similarities between the trace \cite{fulling}
% and the gravitational \cite{robwilczek} anomaly method. Both the approaches uses two inputs: the 
% usual conservation law,  and the trace \cite{fulling} or gravitational 
% \cite{robwilczek} anomaly. Further, since the structure of trace as well as gravitational 
% anomaly, apart from the overall multiplicative factor, is identical for different field species
% (e.g scalar,fermionic etc. ) the methodology of two approaches would not alter for different field
% species. However, the analysis of \cite{fulling} is
% restricted to $1+1$ dimensional conformal fields. In this sense the anomaly cancellation 
% method \cite{robwilczek} is more appealing compared to the trace anomaly approach \cite{fulling}
% \footnote{Comparison among these two approaches and their connection with the $W-infinity$ algebra
% has been discussed in detail by L. Bonora and collaborators \cite{bonora1,bonora2}.}.
The connection of the diffeomorphism anomaly approach with the earlier trace anomaly approach \cite{Christensen:1977jc} was also elaborated \cite{Bonora:2008he,Bonora:2008nk}.

            Interestingly, none of the existing approaches to study Hawking effect, however, corresponds directly to one of the heuristic pictures that visualises the source of radiation as tunneling, first stated in \cite{Hawking:1974sw}. Later on, this picture was mathematically introduced to discuss the Hawking effect \cite{Paddy,Wilczek}. This picture is similar to an electron-positron pair creation in a constant electric field. The idea is that pair production occurs inside the event horizon of a black hole. One member of the pair corresponds to the ingoing mode and other member corresponds to the outgoing mode. The outgoing mode is allowed to follow classically forbidden trajectories, by starting just behind the horizon onward to infinity. So this mode travels back in time, since the horizon is locally to the future of the external region. The actual physical picture is that the tunneling occures by the shrinking of the horizon so that the particle effectively moves out. Thus the classical one particle action becomes complex and so the tunneling amplitude is governed by the imaginary part of this action for the outgoing mode. However, the action for the ingoing mode must be real, since classically a particle can fall behind the horizon. This is an important point of this mechanism as will be seen later. Also, since it is a near horizon theory and the tunneling occures radially, the phenomenon is effectively dominated by the two dimensional ($t-r$) metric. This follows form the fact that near the horizon all the angular part can be neglected and the solution of the field equation corresponds to angular quantum number $l=0$ which is known as $s$-wave \cite{Paddy}. Hence, the essence of tunneling based calculations is, thus, the computation of the imaginary part of the action for the process of $s$-wave emission across the horizon, which in turn is related to the Boltzmann factor for the emission at the Hawking temperature. It also reveals that the presence of the event horizon is necessary and the Hawking effect is a completely quantum mechanical phenomenon. There are two different methods in literature to calculate the imaginary part of the action: one is by Srinivasan et al \cite{Paddy} - {\it the Hamilton-Jacobi (HJ) method} {\footnote{For more elaborative discussions and further development on HJ method see \cite{Shankaranarayanan:2000gb,Shankaranarayanan:2000qv,Shankaranarayanan:2003ya}.}} and another is {\it radial null geodesic method} which was first given by  Parikh - Wilczek \cite{Wilczek} {\footnote{To find the basis of this method see \cite{KW1,KW2,KW3}.}}. Both these approaces will be discussed in this thesis.

       Historically, another phenomenon was discovered by Unruh \cite{Unruh:1976db} - Known as {\it Unruh effect} - in an attempt to understand the physics underlying the {\it Hawking effect} \cite{Hawking:1974sw}. The basic idea of the Unruh effect is based on the {\it equivalence principle} - locally gravitational effect can be ignored by choosing a uniformly accelerated frame and the observers with different notions of positive and negative frequency modes will disagree on the particle content of a given state. A uniformly accelerated observer on the Minkowski space-time percives a horizon. The space-time seen by the observer is known as {\it Rindler space-time} and so the observer is usually called as the {\it Rindler observer}. Although, an inertial observer would describe the Minkowski vacuum as being completely empty, the Rindler observer will detect particles in that vacuum. A detailed calculation tells that the emission spectrum exactly matches with that of the black body with the temperature given by \cite{Unruh:1976db},
\begin{eqnarray}
T = \frac{{\hbar {\tilde a}}}{2\pi}
\label{1.08}
\end{eqnarray}
where $a$ is the accleration of the Rindler observer. The similarity with Hawking temperature is obvious with $a\rightarrow\kappa$. It is now
well understood that Hawking effect is related to the event horizon of a black hole intrinsic to the space-time geometry while Unruh effect connects the horizon associated with a uniformly accelerated observer on the Minkowski space-time.

      A unified description of them was first put forward by Deser and Levin \cite{Deser:1997ri,Deser:1998xb} followed from an earlier attempt \cite{Narnhofer:1996zk}. This is called the global embedding Minkowskian space (GEMS) approach. 
In this approach, the relevant detector in curved space-time (namely Hawking detector) and its event horizon map to the Rindler detector in the corresponding higher dimensional flat embedding space \cite{Goenner,Rosen} and its event horizon. 
Then identifying the acceleration of the Unruh detector and using (\ref{1.08}), the Unruh temperature (or local Hawking temperature) was calculated. Finally, use of the Tolman relation \cite{Tolman} yields the Hawking temperature. Subsequently, this unified approach to determine the Hawking temperature using the Unruh effect was applied for several black hole space-times \cite{Kim:2000ct,Tian:2005yj,Brynjolfsson:2008uc}. However the results were confined to four dimensions and the calculations were done case by case, taking specific black hole metrics. It was not clear whether the technique was applicable to complicated examples like the Kerr-Newman metric which lacks spherical symmetry.

       In the mean time, after the discovery of Hawking effect, it was believed that the black holes may give some hints to find the 
%play a major role in our attempts to shed some light on the nature of 
quantum theory of gravity. It is then natural to consider quantization of a black hole. This was first pioneered by Bekenstein \cite{Bekenstein:1974jk,Bekenstein:1998aw}. The idea was based on the remarkable observation that the horizon area of a non-extremal black hole behaves as a classical {\it{adiabatic invariant}} quantity. In the spirit of the {\it Ehrenfest principle}, any classical adiabatic invariant corresponds to a quantum entity with discrete spectrum, Bekenstein conjectured that the horizon area of a non-extremal black hole should have a discrete eigenvalue spectrum. To elucidate the spacing of the area levels he used Christodoulou's reversible process \cite{Christodoulou:1970wf} - the assimilation of a {\it{neutral point particle}} by a non-extremal black hole. Bekenstein pointed out that the limit of a point particle is not a legal one in quantum theory. Because, according to the {\it Heisenberg's uncertainty principle}, the particle cannot be both at the horizon and at a turning point of its motion. Considering a finite size of the particle - not smaller than the Compton wavelength - he found a lower bound on the increase in the black hole surface area \cite{Bekenstein:1973ur,Bekenstein:1974ax}:
\begin{eqnarray}
(\Delta A)_{min} = 8\pi l_p^2
\label{1.05}
\end{eqnarray}
where $l_p = (\frac{G\hbar}{c^3})^{1/2}$ is the Planck length (we use gravitational units in which $G=c=1$). The independence of the black hole parameters in the lower bound shows its universality and hence it is a strong evidence in favor of a uniformly spaced area spectrum for a quantum black holes.

       These ideas led to a new research direction; namely the derivation of
the area and thus the entropy spectrum of black holes  utilizing  the quasinormal modes (QNM) of black holes \cite{Hod:1998vk}.  According to this method, since QNM frequencies are the characteristic of the black hole itself, the latter must have an adiabatic invariant quantity. Its form is given by energy of the black hole divided by this frequency, as happens in classical mechanics. Hod showed for Schwarzschild black hole that if one considers the real part of the QNM frequency only, then this adiabatic invariant quantity is actually related to area of the black hole horizon. Now use of {\it Bohr-Sommerfield quantization rule} gives the spectrum for the area which is equispaced. Then by the well known Bekenstein-Hawking area law, the entropy spectrum is obtained. In this case the spacing of this entropy spectrum is given by $\Delta S_{bh}=\ln 3$.
Another significant attempt was to fix the Immirzi parameter in the framework of Loop Quantum Gravity  \cite{Dreyer:2002vy} but it was unsuccessful \cite{Domagala:2004jt}.
 Later on Kunstatter \cite{Kunstatter} gave an explicit form of the adiabatic invariant quantity for the black hole:
\begin{eqnarray}
I_{adiab} = \int\frac{dW}{\Delta f(W)}, \,\,\,\ \Delta f = f_{n+1}-f_n
\label{1.06}
\end{eqnarray}
where `$W$' and `$f$' are the energy and the frequency of the QNM respectively. The Borh-Sommerfield quantization rule is given by,
\begin{eqnarray}
I_{adiab}=n\hbar
\label{1.07}
\end{eqnarray}
which is valid for semi-classical (large $n$) limit. For the real part of the frequency of QNM, (\ref{1.06}) can be shown to be related to black hole entropy which, ultimately by (\ref{1.07}), yields the entropy spectrum. For Schwarzschild black hole it yields the same spacing as obtained by Hod \cite{Hod:1998vk}. This, however, disagrees with Bekenstein's result, $\Delta S_{bh}=2\pi$ \cite{Bekenstein:1973ur}. In a recent work \cite{Maggiore:2007nq}, Maggiore told that a black hole behaves like a damped harmonic oscillator whose frequency is given by $f=(f_R^2 + f_I^2)^{\frac{1}{2}}$, where $f_R$ and $f_I$ are the real and imaginary parts of the frequency of the QNM. In the large $n$ limit $f_I>> f_R$. Consequently one has to use $f_I$ rather than $f_R$ in the adiabatic quantity (\ref{1.06}). It then leads to Bekenstein's result. With this new interpretation, entropy spectrum for the most general black hole has been calculated in \cite{Vagenas:2008yi}, which leads to an identical conclusion. In addition, it has been tested that the entropy spectrum is equidistance even for more general gravity theory (e.g. Einstein-Gauss-Bonnet theory), but that of area is not alaways equispaced, particularly, if the entropy is not proportional to area \cite{Daw}. In this sense quantization of entropy is more fundamental than that of area.

       A universal feature for black hole solutions, in a wide class of theories, is that the notions of entropy and temperature can be attributed to them \cite{Bekenstein:1973ur,Hawking:1974rv,Bardeen:1973gs,Jacobson:1993xs}. 
Also, of all forces of nature gravity is clearly the most universal. Gravity influences and is influenced by everything that carries an energy, and is intimately connected with the structure of space-time. The universal nature of gravity is also demonstrated by the fact that its basic equations closely resemble the laws of thermodynamics \cite{Bardeen:1973gs,Jacobson:1993xs,Jacobson:1995ab,Kothawala:2007em}. So far, there has not been a clear explanation for this resemblance. Gravity is also considerably harder to combine with quantum mechanics than all the other forces. The quest for unification of gravity with these other forces of nature, at a microscopic level, may therefore not be the right approach. It is known to lead to many problems, paradoxes and puzzles. 
%String theory has to a certain extent solved some of these, but
%not all. 
Many physicists believe that gravity and space-time geometry are emergent. Also string theory and its related developments have given several indications in this direction. Particularly important clues come from the AdS/CFT correspondence. This correspondence leads to a duality between theories that contain gravity and those that don't. It therfore provides evidence for the fact that gravity can emerge from a microscopic description that doesn't know about its existence {\footnote{Such a prediction was first given long ago by Sakharov \cite{Sakharov:1967pk}.}}. The universality of gravity suggests that its emergence should be understood from general principles that are independent of the specific details of the underlying microscopic theory.

\section{\label{structure} Outline of the thesis}
This thesis, based on the work \cite{Banerjee:2008cf,Banerjee:2008gc,Banerjee:2008ry,Banerjee:2008sn,Banerjee:2009wb,Banerjee:2010ma,Banerjee:2009pf,Banerjee:2010yd}, is focussed towards the applications of field theory, classical as well as quantum, to study black holes -- mainly the Hawking effect. This is discussed by the quantum tunneling mechanism. Here we give a general frame work of the existing tunneling mechanism, both the radial null geodesic and Hamilton -- Jacobi methods.  
 On the radial null geodesic method side, we study the modifications to the tunneling rate, Hawking temperature and the Bekenstein-Hawking area law by including the back reaction as well as non-commutative effects in the space-time.

     A major part of the thesis is devoted to the different aspects of the Hamilton-Jacobi (HJ) method. A reformulation of this method is first introduced. Based on this, a close connection between the quantum tunneling and the gravitational anomaly mechanisms to discuss Hawking effect, is put forwarded. An interesting advantage of this reformulated HJ method is that one can get directly the emission spectrum from the event horizon of the black hole, which was missing in the earlier literature. Also, the quantization of the entropy and area of a black hole is discussed in this method.

      Another part of the thesis is the introduction of a new type of global embedding of curved space-time to higher dimensional Minkowskian space-time (GEMS). Using this a unified description of the Hawking and Unruh effects is given. Advantage of this approach is, it simplifies as well as generalises the conventional embedding. In addition to the spherically symmetric space-times, the Kerr-Newman black hole is exemplified.

     Finally, following the above ideas and the definition of partition function for gravity, it is shown that extremization of entropy leads to the Einstein's equations of motion. In this frame work, a relation between the entropy, energy and the temperature of a black hole is given where energy is shown to be the Komar expression. Interestingly, this relation is the generalized Smarr formula. In this analysis, the GEMS method provides the {\it law of equipartition of energy} as an intermediate step.
\vskip 7mm

  {\it The whole thesis is consists of $9$ - chapters, including this introductory part. Chapter wise summary is given below}.
\vskip 7mm

{\underline{\bf{Chapter}} -\ref{chap:tunneling}: {\bf The tunneling mechanism}:
\vskip 1mm
      In this chapter, we present a general framework of tunneling mechanism within the semi-classical approximation. The black hole is considered to be a general static, spherically symmetric one. First, the HJ method is discussed both in Schwarzschild like coordinates and Painleve coordinates. Then a general methodology of the radial null geodesic method is presented. Here the tunneling rate, which is related to the imaginary part of the action, is shown to be equal to the exponential of the entropy change of the black hole. In both the methods, a general expression for Hawking temperature is obtained, which ultimately reduces to the Hawking expression (\ref{1.01}). Finally, using this general expression, calculation of Hawking temperature for some particular black hole metrics, is explicitly done.       
\vskip 3mm
{\underline{\bf{Chapter}} -\ref{chap:nullgeo}: {\bf Null geodesic approach}:
\vskip 1mm
      In this chapter, we provide an application of the general frame work, discussed in the previous chapter, for the radial null geodesic method, to incorporate back reaction as well as noncommutative effects in the space-time. Here the main motivation is to find their effects on the thermodynamic quantities. First, starting from a modified surface gravity of a black hole due to one loop back reaction effect, the tunneling rate is obtained. From this, the temperature and the area law are derived. The semi-classical Hawking temperature is altered. Interestingly, the leading order correction to the area law is logarithmic of the horizon area of the black hole while the non-leading corrections are the inverse powers of the area. The coefficient of the logarithmic term is related to the trace anomaly. Similar type of corrections were also obtained earlier \cite{Lousto:1988sp,Fursaev:1994te,Mann:1997hm,Kaul:2000kf,Govindarajan:2001ee,Das:2001ic,More:2004hv,Mukherji:2002de,Page:2004xp} by different methods.

      Next, we shall apply our general formulation to discuss various thermodynamic properties of a black hole defined in a noncommutative Schwarzschild space time where back reaction is also taken into account. In particular, we are interested in the black hole temperature when the radius is very small. Such a study is relevant because noncommutativity is supposed to remove the so called ``information paradox'' where for a standard black hole, temperature diverges as the radius shrinks to zero. The Hawking temperature is obtained in a closed form that includes corrections due to noncommutativity and back reaction. These corrections are such that, in some examples, the ``information paradox'' is avoided. Expressions for the entropy and tunneling rate are also found for the leading order in the noncommutative parameter.  Furthermore, in the absence of back reaction, we show that the entropy and area are algebraically related in the same manner as occurs in the standard Bekenstein-Hawking area law.   
       
\vskip 3mm
{\underline{\bf{Chapter}} -\ref{chap:anomaly}: {\bf Tunneling mechanism and anomaly}:
\vskip 1mm
Several existing methods to study Hawking effect yield similar results. The universality of this phenomenon naturally tempts us to find the underlying mechanism which unifies the different approaches. Recently, two widely used approaches – gravitational anomaly method and quantum tunneling method – can be described in a unified picture, since these two have several similarities in their techniques. One of the most important and crucial step in the tunneling approach (in both the methods) is that the tunneling of the particle occurs radially and its a near horizon phenomenon. This enforces that only the near horizon ($t-r$) sector of the original metric is relevant. Also, the ingoing mode is completely trapped inside the horizon. Similar step is also invoked in the gravitational (chiral) anomaly approach \cite{Robinson:2005pd,Iso:2006wa,Iso:2006ut,Banerjee:2007qs}. Here, since near the event horizon the theory is dominated by the two dimensional, ($t-r$) sector of the metric, and the ingoing mode is trapped inside the horizon, the theory is chiral. Hence one should has the gravitational anomaly in the quantum level. Therefore, one might thought that these two approaches - quantum tunneling and anomaly methods - can be discussed in an unified picture.

    We begin this exercise by introducing the chirality conditions on the modes and the energy-momentum tensor in chapter-\ref{chap:anomaly}. The Klein-Gordon equation under the effective ($t-r$) sector of the original metric shows that the there exits a general solution which is a linear combination of two solutions. One is left moving and function of only one null tortoise coordinate ($v$) while other is right moving which is function of the other null tortoise coordinate ($u$). From this information it is easy to find the chirality conditions. Then use of these conditions on the usual expressions for the anomaly in the non-chiral theory in two dimensions leads to the chiral anomaly expression. Finally, following the approach by Banerjee et al \cite{Banerjee:2007qs}, it is easy to find the expression for the Hawking flux. Another portion of this chapter is dedicated to show that the same chirality conditions are enough to find the Hawking temperature in quantum tunneling method. First, the Hamilton-Jacobi equations are obtained from these conditions, which are derived in the usual analysis from the field equations. Then a reformulation of tunneling method is given in which the trapping of the left mode is automatically satisfied. The right mode tunnels through the horizon with a finite probability which is exactly the Boltzmann factor. This immediately leads to the Hawking temperature. Thus, this analysis reflects the crucial role of the chirality to give a unified description of both the approaches to discuss Hawking effect.     

\vskip 3mm
{\underline{\bf{Chapter}} -\ref{chap:spectrum}: {\bf Black body spectrum from tunneling mechanism}:
\vskip 1mm 
   So far, in the tunneling mechanism only the Hawking temperature was obtained by comparing the tunneling rate with the Boltzmann factor. The discussion of the emission spectrum is absent and hence it is not clear whether this temperature really corresponds to the emission spectrum from the black hole event horizon. This shortcoming is addressed in chapter -\ref{chap:spectrum}. Following the modified tunneling approach, introduced in the previous chapter, the reduced density matrix for the outgoing particles, as seen from the asymptotic observer, is constructed. Then determination of the average number of outgoing particles yields the Bose or Fermi distribution depending on the nature of the particles produced inside the horizon. The distributions come out to be exactly similar to those in the case of black body radiation. It is now easy to identify the temperature corresponding to the emission spectrum. The temperature here we obtain is just the Hawking expression. Thereby we provide a complete description of the Hawking effect in the tunneling mechanism.

\vskip 3mm
{\underline{\bf{Chapter}} -\ref{chap:GEMS}: {\bf Global embedding and Hawking - Unruh Effect}:
\vskip 1mm
After Hawking's discovery, Unruh showed that an uniformly accelerated observer on the Minkowski space-time sees a thermal radiation from the Minkowski vacuum. Later on, Levin and Deser gave a unified picture of these two effects by using the globally embedding of the curved space-time in the higher dimensional Minkowski space-time. Such an interesting analysis was done using the embedding of the full curved metric and was confined within the spherically symmetric black hole space-time. The main difficulty to discuss for more general space-times is the finding of the embeddings.

    This issue is addressed in chapter - \ref{chap:GEMS}. Since, the thermodynamic quantities of a black hole are determined by the horizon properties and near the horizon the effective theory is dominated by the two dimensional ($t-r$) metric, it is sufficient to consider the embedding of this two dimensional metric. Considering this fact, a new type of global embedding of curved space-times in higher dimensional flat ones is introduced to present a unified description of Hawking and Unruh effects. Our analysis simplifies as well as generalises the conventional embedding approach.

\vskip 3mm
{\underline{\bf{Chapter}} -\ref{chap:spectroscopy}: {\bf Quantum tunneling and  black hole spectroscopy}:
\vskip 1mm
   The entropy-area spectrum of a black hole has been a long-standing and challenging problem. In chapter - \ref{chap:spectroscopy}, based on the modified tunneling mechanism, introduced in the previous chapters, we obtain the entropy spectrum of a  black hole. In {\it{Einstein's gravity}}, we show that both entropy and area spectrum are evenly spaced. But in more general theories (like {\it{Einstein-Gauss-Bonnet gravity}}), although the entropy spectrum is equispaced, the corresponding area spectrum is not. In this sense, quantization of entropy is more fundamental than that of area.

\vskip 3mm
{\underline{\bf{Chapter}} -\ref{chap:statistical}: {\bf Statistical origin of gravity}:
\vskip 1mm
    Based on the above conceptions and findings, we explore in chapter - \ref{chap:statistical} an intriguing possibility that gravity can be thought as an emergent phenomenon. Starting from the definition of entropy, used in statistical mechanics, we show that it is proportional to the gravity action. For a stationary black hole this entropy is expressed as $S_{bh} = E/ 2T_H$, where $T_H$ is the Hawking temperature and $E$ is shown to be the Komar energy. This relation is also compatible with the generalised Smarr formula for mass.
\vskip 3mm

{\underline{\bf{Chapter}} -\ref{chap:conclusions}: {\bf Conclusions}:
\vskip 1mm
   Finally, in chapter-\ref{chap:conclusions} we present our conclusion and outlook.

%%%%%%%%%%%%%%%%%%%%%%%%%%%%%%%%%%%%%%%%%%%%%%%%%%%%%%%%%%%%%%%%%%
\chapter{\label{chap:tunneling}The tunneling mechanism}
%%%%%%%%%%%%%%%%%%%%%%%%%%%%%%%%%%%%%%%%%%%%%%%%%%%%%%%%%%%%%%%%%%
  Classical general relativity gives the concept of black hole from which nothing can escape. This picture was changed dramatically when Hawking \cite{Hawking:1974rv,Hawking:1974sw} incorporated the quantum nature into this classical problem. In fact he showed that black hole radiates a spectrum of particles which is quite analogous with a thermal black body radiation, popularly known as Hawking effect. Thus Hawking radiation emerges as a nontrivial consequence of combining gravity and quantum mechanics. People then started thinking that this may give some insight towards quantum nature of gravity. Since the original derivation, based on the calculation of Bogoliubov coefficients in the asymptotic states, was technically very involved, several derivations of Hawking radiation were subsequently presented in the literature to give fresh insights. For example, Path integral derivation \cite{Gibbons:1976ue}, Trace anomaly approach \cite{Christensen:1977jc} and chiral (gravitational) anomaly approach \cite{Robinson:2005pd,Iso:2006wa,Iso:2006ut,Banerjee:2007qs,Banerjee:2008az,Banerjee:2007uc,Banerjee:2008wq}, each having its merits and demerits.

       Interestingly, none of the existing approaches to study Hawking effect, however, corresponds directly to one of the heuristic pictures that visualises the source of radiation as tunneling. This picture is similar to an electron-positron pair creation in a constant electric field. The idea is that pair production occurs inside the event horizon of a black hole. One member of the pair corresponds to the ingoing mode and other member corresponds to the outgoing mode. The outgoing mode is allowed to follow classically forbidden trajectories, by starting just behind the horizon onward to infinity. So this mode travels back in time, since the horizon is locally to the future of the external region. Unitarity is not violated since physically it is possible to envisage the tunneling as the shrinking of the horizon forwarded in time rather than the particle travelling backward in time \cite{Wilczek}. The classical one particle action becomes complex and so the tunneling amplitude is governed by the imaginary part of this action for the outgoing mode. However, the action for the ingoing mode must be real, since classically a particle can fall behind the horizon. Another essential fact is that tunneling occurs radially and it is a near horizon phenomenon where the theory is driven by only the effective ($t-r$) metric \cite{Paddy}. Under this circumstance the solution of a field equation corresponds to $l=0$ mode which is actually the $s$ - wave. These are all important points of this mechanism as will be seen later. The essence of tunneling based calculations is, thus, the computation of the imaginary part of the action for the process of $s$-wave emission across the horizon, which in turn is related to the Boltzmann factor for the emission at the Hawking temperature. Also, it reveals that the presence of the event horizon is necessary and the Hawking effect is a completely quantum mechanical phenomenon, determined by properties of the event horizon.

   There are two different methods in the literature to calculate the imaginary part of the action: one is by Parikh-Wilczek \cite{Wilczek} - {\it radial null geodesic method} and another is {\it the Hamilton-Jacobi (HJ) method} which was first used by Srinivasan et. al. \cite{Paddy}. Later, many people \cite{Jiang,Chen} used the radial null geodesic method as well as HJ method to find out the Hawking temperature for different space-time metrics. Also, several issues and aspects of these methods have been discussed extensively \cite{Medved1,Medved2,Borun,Singleton,Akhmedov,Douglas,Pilling,Nakamura,Banerjee:2008fz,Majhi:2008gi,Majhi:2009uk,Modak:2008tg,Banerjee:2009tz,Banerjee:2009sz,Banerjee:2009xx}.

    In this chapter, we will give a short review of both the HJ and radial null geodesic methods. While most of the material is available in the rather extensive literature \cite{Paddy,Wilczek,Shankaranarayanan:2000gb,Shankaranarayanan:2000qv,Shankaranarayanan:2003ya,Jiang,Chen,Medved1,Medved2,Borun,Singleton,Akhmedov,Douglas,Pilling,Nakamura,Banerjee:2008fz,Majhi:2008gi,Majhi:2009uk,Modak:2008tg,Banerjee:2009xx,Mitra:2006qa,Angheben:2005rm,Kerner:2006vu} on tunneling, there are some new insights and clarifications. The organization of the chapter is the following. First we will discuss the HJ method within a semi-classical approximation to find the Hawking temperature both in Schwarzschild like coordinate system and Painleve coordinate system. 
 A general static, spherically symmetric black hole metric will be considered. In the next section, the radial null geodesic method will be introduced. A general derivation of the Hawking temperature of this black hole will be presented. Both these expressions will be shown identical. Then using this obtained expression, the Hawking temperature will be explicitly calculated for some known black hole metrics. Final section will be devoted for the concluding remarks.  
%%%%%%%%%%%%%%%%%%%%%%%%%%%%%%%%%%%%%%%%%%%%%%%%%%%%%%%%%%%%%%%%%%%
\section{Hamilton-Jacobi method}
     Usually, calculations of the Hawking temperature, based on the tunneling formalism, for different black holes conform to the general formula $T_H=\frac{\hbar\kappa}{2\pi}$. This relation is normally understood as a consequence of the mapping of the second law of black hole thermodynamics $dM=\frac{\kappa}{8\pi}dA$ with $dE=T_H dS_{bh}$, coupled with the Bekenstein-Hawking area law $S_{bh}=\frac{A}{4\hbar}$.

    Using the tunneling approach, we now present a derivation of $T_H=\frac{\hbar\kappa}{2\pi}$ where neither the second law of black hole thermodynamics nor the area law are required. In this sense our analysis is general.

    In this section we will briefly discuss about the HJ method \cite{Paddy} to find the temperature of a black hole using the picture of Hawking radiation as quantum tunneling. The analysis will be restricted to the semi-classical limit. Equivalent results are obtained by using either the standard Schwarzschild like coordinates or other types, as for instance, the Painleve ones. We discuss both cases in this section.

\subsection{Schwarzschild like coordinate system}
   First, we consider a general class of static (i.e. invariant under time reversal as well as stationary), spherically symmetric space-time of the form
\begin{eqnarray}
ds^2 = -f(r)dt^2+\frac{dr^2}{g(r)}+r^2 d\Omega^2
\label{2.01}
\end{eqnarray}
where the horizon $r=r_H$ is given by $f(r_H)=g(r_H)=0$.

  Let us consider a massless particle in the space-time (\ref{2.01}) described by the massless Klein-Gordon equation
\begin{eqnarray}
-\frac{\hbar^2}{\sqrt{-g}}\partial_\mu[g^{\mu\nu}\sqrt{-g}\partial_{\nu}]\phi = 0~.
\label{2.02}
\end{eqnarray}
Since tunneling across the event horizon occurs radially, only the radial trajectories will be considered here. Also, it is an near horizon phenomenon and so the theory is effectively dominated by the two dimensional ($r-t$) sector of the full metric. Here the modes corresponds to angular quantum number $l=0$, which is actually the $s$ - wave \cite{Paddy,Robinson:2005pd,Iso:2006wa}. In this regard, only the $(r-t)$ sector of the metric (\ref{2.01}) is important. Therefore under this metric the Klein-Gordon equation reduces to
\begin{eqnarray}
-\frac{1}{\sqrt{f(r)g(r)}}\partial^2_t\phi +\frac{1}{2}\Big(f'(r)\sqrt{\frac{g(r)}{f(r)}}+g'(r)\sqrt{\frac{f(r)}{g(r)}}\Big)\partial_r\phi + \sqrt{f(r)g(r)}\partial_r^2\phi=0~.
\label{2.03}
\end{eqnarray}
The semi-classical wave function satisfying the above equation is obtained by making the standard ansatz for $\phi$ which is
\begin{eqnarray}
\phi(r,t)={\textrm{exp}}\Big[-\frac{i}{\hbar}S(r,t)\Big],
\label{2.04}
\end{eqnarray}  
where $S(r,t)$ is a function which will be expanded in powers of $\hbar$. Substituting into the wave equation (\ref{2.03}), we obtain
\begin{eqnarray}
&&\frac{i}{\sqrt{f(r)g(r)}}\Big(\frac{\partial S}{\partial t}\Big)^2 - i\sqrt{f(r)g(r)}\Big(\frac{\partial S}{\partial r}\Big)^2 - \frac{\hbar}{\sqrt{f(r)g(r)}}\frac{\partial^2 S}{\partial t^2} + \hbar \sqrt{f(r)g(r)}\frac{\partial^2 S}{\partial r^2}
\nonumber
\\
&&+ \frac{\hbar}{2}\Big(\frac{\partial f(r)}{\partial r}\sqrt{\frac{g(r)}{f(r)}}+\frac{\partial g(r)}{\partial r}\sqrt{\frac{f(r)}{g(r)}}\Big)\frac{\partial S}{\partial r}=0~.
\label{2.05}
\end{eqnarray}
Expanding $S(r,t)$ in a powers of $\hbar$, we find,
\begin{eqnarray}
S(r,t)&=&S_0(r,t)+\hbar S_1(r,t)+\hbar^2 S_2(r,t)+...........
\nonumber
\\
&=&S_0(r,t)+\sum_i \hbar^i S_i(r,t).
\label{2.06}
\end{eqnarray}
where $i=1,2,3,......$. In this expansion the terms from ${\cal{O}}(\hbar)$ onwards are treated as quantum corrections over the semi-classical value $S_0$. Here, as mentioned earlier, we will restrict only upto the semi-classical limit, i.e. $\hbar\rightarrow 0$. The effects due to inclusion of higher order terms are discussed in \cite{Banerjee:2008cf,Banerjee:2008fz,Majhi:2008gi,Majhi:2009uk,Modak:2008tg,Banerjee:2009tz,Banerjee:2009sz,Banerjee:2009xx} {\footnote{For extensive literature on the discussion of the higher order terms see \cite{Zhang:2008gt,Siahaan:2008cj}}}.

     Substituting (\ref{2.06}) in (\ref{2.05}) and taking the semi-classical limit $\hbar\rightarrow 0$, we obtain the following equation:
\begin{eqnarray}
\frac{\partial S_0}{\partial t}=\pm \sqrt{f(r)g(r)}\frac{\partial S_0}{\partial r}~.
\label{2.07}
\end{eqnarray}
This is the usual semi-classical Hamilton-Jacobi equation \cite{Paddy}. Now, to obtain a solution for $S_0(r,t)$, we will proceed in the following manner. Since the metric (\ref{2.01}) is static it has a time-like Killing vector. Thus we will look for a solution of (\ref{2.07}) which behaves as 
\begin{eqnarray}
S_0=\omega t + \tilde S_0(r),
\label{2.08}
\end{eqnarray} 
where $\omega$ is the conserved quantity corresponding to the time-like Killing vector. This ultimately is identified as the energy of the particle as seen by an observer at infinity. Substituting this in (\ref{2.07}) and then integrating we obtain,
\begin{eqnarray}
\tilde{S_0}(r) =  \pm \omega\int\frac{dr}{\sqrt{f(r)g(r)}}
\label{2.09}
\end{eqnarray} 
where the limits of the integration are chosen such that the particle just goes through the horizon $r=r_H$. So the one can take the range of integration from $r=r_H-\epsilon$ to $r=r_H+\epsilon$, where $\epsilon$ is a very small constant. The $+ (-)$ sign in front of the integral indicates that the particle is ingoing ($L$) (outgoing ($R$)) (For elaborate discussion to determine the nature of the modes, see Appendix \ref{appendixmode}). Using (\ref{2.09}) in (\ref{2.08}) we obtain
\begin{eqnarray}
S_0(r,t)= \omega t  \pm\omega\int\frac{dr}{\sqrt{f(r)g(r)}}~.
\label{2.10}
\end{eqnarray}  
Therefore the ingoing and outgoing solutions of the Klein-Gordon equation (\ref{2.02}) under the back ground metric (\ref{2.01}) is given by exploiting (\ref{2.04}) and (\ref{2.10}),
\begin{eqnarray}
\phi^{(L)}= {\textrm{exp}}\Big[-\frac{i}{\hbar}\Big(\omega t  +\omega\int\frac{dr}{\sqrt{f(r)g(r)}}\Big)\Big]
\label{2.11}
\end{eqnarray} 
and
\begin{eqnarray}
\phi^{(R)}= {\textrm{exp}}\Big[-\frac{i}{\hbar}\Big(\omega t  -\omega\int\frac{dr}{\sqrt{f(r)g(r)}}\Big)\Big].
\label{2.12}
\end{eqnarray}
In the rest of the analysis we will call $\phi^{(L)}$ as the left mode and $\phi^{(R)}$ as the right mode.

      A point we want to mention here that if one expresses the above modes in terms of null coordinates ($u,v$), then $\phi^{(L)}$ becomes function of ``$v$'' only while $\phi^{(R)}$ becomes that of ``$u$''. These are call holomorphic modes. Such modes satisfies chirality condition. This will be elaborated and used in the later discussions.

     Now for the tunneling of a particle across the horizon the nature of the coordinates change. The time-like coordinate $t$ outside the horizon changes to space-like coordinate inside the horizon and likewise for the outside space-like coordinate $r$. This indicates that `$t$' coordinate may have an imaginary part on crossing the horizon of the black hole and correspondingly there will be a temporal contribution to the probabilities for the ingoing and outgoing particles along with the spacial part. This has similarity with \cite{Akhmedov} where they show for the Schwarzschild metric that two patches across the horizon are connected by a discrete imaginary amount of time.

     The ingoing and outgoing probabilities of the particle are, therefore, given by,
\begin{eqnarray}
P^{(L)}=|\phi^{(L)}|^2= {\textrm{exp}}\Big[\frac{2}{\hbar}(\omega{\textrm{Im}}~t +\omega{\textrm{Im}}\int\frac{dr}{\sqrt{f(r)g(r)}}\Big)\Big]
\label{2.13}
\end{eqnarray}
and
\begin{eqnarray}
P^{(R)}=|\phi^{(R)}|^2= {\textrm{exp}}\Big[\frac{2}{\hbar}\Big(\omega{\textrm{Im}}~t -\omega{\textrm{Im}}\int\frac{dr}{\sqrt{f(r)g(r)}}\Big)\Big]
\label{2.14}
\end{eqnarray}
Now the ingoing probability $P^{(L)}$ has to be unity in the classical limit (i.e. $\hbar\rightarrow 0$) - when there is no reflection and everything is absorbed - instead of zero or infinity \cite{Mitra:2006qa}.Thus, in the classical limit, (\ref{2.13}) leads to,
\begin{eqnarray}
{\textrm{Im}}~t = -{\textrm{Im}}\int\frac{dr}{\sqrt{f(r)g(r)}}~.
\label{2.16}
\end{eqnarray}
It must be noted that the above relation satisfies the classical condition $\frac{\partial S_0}{\partial\omega}=$ constant. This is understood by the following argument. Calculating the left side of this condition from (\ref{2.10}) we obtain,
\begin{eqnarray}
t = {\textrm{constant}}\mp \int\frac{dr}{\sqrt{f(r)g(r)}}
\label{2.17}
\end{eqnarray}
where $-(+)$ sign indicates that the particle is ingoing ($L$) (outgoing ($R$)). So for an ingoing particle this condition immediately yields (\ref{2.16}) considering that ``$\textrm{constant}$'' is always real. On the other hand a naive substitution of `Im$~t$' in (\ref{2.14}) from (\ref{2.17}) for the outgoing particle gives $P^{(R)}=1$. But it must be noted that according to classical general theory of relativity, a particle can be absorbed in the black hole, while the reverse process is forbidden. In this regard, ingoing classical trajectory exists while the outgoing classical trajectory is forbidden. Hence use of the classical condition for outgoing particle is meaningless.

    Now to find out `Im$~t$' for the outgoing particle we will take the help of the Kruskal coordinates which are well behaved throughout the space-time.
The Kruskal time ($T$) and space ($X$) coordinates
inside and outside the horizon are defined in terms of Schwarzschild coordinates as \cite{Ray}
\begin{eqnarray}
&&T_{in}=e^{\kappa r^{*}_{in}} \cosh\!\left(\kappa t_{in}\right)~~;\hspace{4ex}
X_{in} = e^{\kappa r^{*}_{in}} \sinh\!\left(\kappa t_{in}\right)
\label{2.18}
\\
&&T_{out}=e^{\kappa r^{*}_{out}} \sinh\!\left( \kappa t_{out}\right)~~;\hspace{4ex}
X_{out} = e^{\kappa r^{*}_{out}}  \cosh\!\left(\kappa t_{out}\right)
\label{2.19}
\end{eqnarray}
where $\kappa$ is the surface gravity defined by
\begin{eqnarray}
\kappa = \frac{1}{2} \sqrt{f'(r_H)g'(r_H)}~.
\label{2.20}
\end{eqnarray}
Here `$in (out)$' stands for inside (outside) the event horizon while $r^*$ is the tortoise coordinate, defined by
\begin{eqnarray}
r^* = \int\frac{dr}{\sqrt{f(r)g(r)}} ~.
\label{2.21}
\end{eqnarray}
These two sets of coordinates are connected through the following relations
\begin{eqnarray}
&& t_{in} = t_{out}-i\frac{\pi}{2\kappa}
\label{2.22}\\
&& r^{*}_{in} = r^{*}_{out} + i\frac{\pi}{2\kappa}
\label{2.23}
\end{eqnarray}
so that the Kruskal coordinates get identified as $T_{in} = T_{out}$ and $X_{in} = X_{out}$. This indicates that when a particle travels from inside to outside the horizon, `$t$' coordinate picks up an imaginary term $-\frac{\pi}{2{\kappa}}$. This fact will be used elaborately in later chapters. Below we shall show that this is precisely given by (\ref{2.16}). Near the horizon one can expand $f(r)$ and $g(r)$ about the horizon $r_H$:
\begin{eqnarray}
&&f(r)=f'(r_H)(r-r_H)+{\cal{O}}((r-r_H)^2)
\nonumber
\\
&&g(r)=g'(r_H)(r-r_H)+{\cal{O}}((r-r_H)^2)~.
\label{2.28}
\end{eqnarray}
Substituting these in (\ref{2.16}) and using (\ref{2.20}) we obtain,
\begin{eqnarray}
{\textrm{Im}}~ t = -\frac{1}{2\kappa} {\textrm{Im}}~\int_{r_H-\epsilon}^{r_H+\epsilon} \frac{dr}{r-r_H}~.
\label{2.28n1}
\end{eqnarray}
Here we explicitly mentioned the integration limits. Now to evaluate the above integration we make a substitution $r-r_H = \epsilon e^{i\theta}$ where $\theta$ runs from $\pi$ to $2\pi$. Hence,
\begin{eqnarray}
{\textrm{Im}}~ t = -\frac{1}{2\kappa} {\textrm{Im}}~\int_{\pi}^{2\pi} id\theta = -\frac{\pi}{2\kappa}~.
\label{2.28n2}
\end{eqnarray}
For the Schwarzschild space-time, since $\kappa=\frac{1}{4M}$, one can easily show that ${\textrm{Im}}~t = -2\pi M$ which is precisely the value given in \cite{Akhmedov}.

   Therefore, substituting (\ref{2.16}) in (\ref{2.14}), the probability of the outgoing particle is
\begin{eqnarray}
P^{(R)}={\textrm{exp}}\Big[-\frac{4}{\hbar}\omega{\textrm{Im}}\int\frac{dr}{\sqrt{f(r)g(r)}}\Big].
\label{2.24}
\end{eqnarray}
Now using the principle of ``detailed balance'' \cite{Paddy}
\begin{eqnarray}
P^{(R)}= {\textrm {exp}}\Big(-\frac{\omega}{T_H}\Big)P^{(L)}={\textrm{exp}} \Big(-\frac{\omega}{T_H}\Big),
\label{2.25}
\end{eqnarray}
we obtain the temperature of the black hole as
\begin{eqnarray}
T_H = \frac{\hbar}{4}\Big({\textrm{Im}}\int\frac{dr}{\sqrt{f(r)g(r)}}\Big)^{-1}~.
\label{2.26}
\end{eqnarray}
This is the standard semi-classical Hawking temperature of the black hole. Using this expression and knowing the metric coefficients $f(r)$ and $g(r)$ one can easily find out the temperature of the corresponding black hole.

      Some comments are now in order. The first point is that (\ref{2.26}) yields a novel form of the semi-classical Hawking temperature. Also, we will later show that (\ref{2.26}) can be applied to non-spherically symmetric metrics. This will be exemplified in the case of the Kerr metric. For a spherically symmetric metric it is possible to show that (\ref{2.26}) reproduces the familiar form 
\begin{eqnarray}
T_H=\frac{\hbar\kappa}{2\pi}~.
\label{2.27} 
\end{eqnarray}
This can be done in the following way. The near horizon expansions for $f(r)$ and $g(r)$ are given by (\ref{2.28}).
Inserting these in (\ref{2.26}) and performing the contour integration, as done earlier, (\ref{2.27}) is obtained. Note that this form is the standard Hawking temperature found \cite{Angheben:2005rm,Kerner:2006vu} by the Hamilton-Jacobi method. There is no ambiguity regarding a factor of two in the Hawking temperature as reported in the literature \cite{Angheben:2005rm,Kerner:2006vu,Stotyn:2008qu}. This issue is completely avoided in the present analysis where the standard expression for the Hawking temperature is reproduced.

       The other point is that the form of the solution (\ref{2.08}) of (\ref{2.07}) is not unique, since any constant multiple of `$S_0$' can be a solution as well. For that case one can easily see that the final expression (\ref{2.26}) for the temperature still remains unchanged. It is only a matter of rescaling the particle energy `$\omega$'. This shows the uniqueness of the expression (\ref{2.26}) for the Hawking temperature.

\subsection{Painleve coordinate system}
     Here we will discuss the Hamilton-Jacobi method in Painleve coordinates and explicitly show how one can obtain the standard Hawking temperature.
Consider a metric of the form (\ref{2.01}),
which describes a general class of static, spherically symmetric space time. There is a coordinate singularity in this metric at the horizon $r=r_H$ where $f(r_H)=g(r_H)=0$. This singularity is avoided by the use of Painleve coordinate transformation \cite{Painleve}, 
\begin{eqnarray}
dt\to dt-\sqrt{\frac{1-g(r)}{f(r)g(r)}}dr~.
\label{2.29} 
\end{eqnarray}
Under this transformation, the metric (\ref{2.01}) takes the following form,
\begin{eqnarray}
ds^2 = -f(r)dt^2+2f(r)\sqrt{\frac{1-g(r)}{f(r)g(r)}} dt dr+dr^2+r^2d\Omega^2.
\label{2.30}
\end{eqnarray}
Note that the metric (\ref{2.01}) looks both stationary and static, whereas the transformed metric (\ref{2.30}) is stationary but not static which reflects the correct nature of the space time. 

   As before, consider a massless scalar particle in the spacetime metric (\ref{2.30}) described by the Painleve coordinates. Since the Klein-Gordon equation (\ref{2.02}) is in covariant form, the scalar particle in the background metric (\ref{2.30}) also satisfies (\ref{2.02}). Therefore under this metric the Klein-Gordon equation reduces to
\begin{eqnarray}
&-&(\frac{g}{f})^{\frac{3}{2}}\partial^2_t\phi+\frac{2g\sqrt{1-g}}{f}\partial_t\partial_r \phi-\frac{gg'}{2f\sqrt{1-g}}\partial_t\phi+g\sqrt{\frac{g}{f}}\partial^2_r\phi
\nonumber
\\
&+&\frac{1}{2}\sqrt{\frac{g}{f}}(3g'-\frac{f'g}{f})\partial_r\phi=0.
\label{2.30n1}
\end{eqnarray}
As before, substituting the standard ansatz (\ref{2.04}) for $\phi$ in the above equation, we obtain,
\begin{eqnarray}
&&-(\frac{g}{f})^{\frac{3}{2}}\Big[-\frac{i}{\hbar}\Big(\frac{\partial S}{\partial t}\Big)^2+\frac{\partial^2S}{\partial t^2}\Big] + \frac{2g\sqrt{1-g}}{f}\Big[-\frac{i}{\hbar}\frac{\partial S}{\partial t}\frac{\partial S}{\partial r}+\frac{\partial^2S}{\partial r\partial t}\Big]-\frac{gg'}{2f\sqrt{1-g}}\frac{\partial S}{\partial t}
\nonumber
\\
&&+g\sqrt{\frac{g}{f}}\Big[-\frac{i}{\hbar}\Big(\frac{\partial S}{\partial r}\Big)^2+\frac{\partial^2S}{\partial r^2}\Big]+\frac{1}{2}(3g'-\frac{f'g}{f})\frac{\partial S}{\partial r}=0.
\label{2.30n2}
\end{eqnarray}
Substituting (\ref{2.06}) in the above and then neglecting the terms of order $\hbar$ and greater we find to the lowest order,
\begin{eqnarray}
(\frac{g}{f})^{\frac{3}{2}}\Big(\frac{\partial S_0}{\partial t}\Big)^2-\frac{2g\sqrt{1-g}}{f}\frac{\partial S_0}{\partial t}\frac{\partial S_0}{\partial r}-g\sqrt{\frac{g}{f}}\Big(\frac{\partial S_0}{\partial r}\Big)^2=0.
\label{2.30n3}
\end{eqnarray}
It has been stated earlier that the metric (\ref{2.30}) is stationary. Therefore following the same argument as before it has a solution of the form (\ref{2.08}). Inserting this in (\ref{2.30n3}) yields,
\begin{eqnarray}
\frac{d\tilde S_0(r)}{dr}=\omega\sqrt{\frac{1-g(r)}{f(r)g(r)}}\Big(-1\pm\frac{1}{\sqrt{1-g(r)}}\Big)
\label{2.30n4}
\end{eqnarray}
Integrating,
\begin{eqnarray}
\tilde S_0(r)=\omega\int\sqrt{\frac{1-g(r)}{f(r)g(r)}}\Big(-1\pm\frac{1}{\sqrt{1-g(r)}}\Big)dr.
\label{2.30n5}
\end{eqnarray}
The $+(-)$ sign in front of the integral indicates that the particle is ingoing (outgoing). Therefore the actions for ingoing and outgoing particles are
\begin{eqnarray}
S_0^{(L)}(r,t)=\omega t +\omega\int\frac{1-\sqrt{1-g}}{\sqrt{fg}}dr
\label{2.30n6}
\end{eqnarray}
and
\begin{eqnarray}
S_0^{(R)}(r,t)=\omega t -\omega\int\frac{1+\sqrt{1-g}}{\sqrt{fg}}dr
\label{2.30n7}
\end{eqnarray}
Since in the classical limit (i.e. $\hbar\rightarrow 0$) the probability for the ingoing particle ($P^{(L)}$) has to be unity, $S_0^{(L)}$ must be real. Following identical steps employed in deriving (\ref{2.16}) we obtain, starting from (\ref{2.30n6}), the analogous condition,
\begin{eqnarray}
{\textrm{Im}}~t=-{\textrm{Im}}\int\frac{1-\sqrt{1-g}}{\sqrt{fg}}dr
\label{2.30n8}
\end{eqnarray}
Substituting this in (\ref{2.30n7}) we obtain the action for the outgoing particle:
\begin{eqnarray}
S_0^{(R)}(r,t)=\omega{\textrm{Re}}~t -\omega{\textrm{Re}}\int\frac{1+\sqrt{1-g}}{\sqrt{fg}}dr-2i\omega{\textrm{Im}}\int\frac{dr}{\sqrt{fg}}
\label{2.30n9}
\end{eqnarray}
Therefore the probability for the outgoing particle is
\begin{eqnarray}
P^{(R)}=|e^{-\frac{i}{\hbar}S_0^{(R)}}|^2=e^{-\frac{4}{\hbar}\omega{\textrm{Im}}\int\frac{dr}{\sqrt{f(r)g(r)}}}
\label{2.30n10}
\end{eqnarray}
Now using the principle of ``detailed balance'' (\ref{2.25}) we obtain the same expression (\ref{2.26}) for the standard Hawking temperature which was calculated in Schwarzschild like coordinates by the Hamilton-Jacobi method. 
%%%%%%%%%%%%%%%%%%%%%%%%%%%%%%%%%%%%%%%%%%%%%%%%%%%%%%%%%%%%%
\section{Radial null geodesic method}
    So far, we gave a general discussion on the HJ method both in Scwarzschild like coordinates as well as Painleve coordinates and obtained the expression of the temperature for a static, spherically symmetric black hole. Also, this has been reduced to the famous Hawking expression - temperature is proportional to the surface gravity.

    In this section, we will give a general discussion on the radial null geodesic method. A derivation of the Hawking temperature by this method will be explicitly performed for the metric (\ref{2.01}).

   In this method, the first step is to find the radial null geodesic. To do that it is necessary to remove the apparent singularity at the event horizon. This is done by going to the Painleve coordinates. In these coordinates, the metric (\ref{2.01}) takes the form (\ref{2.30}). Then the radial null geodesics are obtained by setting $ds^2=d\Omega^2=0$ in (\ref{2.30}),
\begin{eqnarray}
\dot{r}\equiv\frac{dr}{dt}=\sqrt{\frac{f(r)}{g(r)}}\Big(\pm 1-\sqrt{1-g(r)}\Big)
\label{2.31}
\end{eqnarray}
where the positive (negative) sign gives outgoing (incoming) radial geodesics. At the neighbourhood of the black hole horizon, the trajectory (\ref{2.31}) of an outgoing shell is written as,
\begin{eqnarray}
\dot{r}=\frac{1}{2}\sqrt{f'(r_H)g'(r_H)}(r-r_H)+{\mathcal O}((r-r_H)^2)
\label{2.32}
\end{eqnarray}
where we have used the expansions (\ref{2.28}) of the functions $f(r)$ and $g(r)$. Now we want to write (\ref{2.32}) in terms of the surface gravity of the black hole. The reason is that in some cases, for example in the presence of back reaction, one may not know the exact form of the metric but what one usually knows is the surface gravity of the problem. Also, the Hawking temperature is eventually expressed in terms of the surface gravity. The form of surface gravity for the transformed metric (\ref{2.30}) at the horizon is given by, 
\begin{eqnarray}
\kappa=\Gamma{^0}{_{00}}|_{r=r_H}=\frac{1}{2}\Big[\sqrt{\frac{1-g(r)}{f(r)g(r)}}g(r)\frac{df(r)}{dr}\Big]|_{r=r_H}.
\label{2.33}
\end{eqnarray}
Using the Taylor series (\ref{2.28}), the above equation reduces to the familiar form of surface gravity (\ref{2.20}). This expression of surface gravity is used to write (\ref{2.32}) in the form,
\begin{eqnarray}
\dot{r}=\kappa (r-r_H)+{\mathcal O}((r-r_H)^2).
\label{2.34}
\end{eqnarray}

We consider a positive energy shell which crosses the horizon in the outward direction from $r_{{\textrm{in}}}$ to $r_{{\textrm{out}}}$. The imaginary part of the action for that shell is given by \cite{Wilczek},
 \begin{eqnarray}
\textrm{Im}~ {\cal{S}} =\textrm{Im} \int_{r_{{\textrm{in}}}}^{r_{{\textrm{out}}}} p_r dr = \textrm{Im}  \int_{r_{in}}^{r_{out}} \int_{0}^{p_r} dp_r'dr.
\label{2.35}
\end{eqnarray}
Using the Hamilton's equation of motion $\dot{r}=\frac{dH}{dp_r}|_r$ the last equality of the above equation is written as,
\begin{eqnarray}
\textrm{Im}~ {\cal{S}} 
&=& \textrm{Im} \int_{r_{{\textrm{in}}}}^{r_{{\textrm{out}}}} \int_{0}^{H} \frac{dH'}{\dot{r}} dr
\label{2.36}
\end{eqnarray}
where, instead of momentum, energy is used as the variable of integration.

     Now we consider the self gravitation effect \cite{KW1} of the particle itself, for which (\ref{2.34}) and (\ref{2.36}) will be modified. Following \cite{Wilczek}, under the $s$- wave approximation, we make the replacement $M\rightarrow M-\omega$ in (\ref{2.34}) to get the following expression
\begin{equation}
\dot{r}=(r-r_H)\kappa[M-\omega]
\label{2.37}
\end{equation}
where $\omega$ is the energy of a shell moving along the geodesic of space-time {\footnote{Here $\kappa[M-\omega]$ represents that $\kappa$ is a function of ($M-\omega$). This symbol will be used in the later part of the chapter for a similar purpose.}}.

   Now we use the fact \cite{Wilczek}, for a black hole of mass $M$, the Hamiltonian $H=M-\omega$. Inserting in (\ref{2.36}) the modified expression due to the self gravitation effect is obtained as,
\begin{eqnarray}
\textrm{Im} ~{\cal{S}} = \textrm{Im} \int_{r_{{\textrm{in}}}}^{r_{{\textrm{out}}}} \int_{M}^{M-\omega} \frac{d(M-\omega')}{\dot{r}} dr  =-\textrm{Im} \int_{r_{{\textrm{in}}}}^{r_{{\textrm{out}}}} \int_{0}^{\omega} \frac{d\omega'}{\dot{r}} dr
\label{2.38}
\end{eqnarray}
where in the final step we have changed the integration variable from $H'$ to $\omega'$. Substituting the expression of $\dot r$ from (\ref{2.37}) into (\ref{2.38}) we find,
\begin{eqnarray}
\textrm{Im} ~{\cal{S}} =  -\textrm{Im} \int_{0}^{\omega} \frac{d\omega'}{{\kappa}[M-\omega']}\int_{r_{{\textrm{in}}}}^{r_{{\textrm{out}}}}\frac{dr}{r-r_H}~.
\label{2.39}
\end{eqnarray}
The $r$-integration is done by deforming the contour. Ensuring that the positive energy solutions decay in time (i.e. into the lower half of $\omega'$ plane and $r_{{\textrm{in}}}>r_{{\textrm{out}}}$) we have after $r$ integration{\footnote{One can also take the contour in the upper half plane with the replacement $M\rightarrow M+\omega$ \cite{KW1}.}}, 
\begin{eqnarray}
\textrm{Im}~ {\cal{S}}= \pi\int_{0}^{\omega} \frac{d\omega'}{{\kappa}[M-\omega']}~. 
\label{2.40}
\end{eqnarray}
To understand the ordering $r_{{\textrm{in}}}>r_{{\textrm{out}}}$ - which supplies the correct sign, let us do the following analysis. For simplicity, we consider the Schwarzschild black hole whose surface gravity is given by $\kappa[M] = \frac{1}{4M}$. Substituting this in (\ref{2.40}) and performing the $\omega'$ integration we obtain 
\begin{eqnarray}
{\textrm{Im}}~{\cal{S}} = 4\pi\omega (M-\frac{\omega}{2}).
\label{2.40n1}
\end{eqnarray}
Now let us first perform the $\omega'$ integration before $r$ integration in (\ref{2.39}). For Schwarzschild black hole this will give
\begin{eqnarray}
{\textrm{Im}}~{\cal{S}} = 4 ~{\textrm{Im}}~ \int_{r_{\textrm{in}}}^{r_{\textrm{out}}} dr \int_M^{M-\omega}\frac{M'}{r-2M'}dM'
\label{2.40n2}
\end{eqnarray}
where substitution of $M'=M-\omega'$ has been used.
Evaluation of $M'$ integration and then $r$ integration in the above yields,
\begin{eqnarray}
{\textrm{Im}}~{\cal{S}} = \frac{\pi}{2} (r_{\textrm{in}}^2 - r_{\textrm{out}}^2)
\label{2.40n3}
\end{eqnarray}
Hence (\ref{2.40n1}) and (\ref{2.40n3}) to be equal we must have $r_{\textrm{in}} = 2M$ and $r_{\textrm{out}} = 2(M-\omega)$, which clearly shows that $r_{{\textrm{in}}}>r_{{\textrm{out}}}$.

    The tunneling amplitude following from the WKB approximation is given by,
\begin{eqnarray}
\Gamma\sim e^{-\frac{2}{\hbar}{\textrm{Im}}~{\cal{S}}}=e^{\Delta S_{bh}}
\label{2.41}
\end{eqnarray}
where the result is expressed more naturally in terms of the black hole entropy change \cite{Wilczek}. To understand the last identification ($\Gamma=e^{\Delta S_{bh}}$),
consider a process where a black hole emits a shell of energy. We denote the initial state and final state by the levels $i$ and $f$. In thermal equilibrium,
\begin{eqnarray}
\frac{dP_i}{dt}=P_i P_{i\rightarrow f}-P_f P_{f\rightarrow i}=0
\label{2.42}
\end{eqnarray}
where $P_a$ denotes the probability of getting the system in the macrostate $a (a=i,f)$ and $P_{a\rightarrow b}$ denotes the transition probability from the state $a$ to $b$ ($a,b=i,f$). According to statistical mechanics, the entropy of a given state (specified by its macrostates) is a logarithmic function of the total number of microstates ($S_{bh}={\textrm {log}}\Omega$). So the number of microstates $\Omega$ for a given black hole is $e^{S_{{bh}}}$. Since the probability of getting a system in a particular macrostate is proportional to the number of microstates available for that configuration, we get from (\ref{2.42}), 
\begin{eqnarray}
e^{S_i} P_{{\textrm {emission}}}=e^{S_f}P_{{\textrm {absorption}}}
\label{2.43}
\end{eqnarray}  
where $P_{{\textrm {emission}}}$ is the emission probability $P_{i\rightarrow f}$ and $P_{{\textrm {absorption}}}$ is the absorption probability $P_{f\rightarrow i}$. So the tunneling amplitude is given by,
\begin{eqnarray}
\Gamma = \frac{P_{{\textrm {emission}}}}{P_{{\textrm {absorption}}}} = e^{S_f-S_i}=e^{\Delta S_{{bh}}}
\label{2.44}
\end{eqnarray}
thereby leading to the correspondence,
\begin{eqnarray}
\Delta S_{bh} = -\frac{2}{\hbar}{\textrm{Im}}~ {\cal{S}} 
\label{2.45}
\end{eqnarray}
that follows from (\ref{2.41}). We mention that the above relation (\ref{2.45}) has also been shown using semi-classical arguments based on the second law of thermodynamics \cite{Sarkar:2007sx} or on the assumption of entropy being proportional to area \cite{KW3,Pilling:2007cn}. But such arguments are not used in our derivation. Rather our analysis has some points of similarity with the physical picture suggested in \cite{Wilczek} leading to a general validity of (\ref{2.44}).  
This implies that when quantum effects are taken into consideration, both sides of (\ref{2.45}) are modified keeping the functional relationship identical. In our analysis we will show that self consistency is preserved by (\ref{2.45}).

    In order to write the black hole entropy in terms of its mass alone we have to substitute the value of $\omega$ in terms of $M$ for which the black hole is stable i. e.
\begin{eqnarray}
\frac{d(\Delta S_{bh})}{d\omega}=0~.
\label{2.46}
\end{eqnarray}
Using (\ref{2.40}) and (\ref{2.45}) in the above equation we get,
\begin{eqnarray}
\frac{1}{{\kappa}[M-\omega]}=0~.
\label{2.47}
\end{eqnarray}
The roots of this equation are written in the form
\begin{eqnarray}
\omega=\psi[M]
\label{2.48}
\end{eqnarray}
which means
\begin{eqnarray}
\frac{1}{{\kappa}[M-\psi[M]]}=0.
\label{2.49}
\end{eqnarray}
This value of $\omega$ from eq. (\ref{2.48}) is substituted back in the expression of $\Delta S_{bh}$ to yield,
\begin{eqnarray}
\Delta S_{bh}=-\frac{2\pi}{\hbar}\int_{0}^{\psi[M]}\frac{d\omega'}{{\kappa}[M-\omega']}~.
\label{2.50}
\end{eqnarray}
Having obtained the form of entropy change, we are now able to give an expression of entropy for a particular state. We recall the simple definition of entropy change
\begin{eqnarray}
\Delta S_{bh}=S_{{\textrm{final}}}-S_{{\textrm{initial}}}~.
\label{2.51}
\end{eqnarray}
Now setting the black hole entropy at the final state to be zero we get the expression of entropy as
\begin{eqnarray}
S_{bh}=S_{{\textrm{initial}}}=-\Delta S_{bh}=\frac{2\pi}{\hbar}\int_{0}^{\psi[M]} \frac{d\omega'}{{\kappa}[M-\omega']}~.
\label{2.52} 
\end{eqnarray}
From the law of thermodynamics,
we write the inverse black hole temperature as,
\begin{eqnarray}
\frac{1}{T_H}&=&\frac{dS_{bh}}{dM}
\label{2.53}
\\&=&\frac{2\pi}{\hbar}\frac{d}{dM}\int_0^{\psi[M]}\frac{d\omega'}{{\kappa}[M-\omega']}~.
\end{eqnarray}
Using the identity,
\begin{eqnarray}
\frac{dF[x]}{dx}=f[x,b[x]]b'[x]-f[x,a[x]]a'[x]+\int_{a[x]}^{b[x]}\frac{\partial}{\partial x}f[x,t]dt
\label{2.54}
\end{eqnarray}
for,
\begin{eqnarray}
F[x]=\int_{a[x]}^{b[x]}f[x,t]dt
\label{2.55}
\end{eqnarray}
we find,
\begin{eqnarray}
\frac{1}{T_H}=\frac{2\pi}{\hbar}\big[\frac{1}{{\kappa}[M-\psi[M]]}\psi'[M]-\int_{0}^{\psi[M]}\frac{1}{[{\kappa}[M-\omega']]^2}\frac{\partial {\kappa}[M-\omega']}{\partial (M-\omega')}d\omega'
\big]~.
\label{2.56}
\end{eqnarray}
Making the change of variable $x=M-\omega'$ in the second integral we obtain,
\begin{eqnarray}
\frac{1}{T_H}=\frac{2\pi}{\hbar}\big[\frac{\psi'[M]-1}{{\kappa}[M-\psi[M]]}+\frac{1}{{\kappa}[M]}\big]~.
\label{2.57}
\end{eqnarray}
Finally, making use of (\ref{2.49}), the cherished expression (\ref{2.27}) for the Hawking temperature follows.

    For a consistency check, consider the second law of thermodynamics which is now written as,
\begin{eqnarray}
dM=d\omega'=T_hdS_{bh}=\frac{\hbar{\kappa}[M]}{2\pi}dS_{bh}~.
\label{2.58}
\end{eqnarray}
Inserting in (\ref{2.40}), yields,
\begin{eqnarray}
{\textrm {Im}}~{\cal S}=\frac{\hbar}{2}\int_{S_{bh}[M]}^{S_{bh}[M-\omega]}dS_{bh}=-\frac{\hbar}{2}\Delta S_{bh}
\label{2.59}
\end{eqnarray}
thereby reproducing (\ref{2.45}). This shows the internal consistency of the tunneling approach.
%%%%%%%%%%%%%%%%%%%%%%%%%%%%%%%%%%%%%%%%%%%%%%%%%%%%%%%%%%%%%
\section{Calculation of Hawking temperature}
   In this section we will consider some standard metrics to show how the semi-classical Hawking temperature can be calculated from (\ref{2.26}). For instance we consider a spherically symmetric space-time, the Schwarzschild metric and a non-spherically symmetric space-time, the Kerr metric.
\subsection{Schwarzschild black hole}
The spacetime metric is given by
\begin{eqnarray}
ds^2=-(1-\frac{2M}{r})dt^2+(1-\frac{2M}{r})^{-1}dr^2+r^2d\Omega^2.
\label{2.60}
\end{eqnarray}
So the metric coefficients are
\begin{eqnarray}
f(r)=g(r)=(1-\frac{r_H}{r});\,\,\,r_H = 2M.
\label{2.61}
\end{eqnarray}
Since this metric is spherically symmetric we use the formula (\ref{2.26}) to compute the semi-classical Hawking temperature. This is found to be, 
\begin{eqnarray}
T_H=\frac{\hbar}{4\pi r_H}=\frac{\hbar}{8\pi M}.
\label{2.62}
\end{eqnarray}
which is the standard expression (\ref{2.27}) where the surface gravity, calculated by (\ref{2.20}), is $\kappa=1/4M$.
\subsection{Kerr black hole}
This example provides a nontrivial application of our formula (\ref{2.26}) for computing the semi-classical Hawking temperature. Here the metric is not spherically symmetric, invalidating the use of (\ref{2.27}).

     In Boyer-Linquist coordinates the form of the Kerr metric is given by
\begin{eqnarray}
ds^2&=&-\Big(1-\frac{2Mr}{\rho^2}\Big)dt^2-\frac{2Mar~{\textrm {sin}}^2\theta}{\rho^2}(dt d\phi+d\phi dt)
\nonumber
\\
&+&\frac{\rho^2}{\Delta}dr^2+\rho^2d\theta^2+\frac{{\textrm {sin}}^2\theta}{\rho^2}~\Big[(r^2+a^2)^2-a^2\Delta~{\textrm {sin}}^2\theta\Big]d\phi^2
\label{2.63}
\end{eqnarray}
where
\begin{eqnarray}
\Delta(r)&=&r^2-2Mr+a^2;\,\,\,\rho^2(r,\theta)=r^2+a^2~{\textrm{cos}}^2\theta
\nonumber
\\
a&=&\frac{J}{M}
\label{2.64}
\end{eqnarray}
and $J$ is the Komar angular momentum. We have chosen the coordinates for Kerr metric such that the event horizons occur at those fixed values of $r$ for which $g^{rr}=\frac{\Delta}{\rho^2}=0$. Therefore the event horizons are
\begin{eqnarray}
r_\pm = M\pm\sqrt{M^2-a^2}.
\label{2.65}
\end{eqnarray}
This metric is not spherically symmetric and static but stationary. So it must have time-like Killing vectors. Subtleties in employing the tunneling mechanism for such (rotating) black holes were first discussed in \cite{Angheben:2005rm,Kerner:2006vu}. In the present discussion we will show that, although the general formulation was based only on the static, spherically symmetric metrics, it is still possible to apply this methodology for such a metric. The point is that for radial trajectories, the Kerr metric simplifies to the following form
\begin{eqnarray}
ds^2=-\Big(\frac{r^2+a^2-2Mr}{r^2+a^2}\Big)dt^2+\Big(\frac{r^2+a^2}{r^2+a^2-2Mr}\Big)dr^2
\label{2.67}
\end{eqnarray}
where, for simplicity, we have taken $\theta=0$ (i.e. particle is going along $z$-axis). This is exactly the form of the $(r-t)$ sector of the metric (\ref{2.01}). Since in our formalism only the $(r-t)$ sector is important, our results are applicable here. In particular if the metric has no terms like $(dr dt)$ then we can apply (\ref{2.26}) to find the semi-classical Hawking temperature. Here,
\begin{eqnarray}
f(r)=g(r)=\Big(\frac{r^2+a^2-2Mr}{r^2+a^2}\Big)
\label{2.68}
\end{eqnarray}
Substituting these in (\ref{2.26}) we obtain,
\begin{eqnarray}
T_H=\frac{\hbar}{4}\Big({\textrm{Im}}\int\frac{r^2+a^2}{(r-r_+)(r-r_-)}\Big)^{-1}.
\label{2.69}
\end{eqnarray}
The integrand has simple poles at $r=r_+$ and $r=r_-$. Since we are interested only with the event horizon at $r=r_+$, we choose the contour as a small half-loop going around this pole from left to right. Integrating, we obtain the value of the semi-classical Hawking temperature as
\begin{eqnarray}
T_H=\frac{\hbar}{4\pi}\frac{r_+ - r_-}{r_+^2+a^2}.
\label{2.70}
\end{eqnarray}
which is the result quoted in the literature \cite{Carrol}. This can also be expressed in standard expression (\ref{2.27}) where $\kappa =  \frac{r_+ - r_-}{2(r_+^2 + a^2)}$. 
%%%%%%%%%%%%%%%%%%%%%%%%%%%%%%%%%%%%%%%%%%%%%%%%%%%%%%%%%%%%%%%%
\section{Discussions}
      In this chapter, we introduced the tunneling method to study the Hawking effect within the semi-classical limit (i.e. $\hbar\rightarrow 0$), particularly to find the familiar form of the semi-classical Hawking temperature. There exist two approaches: Hamilton-Jacobi method \cite{Paddy} (HJ) and radial null geodesic method \cite{Wilczek}. For simplicity, a general form of the static, spherically symmetric black hole metric was considered.

     First, discussions on HJ method in both the Schwarzschild like coordinates and Painleve coordinates have been done. In both coordinate systems, we obtained identical results. A general expression (\ref{2.26}) for the semi-classical Hawking temperature was obtained. For the particular case of a spherically symmetric metric, our expression reduces to the standard form (\ref{2.27}). The factor of two problem in the Hawking temperature has been taken care of by considering the contribution from the imaginary part of the temporal coordinate since it changes its nature across the horizon. Also, this method is free of the rather ad hoc way of introducing an integration constant, as reported in \cite{Mitra:2006qa}. Our approach, on the other hand, is similar in spirit to \cite{Akhmedov} where it has been shown that `$t$' changes by an imaginary discrete amount across the horizon. Indeed, the explicit expression for this change, in the case of Schwarzschild metric, calculated from our general formula (\ref{2.16}), agrees with the findings of \cite{Akhmedov}. Then, a general discussion on the other method, the radial null geodesic method, was given. In this method, again the standard form of the Hawking temperature was obtained. Finally, as an application, we calculated the semi-classical temperature of the Schwarzschild black hole from the general expression (\ref{2.26}). Also, use of this expression to find the temperature of a non-spherically symmetric metric, for instance Kerr metric, has been shown.

   As a final remark, we want to mention that our derivation of Hawking temperature in terms of the surface gravity by considering the action of an outgoing particle crossing the black hole horizon due to quantum mechanical tunneling is completely general. The expression of temperature was known long before \cite{Bekenstein:1972tm,Bardeen:1973gs,Bekenstein:1973ur,Bekenstein:1974ax} from a comparison between two classical laws. One is the law of black hole thermodynamics which states that the mass change is proportional to the change of horizon area multiplied by surface gravity at the horizon. The other is the area law according to which the black hole entropy is proportional to the surface area of the horizon. The important point of our derivation is that it is not based on either of these two classical laws. The only assumption is that the metric is static and spherically symmetric. Hence it is useful to apply this method to study the Hawking effect for the black holes which incorporates both the back reaction and noncommutative effects but still are in static, spherically symmetric form. This will be done in the next chapter.
%%%%%%%%%%%%%%%%%%%%%%%%%%%%%%%%%%%%%%%%%%%%%%%%%%%%%%%%%%%
\begin{subappendices}
\chapter*{Appendix}
\section{\label{appendixmode}Ingoing and outgoing modes}
\renewcommand{\theequation}{2a.\arabic{equation}}
\setcounter{equation}{0}  % reset counter
    Our convention is such that, a mode will be called ingoing (outgoing) if its radial momentum eigenvalue is negative (positive). For a wave function $\phi$, the momentum eigenvalue equation is
\begin{eqnarray}
{\hat{p}_r} \phi = p_r \phi,
\label{app2.01}
\end{eqnarray}
where ${\hat{p}_r} = -i\hbar\frac{\partial}{\partial r}$. So according to our convention, if $p_r < 0$ for a mode, then it is ingoing and vice versa.

      Now the mode solutions are given by (\ref{2.11}) and (\ref{2.12}). So according to (\ref{app2.01}), the momentum eigenvalue for $\phi^{(L)}$ is $p_r^{(L)} = -\frac{\omega}{\sqrt{fg}}$ which is negative. So this mode is ingoing. Similarly, the momentum eigenvalue for $\phi^{(R)}$ mode comes out to be positive and hence it is a outgoing mode.
\end{subappendices}

%%%%%%%%%%%%%%%%%%%%%%%%%%%%%%%%%%%%%%%%%%%%%%%%%%%%%%%%%%%%%%%%%%
\chapter{\label{chap:nullgeo}Null geodesic approach}
%%%%%%%%%%%%%%%%%%%%%%%%%%%%%%%%%%%%%%%%%%%%%%%%%%%%%%%%%%%%%%%%%%
       In the previous chapter, a systematic analysis on tunneling mechanism, both by HJ and radial null geodesic methods, to find the Hawking temperature has been presented. The temperature was found to be proportional to the surface gravity of a black hole represented by a general static, spherically symmetric metric. This indicates that such an analysis can be extended to the cases in which the space-time metric is modified by effects like back reaction and noncommutivity, provided these are still in the static, spherically symmetric form.

       To investigate the last stage evolution of black hole evaporation back reaction in space-time has a significant influence. An approach to this problem
is to solve the semiclassical Einstein equations in which the matter fields including the graviton, are quantized at the one-loop level and coupled to (c -number) gravity through Einstein's equation. The space-time geometry $g_{\mu\nu}$, generates a non-zero vacuum expectation value of the energy-momentum
tensor ($<T_{\mu\nu}>$) which in turn acts as a source of curvature (this is the so-called "back-reaction problem"). With this energy momentum tensor and an ansatz for the metric, the solutions of Einstein's equation yields the metric solution, which is static and spherically symmetric \cite{Lousto:1988sp}. 
 Using the conformal anomaly method the modifications to the space-time metric by the one loop back reaction was computed \cite{York:1984wp,Lousto:1988sp}. Later it was shown \cite{Fursaev:1994te,Mann:1997hm} that the Bekenstein-Hawking area law was modified, in the leading order, by logarithmic corrections. Similar conclusions were also obtained by using quantum gravity techniques \cite{Kaul:2000kf,Page:2004xp,Ghosh:2004wq}. Likewise, corrections to the semi-classical Hawking temperature were derived \cite{Govindarajan:2001ee,Das:2001ic,More:2004hv,Mukherji:2002de}. 
%Therefore, a natural question that arises in the context of tunneling mechanism to study Hawking effect is the feasibility of this approach to include quantum corrections.

     It is known that for the usual cases, the Hawking temperature diverges as the radius of the event horizon decreases. This uncomfortable situation leads to the ``information paradox''. To avoid this one of the attempts is inclusion of the noncommutative effect in the space-time. There exits two methods: (i) directly take the space-time as noncommutative, $[x_\mu,x_\nu] = i\theta_{\mu\nu}$ and use Seibarg-Witten map to recast the gravitational theory (in noncommutative space) in terms of the corresponding theory in usual (commutative space) variables, and (ii) incorporate the effect of noncommutativity in the mass term of the gravitating object.

   In this chapter, we shall include the back reaction as well as noncommutative effects in the space-time metric. Following the radial null geodesic method presented in the previous chapter, the thermodynamic entities will be calculated. Although there have been sporadic attempts in this direction \cite{Medved1,Medved2} a systematic, thorough and complete analysis was lacking.

    The organization of the chapter is as follows. In the first section, we compute the corrections to the semi-classical tunneling rate by including the effects of self gravitation and back reaction. The usual expression found in \cite{Wilczek}, given in the Maxwell-Boltzmann form $e^{-\frac{\omega}{T_{H}}}$, is modified by a prefactor. This prefactor leads to a modified Bekenstein-Hawking entropy. The semi-classical Bekenstein-Hawking area law connecting the entropy to the horizon area is altered. As obtained in other approaches \cite{Fursaev:1994te,Mann:1997hm,Kaul:2000kf,Govindarajan:2001ee,Das:2001ic,More:2004hv,Mukherji:2002de,Page:2004xp}, the leading correction is found to be logarithmic while the nonleading one is a series in inverse powers of the horizon area (or Bekenstein-Hawking entropy). We also compute the appropriate modification to the Hawking temperature. Explicit results are given for the Schwarzschild black hole.

    Next, we shall apply our general formulation to discuss various thermodynamic properties of a black hole defined in a noncommutative Schwarzschild space time where back reaction is also taken into account. A short introduction of the noncommutative Schwarzschild black hole is presented at the beginning of the section (3.2). In particular we are interested in the black hole temperature when the radius is of the order $\sqrt{\theta}$, where $\theta$ is the noncommutative parameter. Such a study is relevant because noncommutativity is supposed to remove the so called information paradox where for a standard black hole, temperature diverges as the radius shrinks to zero. The Hawking temperature is obtained in a closed form that includes corrections due to noncommutativity and back reaction. These corrections are such that, in some examples, the information paradox is avoided. Expressions for the entropy and tunneling rate are also found for the leading order in the noncommutative parameter.  Furthermore, in the absence of back reaction, we show that the entropy and area are algebraically related in the same manner as occurs in the standard Bekenstein-Hawking area law.   

\section{Back reaction effect}
    In this section we shall derive the modifications in the Hawking temperature and Bekenstein-Hawking area law due to the one loop back reaction effect in the space-time. Back reaction is essentially the effect of the Hawking radiation on the horizon. For simplicity, only the Schwarzschild black hole will be considered. One way to include the back reaction effect into the problem is to solve Einstein's equation with an appropriate source. In this case one considers the renormalized energy-momentum tensor due to one loop back reaction effect on the right hand side of the Einstein's equation. Then solution of this equation gives the black hole metric given by the form (\ref{2.01}) \cite{Lousto:1988sp}. Therefore it is feasible to apply the tunneling method developed in the previous chapter for this case to find the modifications to the usual thermodynamical entities. Here our discussions will be based on the radial null geodesic method.

   Here our starting point is the expression for the imaginary part of the action (\ref{2.40}), since in the present problem the form of the modified surface gravity of the black hole is known. The modified surface gravity due to one loop back reaction effects is given by \cite{Lousto:1988sp},
\begin{eqnarray}
\kappa[M] = \kappa_0[M]\Big(1+\frac{\alpha}{M^2}\Big)
\label{3.01}
\end{eqnarray}
where $\kappa_0$ is the classical surface gravity at the horizon of the black hole. Such a form is physically dictated by simple scaling arguments. As is well known, a loop expansion is equivalent to an expansion in powers of the Planck constant $\hbar$. Therefore, the one loop back reaction effect in the surface gravity is written as,
\begin{eqnarray}
\kappa = \kappa_0 + \xi \kappa_0
\label{3.02}
\end{eqnarray}
where $\xi$ is a dimensionless constant having magnitude of the order $\hbar$.  Now, in natural units $G=c=k_B=1$, Planck lenght $l_p =$ Planck mass $M_p=\sqrt{\hbar}$ {\footnote{Planck length $l_p = \sqrt{\frac{\hbar G}{c^3}}$, Planck mass $M_p = \sqrt{\frac{\hbar c}{G}}$}}.  On the other hand, for Schwarzschild black hole, mass $M$ is the only macroscopic parameter. Therefore, $\xi$ must be function of $\frac{M_p}{M}$ which vanishes in the limit $M_p<<M$. Since, as stated earlier, $\xi$ is a dimensionless constant with magnitude of order $\hbar$, the leading term has the following quadratic form,
\begin{eqnarray}
\xi = \beta\frac{M_p^2}{M^2}~.
\label{3.03}
\end{eqnarray}
In the above, $\beta$ is a pure numerical factor. Taking $\alpha = \beta M_p^2$ and then substituting (\ref{3.03}) in (\ref{3.02}) we obtain (\ref{3.01}). The constant $\beta$ is related to the trace anomaly coefficient taking into account the degrees of freedom of the fields \cite{Christensen:1978gi,Lousto:1988sp,Fursaev:1994te}. Its explicit form is given by \cite{Christensen:1978gi,Fursaev:1994te}, 
\begin{eqnarray}
\beta=\frac{1}{360\pi}\Big(-N_0 - \frac{7}{4}N_{\frac{1}{2}} +13N_1 + \frac{233}{4}N_{\frac{3}{2}} - 212 N_2) 
\label{3.04}
\end{eqnarray} 
where $N_s$ denotes the number of fields with spin `$s$'.

    For the classical Schwarzschild space-time the metric coefficients are given by (\ref{2.61})
and so by equation (\ref{2.20})
the value of $\kappa_0[M]$ is
\begin{eqnarray}
\kappa_0[M] = \frac{f'(r_H=2M)}{2}=\frac{1}{4M}~.
\label{3.05}
\end{eqnarray}
Substituting (\ref{3.01}) with $\kappa_0$ is given by (\ref{3.05}) in (\ref{2.40}) and then integrating over $\omega'$ we have 
\begin{eqnarray}
Im~ {\cal{S}} = 4\pi\omega(M-\frac{\omega}{2})+2\pi\alpha \ln{\Big[\frac{(M-\omega)^2+\alpha}{M^2+\alpha}\Big]}~.
\label{3.06}
\end{eqnarray}
Now according to the WKB-approximation method the tunneling probability is given by (\ref{2.41}).
So the modified tunneling probability due to back reaction effects is, 
\begin{eqnarray}
\Gamma\sim \Big[1-\frac{2\omega(M-\frac{\omega}{2})}{M^2+\alpha}\Big]^{-\frac{4\pi\alpha}{\hbar}} e^{-\frac{8\pi\omega}{\hbar}(M-\frac{\omega}{2})} 
\label{3.07}
\end{eqnarray}
The exponential factor of the tunneling probability was previously obtained by Parikh and Wilczek \cite{Wilczek}. The factor before the exponential is new. It is actually due the effect of back reaction. It will eventually give the correction to the Bekenstein-Hawking entropy, area law and the Hawking temperature as will be shown below.

    It was shown in the previous chapter and also in the literature \cite{Wilczek,Pilling,Sarkar:2007sx} that a change in the Bekenstein-Hawking entropy due to the tunneling through the horizon is related to $Im~ {\cal{S}}$ by the relation (\ref{2.45}). 
Therefore the corrected change in Bekenstein-Hawking entropy is 
\begin{eqnarray}
\Delta S_{bh}=-\frac{8\pi\omega}{\hbar}(M-\frac{\omega}{2})-\frac{4\pi\alpha}{\hbar}\ln\Big[(M-\omega)^2+\alpha\Big]+\frac{4\pi\alpha}{\hbar}\ln(M^2+\alpha)
\label{3.08}
\end{eqnarray}
Next using the stability criterion $\frac{d(\Delta S_{bh})}{d\omega}=0 $ for the black hole, one obtains the following condition
\begin{eqnarray}
(\omega-M)^3=0 
\label{3.09}
\end{eqnarray}
which gives the only solution as $\omega=M$. Substituting this value of $\omega$ in (\ref{3.08}) we will have the change in entropy of the black hole from its initial state to final state:
\begin{eqnarray}
S_{final}-S_{initial}= -\frac{4\pi M^2}{\hbar}+\frac{4\pi\alpha}{\hbar}\ln{(\frac{M^2}{\alpha}+1)}~.
\label{3.10}
\end{eqnarray}
Setting $S_{final} = 0$, the Bekenstein-Hawking entropy of the black hole with mass $M$ is 
\begin{eqnarray}
S_{bh} = S_{initial} &=& \frac{4\pi M^2}{\hbar}-4\pi\beta\ln{(\frac{M^2}{\beta\hbar}+1)}
\label{3.11}
\end{eqnarray}
where we have substituted $\alpha=\beta M_p^2 = \beta\hbar$.

  Now the area of the black hole horizon given by
\begin{eqnarray}
A=4\pi r^2_H=16\pi M^2~.
\label{3.12}
\end{eqnarray}
Putting (\ref{3.12}) in (\ref{3.11}) and expanding the logarithm, we obtain the final form,
\begin{eqnarray}
S_{bh}&=&\frac{A}{4\hbar}-4\pi\beta\ln\frac{A}{4\hbar}-64\pi^2\hbar\beta^2\Big[\frac{1}{A}-\frac{16\pi\hbar\beta}{2A^2}+\frac{(16\pi\hbar\beta)^2}{3A^3}-.....\Big]
\nonumber
\\
&+&\textrm{const.(independent~ of~ $A$)}~.
\label{3.13}
\end{eqnarray}
The first term is the usual semi-classical area law \cite{Bekenstein:1973ur,Hawking:1974sw} and other terms are the corrections due to the one loop back reaction effect. The leading correction is the well known logarithmic correction \cite{Fursaev:1994te,Mann:1997hm,Kaul:2000kf,Govindarajan:2001ee,Das:2001ic,More:2004hv,Mukherji:2002de,Page:2004xp}. Quantum gravity calculations lead to a prefactor $-\frac{1}{2}$ for the $\ln\frac{A}{4\hbar}$ term which would correspond to choosing $\beta=\frac{1}{8\pi}$. But here on the contrary $\beta$ is given by (\ref{3.04}). Also, the nonleading corrections are found to be expressed as a series in inverse powers of $A$, exactly as happens in quantum gravity inspired analysis \cite{Kaul:2000kf,Page:2004xp}.
Now using the first law of black hole mechanics, 
$T_H dS_{bh}=dM$,
or the relation (\ref{2.27}) between the Hawking temperature and surface gravity,
we can find the corrected form of the Hawking temperature $T_H$ due to back reaction. This is obtained from (\ref{3.01}) as, 
\begin{eqnarray}
T_H = T_0\Big(1+\frac{\beta\hbar}{M^2}\Big) 
\label{3.14}
\end{eqnarray}
where $T_0 = \frac{\hbar\kappa_0}{2\pi} = \frac{\hbar}{8\pi M}$ is the semi-classical Hawking temperature and the other term is the correction due to the back reaction. A similar expression was obtained previously in \cite{Fursaev:1994te} by the conformal anomaly method.

   It is also possible to obtain the corrected Hawking temperature (\ref{3.14}) in the standard tunneling method to leading order \cite{Wilczek} where this temperature is read off from the coefficient of `$\omega$' in the exponential of the probability amplitude (\ref{3.07}). Recasting this amplitude as,
\begin{eqnarray}
\Gamma\sim e^{-\frac{8\pi\omega}{\hbar}(M-\frac{\omega}{2})-\frac{4\pi\alpha}{\hbar} \ln(1-\frac{2\omega(M-\frac{\omega}{2})}{M^2+\alpha})}
\label{3.15} 
\end{eqnarray} 
and retaining terms upto leading order in $\omega$, we obtain,
\begin{eqnarray}
\Gamma &\sim& e^{-\frac{8\pi M \omega}{\hbar}+4\pi\beta (\frac{2M\omega}{M^2+\beta\hbar})}
\nonumber
\\
&=&e^{-(\frac{8\pi M^3}{\hbar(M^2+\beta\hbar)})\omega}=e^{-\frac{\omega}{T_H}}. 
\end{eqnarray}
The inverse Hawking temperature, indentified with the coefficient of `$\omega$',
reproduces (\ref{3.14}).

   The above analysis showed how the effects of back reaction in the space-time can be discussed in a general frame work of tunneling mechanism. The only assumption was that the modified metric must be static, spherically symmetric. In particular, the modifications to the temperature and entropy for the Schwarzschild case were explicitly evaluated. The results agree with earlier findings by different methods. Next, we shall discuss the noncommutative effect in addition to the back reaction effect in the space-time using our general frame work. 
%%%%%%%%%%%%%%%%%%%%%%%%%%%%%%%%%%%%%%%%%%%%%%%%%%%%%%%
\section{Inclusion of noncommutativity}
   Here we shall apply our previous formulations to find the modifications to the Hawking temperature and Bekenstein-Hawking area law due to noncommutative as well as back reaction effects. In the vanishing limit of noncommutative parameter, the results reduce to those obtained in the previous section. First a short introduction on the noncommutative Schwarzschild black hole will be given. Then the modifications to the thermodynamic entities will be calculated.

\subsection{Schwarzschild black hole in noncommutative space}
       The fact is that gravitation is a manifestation of the structure of spacetime as dictated by the presence of gravitating objects. Therefore, inclusion of noncommutative effects in gravity can be done in two ways. Directly take the spacetime as noncommutative, $[x_\mu,x_\nu]=i\theta_{\mu\nu}$ and use the Seibarg-Witten map to recast the gravitational theory (in noncommutative space) in terms of the corresponding theory in usual (commutative space) variables. This leads to correction terms (involving powers of $\theta{\mu\nu}$) in the various expressions like the metric, Riemann tensor etc. This approach has been adopted in \cite{LopezDominguez:2006wd,Chaichian:2007we,Mukherjee:2007fa,Kobakhidze:2007jn,Banerjee:2007th} {\footnote{For a detailed discussions of this approach and a list of references see \cite{Banerjee:2009gr}.}}. Alternatively,  incorporate the effect of noncommutativity in the mass term of the gravitating object. Here the mass density, instead of being represented by a Dirac delta function, is replaced by a Gaussian distribution. This approach has been adopted in \cite{Nicolini:2005vd,Ansoldi:2006vg,Myung:2006mz,Banerjee:2008du,Nozari:2008rc,Nozari:2009nr} {\footnote{For a review and list of references, see \cite{Nicolini:2008aj}.}}. The two ways of incorporating noncommutative effects in gravity are, in general, not equivalent. Here we follow the second approach, for our investigation on the computation of thermodynamic entities and area law for the noncommutative Schwarzschild black hole.

     The usual definition of mass density in terms of the Dirac delta function in commutative space does not hold good in noncommutative space because of the position-position uncertainty relation. In noncommutative space mass density is defined by replacing the Dirac delta function by a Gaussian distribution of minimal width $\sqrt\theta$  in the following way \cite{Nicolini:2005vd}
\begin{eqnarray}    
\rho_{\theta}(r) = \dfrac{M}{{(4\pi\theta)}^{3/2}}e^{-{\frac{r^2}{4\theta}}};\,\,\,\ {\displaystyle{Lim}}_{\theta\rightarrow 0} \rho_{\theta}(r) = \frac{M\delta(r)}{4\pi r^2}
\label{3.17}
\end{eqnarray}
where the noncommutative parameter $\theta$ is a small ($\sim$ Plank length$^2$) positive number. This mass distribution is inspired from the coherent state approach, where one has to consider the Voros star product instead of the Moyal star product \cite{Banerjee:2009xx}. Using this expression one can write the mass of the black hole of radius $r$ in the following way
\begin{eqnarray}
m_\theta(r) = \int_0^r{4\pi r'^{2}\rho_{\theta}(r')dr' }=\frac{2M}{\sqrt\pi}\gamma(3/2 , r^2/4\theta) 
\label{3.18}
\end{eqnarray}
where $\gamma(3/2 , r^2/4\theta)$ is the lower incomplete gamma function, which is discussed in the appendix. 
In the limit $\theta\rightarrow 0$ it becomes the usual gamma function $(\Gamma_{{\textrm{total}}})$. Therefore $m_\theta(r)\rightarrow M$ is the commutative limit of the noncommutative mass $m_\theta(r)$.

   To find a solution of Einstein equation with the noncommutative mass density of the type (\ref{3.17}), the temporal component of the energy momentum tensor ${(T_{\theta})}_\mu^\nu$ is identified as, ${(T_{\theta})}_t^t=-\rho_\theta$. Now demanding the condition on the metric coefficients ${(g_\theta)}_{tt}=-{(g_\theta)}^{rr}$ for the noncommutative Schwarzschild metric and using the covariant conservation of energy momentum tensor ${(T_{\theta})}_\mu^\nu~_{;\nu}=0$, the energy momentum tensor can be fixed to the form,
\begin{eqnarray}
{(T_\theta)}_\mu^\nu={\textrm {diag}}{[-\rho_\theta,p_r,p',p']},
\label{3.19}
\end{eqnarray}
 where, $p_r=-\rho_\theta$ and $p'=p_r-\frac{r}{2}\partial_r\rho_\theta$. This form of energy momentum tensor is different from the perfect fluid because here $p_r$ and $p'$ are not same,
\begin{eqnarray}
p'=\Big[\frac{r^2}{4\theta}-1\Big]\frac{M}{(4\pi\theta)^{\frac{3}{2}}}e^{-\frac{r^2}{4\theta}}
\label{3.20}
\end{eqnarray}
i.e. the pressure is anisotopic. 
%But for $r<<\sqrt\theta$, the first term in (\ref{3.20}) drops out and $p'=-\rho_\theta=p_r$, i.e. the energy-momentum tensor takes the isotropic form. When $r\rightarrow 0$ the energy density tends to a constant value $-\frac{M}{(4\pi\theta)^{\frac{3}{2}}}$. On the other hand, at the large values of $r$ ($r>>\sqrt\theta$) all the components of the energy-momentum tensor very quickly tend to zero and so the pressure is again isotropic and the Schwarzschild vacuum solution is well applicable. 

     The solution of Einstein equation (in $c=G=1$ unit) ${(G_\theta)}^{\mu\nu}=8\pi{{(T_\theta)}^{\mu\nu}}$, using (\ref{3.19}) as the matter source, is given by the line element \cite{Nicolini:2005vd},   
\begin{eqnarray}
ds^2=-f_\theta(r)dt^2+\frac{dr^2}{f_\theta(r)}+r^2d\Omega^2;\,\,\, f_\theta(r)=-{(g_\theta)}_{tt}=\left(1-\frac{4M}{r\sqrt\pi}\gamma(\frac{3}{2},\frac{r^2}{4\theta})\right)~. 
\label{3.21}
\end{eqnarray}
 Incidentally, this is same if one just replaces the mass term in the usual commutative Schwarzschild space-time by the noncommutative mass $m_\theta(r)$ from (\ref{3.18}). Also observe that for $r>>\sqrt\theta$ the above noncommutative metric reduces to the standard Schwarzschild form.

%     The metric (\ref{3.21}) represents a self-gravitating, anisotropic fluid type matter. The existence of the radial pressure in the small length scale ($r<<\sqrt\theta$) is due to the quantum vacuum fluctuation and it balances the inward gravitational pull to prevent the collapse of the matter to a point. This is reminiscent of earlier works \cite{Frolov,Dymnikova:1992ux} where such a phenomenon is associated with the occurrence of a de-Sitter metric inside the black hole $(f_\theta(r)<0)$. As we now show the introduction of noncommutativity naturally induces a de-Sitter metric for $r<<\sqrt\theta$. In this limit the metric coefficient $f_\theta(r)$ in (\ref{3.21}) reduces to,
%\begin{eqnarray}
%f_\theta(r)\simeq 1-\frac{Mr^2}{3\sqrt{\pi}\theta^{\frac{3}{2}}}.
%\label{3.22}
%\end{eqnarray}
%Therefore in this limit the metric (\ref{3.21}) reduces to a de-Sitter metric with cosmological constant 
%\begin{eqnarray}
%\Lambda_\theta=\frac{M}{3\sqrt\pi\theta^{3/2}}
%\label{3.23}
%\end{eqnarray}
%which has a constant scalar curvature, given by, 
%\begin{eqnarray}
%R_\theta=\frac{4M}{\sqrt\pi\theta^{3/2}}.
%\label{3.24}
%\end{eqnarray}
%Consequently there is no curvature singularity present any more, instead one finds a de-Sitter core of constant positive curvature surrounding the close vicinity of the singularity at $r=0$. This is in agreement with \cite{Frolov,Dymnikova:1992ux} where the existence of the inner de-Sitter core was mentioned. Taking the commutative limit $\theta\rightarrow 0$ in (\ref{3.24}) immediately manifests the singularity.         

    It is interesting to note that the noncommutative metric (\ref{3.21}) is still stationary, static and spherically symmetric as in the commutative case. One or more of these properties is usually violated for other approaches \cite{Chaichian:2007we,Mukherjee:2007fa,Kobakhidze:2007jn} of introducing noncommutativity, particularly those based on Seiberg-Witten maps that relate commutative spaces with noncommutative ones.

The event horizon can be found where $g^{rr}(r_H)=0$, that is
\begin{eqnarray}
r_H = \frac{4M}{\sqrt{\pi}}\gamma\Big(\frac{3}{2},\frac{r_H^2}{4\theta}\Big).
\label{3.25}
\end{eqnarray}
This equation cannot be solved for $r_H$ in a closed form. In the large radius regime ($\frac{r_H^2}{4\theta}>>1$) we use the expanded form of the incomplete $\gamma$ function given in the Appendix (eq. (\ref{app4})) to solve eq. (\ref{3.25}) by iteration. Keeping upto the order $\frac{1}{\sqrt{\theta}}e^{-\frac{M^2}{\theta}}$, we find
\begin{eqnarray}
r_H\simeq2M\Big(1-\frac{2M}{\sqrt{\pi\theta}}e^{-\frac{M^2}{\theta}}\Big).
\label{3.26}
\end{eqnarray}

\subsection{Noncommutative Hawking temperature, tunneling rate and entropy in the presence of back reaction}
  Here the one loop back reaction effect on the space-time will be considered. As explained earlier the modified surface gravity will be of the form given by (\ref{3.02}). But in this case since the only macroscopic parameter is $m_\theta$, $\xi$ will has the following structure:
\begin{eqnarray}
\xi=\beta\frac{M_{\textrm p}^2}{m^2_\theta}
\label{3.27}
\end{eqnarray}
where, as earlier, $\beta$ is a pure numerical factor. In the commutative picture $\beta$ is known to be related to the trace anomaly coefficient \cite{Lousto:1988sp,Fursaev:1994te}. Putting this form of $\xi$ in (\ref{3.02}) we get,
\begin{eqnarray}
\kappa = \kappa_0[r_H]\Big(1+\beta\frac{M_{\textrm p}^2}{m^2_\theta}\Big).
\label{3.28}
\end{eqnarray}
Equation (\ref{3.28}) is recast as,
\begin{eqnarray}
\kappa = \kappa_0[r_H]\Big(1+\frac{\alpha}{m^2_\theta(r_h)}\Big)
\label{3.29}
\end{eqnarray}
where $\alpha=\beta M_{\textrm p}^2$. Since as mentioned already, the noncommutative parameter $\theta$ is of the order of $l_{\textrm p}^2$, $\alpha$ and $\theta$ are of the same order. This fact will be used later when doing the graphical analysis.

   In order to calculate the right hand side of (\ref{3.29}), we need to obtain an expression for noncommutative classical surface gravity at the horizon of the black hole $(\kappa_0[r_H])$. This is done by using (\ref{2.20}). For the classical noncommutative Schwarzschild spacetime the metric coefficients are given by (\ref{3.21}).
The value of $\kappa_0[r_H]$ is thus found to be,
\begin{eqnarray}
\kappa_0[r_H] = \frac{f'(r_H)}{2}= \frac{1}{2}\Big[\frac{1}{r_H}-\frac{r^2_H}{4\theta^{\frac{3}{2}}} \frac{e^{-\frac{r^2_H}{4\theta}}}{\gamma\Big(\frac{3}{2},\frac{r^2_H}{4\theta}\Big)}\Big].
\label{3.30}
\end{eqnarray}
Inserting (\ref{3.30}) in (\ref{3.29}) we get,
\begin{eqnarray}
\kappa = \frac{1}{2}\Big[\frac{1}{r_H}-\frac{r^2_H}{4\theta^{\frac{3}{2}}} \frac{e^{-\frac{r^2_H}{4\theta}}}{\gamma\Big(\frac{3}{2},\frac{r^2_H}{4\theta}\Big)}\Big]\Big(1+\frac{\alpha}{m^2_\theta(r_H)}\Big).
\label{3.31}
\end{eqnarray}
In order to write the above equation completely in terms of $r_H$ we have to express the mass $m_\theta$ in terms of $r_H$. For that we compare equations (\ref{3.18}) and (\ref{3.25}) to get,
\begin{eqnarray}
m_{\theta}(r_H)=\frac{r_H}{2}.
\label{3.32}
\end{eqnarray}
This relation is the noncommutative deformation of the standard radius-mass relation for the usual (commutative space) Schwarzschild black hole. Expectedly in the limit $\theta\rightarrow 0$ eq. (\ref{3.32}) reduces to its commutative version $r_H=2M$.

   Substituting (\ref{3.32}) in (\ref{3.31}) we get the value of modified noncommutative surface gravity
\begin{eqnarray}
\kappa = \frac{1}{2}\Big[\frac{1}{r_H}-\frac{r^2_H}{4\theta^{\frac{3}{2}}} \frac{e^{-\frac{r^2_H}{4\theta}}}{\gamma\Big(\frac{3}{2},\frac{r^2_H}{4\theta}\Big)}\Big]\Big(1+\frac{4\alpha}{r^2_H}\Big).
\label{3.33}
\end{eqnarray} 
So from (\ref{2.27}), the modified noncommutative Hawking temperature including the effect of back reaction is given by,
\begin{eqnarray}
T_H=\frac{\hbar\kappa}{2\pi}=\frac{\hbar}{4\pi}\Big[\frac{1}{r_H}-\frac{r^2_H}{4\theta^{\frac{3}{2}}} \frac{e^{-\frac{r^2_H}{4\theta}}}{\gamma\Big(\frac{3}{2},\frac{r^2_H}{4\theta}\Big)}\Big]\Big(1+\frac{4\alpha}{r^2_H}\Big).
\label{3.34}
\end{eqnarray}
If the back reaction is ignored (i. e. $\alpha =0$), the expression for the Hawking temperature agrees with that given in \cite{Nicolini:2005vd}. Also for the $\theta\rightarrow 0$ limit, one can recover the standard result (\ref{3.14}) \cite{Lousto:1988sp,Fursaev:1994te}.

    In the standard (commutative) case $T_H$ diverges as $M\rightarrow 0$ and this puts a limit on the validity of the conventional description of Hawking radiation. Against this scenario, temperature (\ref{3.34}) includes noncommutative and back reaction effects which are relevant at distances comparable to $\sqrt\theta$. The behaviour of the temperature $T_H$ as a function of horizon radius $r_H$ is plotted in fig.(\ref{fig1}) (with positive $\alpha$) and in fig.(\ref{fig2}) (with negative $\alpha$).

\begin{figure}[t] 
\centering
%\rotatebox{270}{
\includegraphics[width=0.9\textwidth]{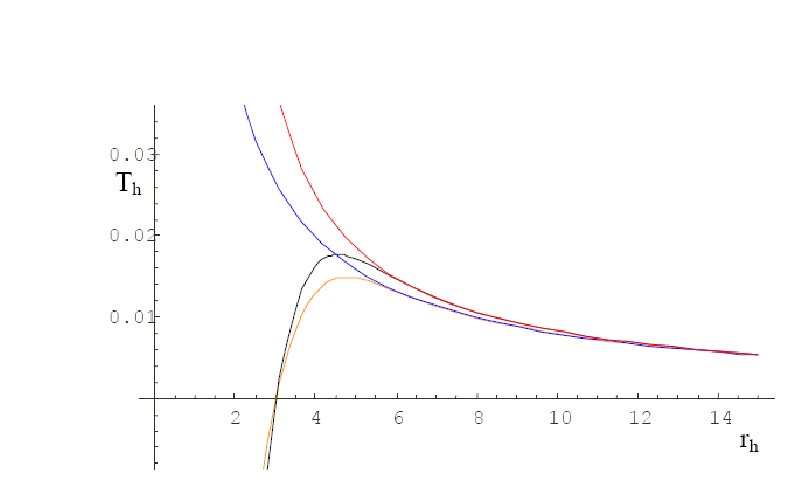}
\caption[]{\it{{$T_H$ Vs. $r_H$ plot (Here $\alpha=\theta$, $\alpha$ and $\theta$ are positive).\\ 
$r_H$ is plotted in units of $\sqrt\theta$ and $T_H$ is plotted in units of $\frac{1}{\sqrt\theta}$.\\
Red curve: $\alpha\neq 0, \theta=0$.\\
Blue curve: $\alpha= 0, \theta= 0$.\\
Black curve: $\alpha\neq 0, \theta\neq 0$.\\
Yellow curve: $\alpha= 0, \theta\neq 0$.}}}
\label{fig1}
\end{figure}

    Fig.(\ref{fig1}) shows that in the region $r_H\simeq\sqrt\theta$, the effect of noncommutativity significantly changes the nature of commutative space curves. Interestingly two noncommutative curves, whether including back reaction or not are qualitatively same. Both of them attain a maximum value at $r_H={\tilde{r}}_0\simeq 4.7\sqrt\theta$ and then sharply drop to zero forming an extremal black hole. In the region $r_H<r_0 \simeq 3.0 \sqrt{\theta}$ there is no black hole, because physically $T_H$ cannot be negative. The only difference between them is that the back reaction effect increases the maximum temperature by $20\%$. Infact, in the commutative space also, back reaction effect increases the value of Hawking temperature. But quite contrary to the noncommutative curves, both of them diverge as $r_H\rightarrow 0$. As easily observed, the Hawking paradox is circumvented by noncommutativity, with or without back reaction. This was also noted in \cite{Nicolini:2005vd} where, however, the quantitative effects of back reaction were not considered.  

     On the other hand fig.(\ref{fig2}) shows that if any of the two effects (i.e. either noncommutativity or back reaction) is present $T_H$ drops to zero. For $\alpha=0, \theta\neq 0$ (yellow curve) $T_H$ becomes zero at $r_H=r_0\simeq 3.0\sqrt\theta$ and for $\alpha\neq 0, \theta=0$ (red curve) it becomes zero at $r_H=r_0\simeq 2.0\sqrt\theta$ . These cases therefore bypass the Hawking paradox. But for noncommutative black hole with back reaction ($\alpha\neq0,\theta\neq0$), $T_H$ is zero for two values of $r_H$: $r_H\simeq 3.0\sqrt\theta$ and $r_H=2.0\sqrt\theta$ and then it diverges towards positive infinity. This is not physically possible since after entering the forbidden zone it resurfaces on the allowed sector. So for both noncommutativity and back reaction effect, $\alpha$ can never be negative.

%\begin{figure}[t]
%  \begin{center}
%    \includegraphics[width=\textwidth]{Graph4_new.jpg}
%  \end{center}

% \caption{\small Figure caption. To get a figure to span two
%     columns, use the environment figure* rather than figure.}
% \label{fig-label}
%\end{figure}

\begin{figure}[t] 
\centering
%\rotatebox{270}{
\includegraphics[angle=0,width=0.9\textwidth]{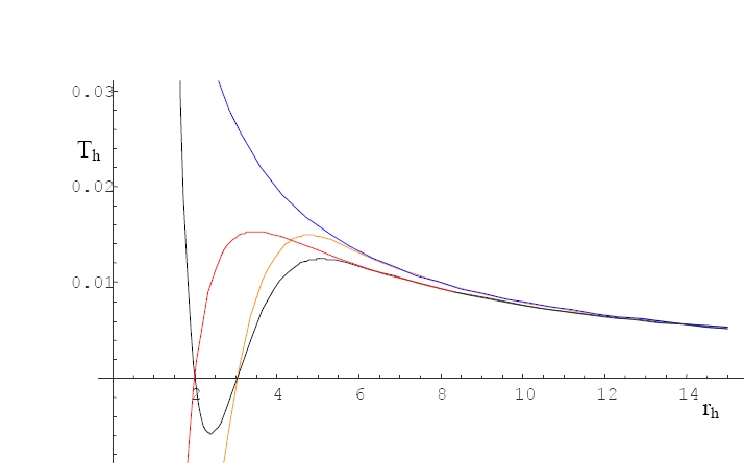}
\caption[]{{\it{$T_H$ Vs. $r_H$ plot (Here $|\alpha|=\theta$, $\alpha$ is negative but $\theta$ is positive).\\
$r_H$ is plotted in units of $\sqrt\theta$ and $T_H$ is plotted in units of $\frac{1}{\sqrt\theta}$.\\
Red curve: $\alpha\neq 0, \theta=0$.\\
Blue curve: $\alpha= 0, \theta= 0$.\\
Black curve: $\alpha\neq 0, \theta\neq 0$.\\
Yellow curve: $\alpha= 0, \theta\neq 0$.}}}
\label{fig2}
\end{figure}

      Having obtained the Hawking temperature of the black hole we calculate the Bekenstein-Hawking entropy. The expression of entropy can be obtained from the second law of thermodynamics. But instead of using it we employ the formula (\ref{2.45}) to calculate the entropy. Using (\ref{3.26}) the modified surface gravity (\ref{3.33}) can be approximately expressed in terms of $M$. To the leading order, we obtain, 
\begin{eqnarray}
\kappa(M)&=&\frac{M^2+\alpha}{4M^3}\Big[1-\frac{4M^5}{(M^2+\alpha)\theta\sqrt{\pi\theta}}e^{-\frac{M^2}{\theta}}\Big]+{\cal{O}}(\frac{1}{\sqrt\theta}e^{-\frac{M^2}{\theta}}).
\label{3.35}
\end{eqnarray}
Substituting this in (\ref{2.40}) and then integrating over $\omega'$ we have, 
\begin{eqnarray}
{\textrm{Im}}~{\cal{S}} &=& 4\pi\omega(M-\frac{\omega}{2})+2\pi\alpha \ln{\Big[\frac{(M-\omega)^2+\alpha}{M^2+\alpha}\Big]}- 8\sqrt{\frac{\pi}{\theta}}M^3 e^{-\frac{M^2}{\theta}}
\nonumber
\\
&+& 8\sqrt{\frac{\pi}{\theta}}(M-\omega)^3 e^{-\frac{(M-\omega)^2}{\theta}}
\nonumber
\\
&+&\textrm{const.(independent of $M$)}+{\cal O}(\sqrt\theta e^{-\frac{M^2}{\theta}}).
\label{3.36}
\end{eqnarray}
So by the relation (\ref{2.41}) the modified tunneling probability due to noncommutativity and back reaction effects is, 
\begin{eqnarray}
\Gamma&\sim& \Big[1-\frac{2\omega(M-\frac{\omega}{2})}{M^2+\alpha}\Big]^{-\frac{4\pi\alpha}{\hbar}} \textrm{exp}\Big[\frac{16}{\hbar}\sqrt{\frac{\pi}{\theta}}M^3e^{-\frac{M^2}{\theta}}-\frac{16}{\hbar}\sqrt{\frac{\pi}{\theta}}(M-\omega)^3e^{-\frac{(M-\omega)^2}{\theta}}\nonumber
\\
&+&\textrm{const.(independent of $M$)}\Big]
\textrm{exp}\Big[-\frac{8\pi\omega}{\hbar}(M-\frac{\omega}{2})\Big]. 
\label{3.37}
\end{eqnarray}
The last exponential factor of the tunneling probability was previously obtained by Parikh and Wilczek \cite{Wilczek} where neither noncommutativity nor back reaction effects were considered. The factors before this exponential are actually due the effect of back reaction and noncommutativity. It will eventually give the correction to the Bekenstein-Hawking entropy and the Hawking temperature as will be shown below. Taking $\theta\rightarrow 0$ limit we can immediately reproduce the commutative tunneling rate for Schwarzschild black hole with back reaction effect \cite{Banerjee:2008ry}.

     We are now in a position to obtain the noncommutative deformation of the Bekenstein-Hawking area law. The first step is to compute the entropy change $\Delta S_{bh}$. Using (\ref{2.41}) and (\ref{3.37}) we obtain, to the leading order,
\begin{eqnarray}
\Delta S_{bh} = S_{final} - S_{initial}&\simeq& -\frac{8\pi\omega}{\hbar}(M-\frac{\omega}{2})-\frac{4\pi\alpha}{\hbar} \ln{\Big[\frac{(M-\omega)^2+\alpha}{M^2+\alpha}\Big]}+\frac{16}{\hbar} \sqrt{\frac{\pi}{\theta}}M^3 e^{-\frac{M^2}{\theta}}
\nonumber
\\
&-& \frac{16}{\hbar}\sqrt{\frac{\pi}{\theta}}(M-\omega)^3 e^{-\frac{(M-\omega)^2}{\theta}}+\textrm{const.(independent of $M$)}.
\label{3.38}
\end{eqnarray}
Next using the stability criterion $\frac{d(\Delta S_{bh})}{d\omega}=0 $ for the black hole, one obtains the only physically possible solution for $\omega$ as
 $\omega=M$. Substituting this value of $\omega$ in (\ref{3.38}) and setting $S_{{\textrm{final}}}=0$ we have the Bekenstein-Hawking entropy
\begin{eqnarray}
S_{bh}=S_{{\textrm{initial}}}&\simeq& \frac{4\pi M^2}{\hbar}-\frac{4\pi\alpha}{\hbar}\ln{(\frac{M^2}{\alpha}+1)}
\nonumber
\\
&-& \frac{16}{\hbar}\sqrt{\frac{\pi}{\theta}}M^3 e^{-\frac{M^2}{\theta}}+\textrm{const.(independent of $M$)}.
\label{3.39}
\end{eqnarray}
Neglecting the back reaction effect ($\alpha=0$) the above expression of black hole entropy is written as 
\begin{eqnarray}
S_{bh}\simeq \frac{4\pi M^2}{\hbar}- \frac{16}{\hbar}\sqrt{\frac{\pi}{\theta}}M^3 e^{-\frac{M^2}{\theta}}.
\label{3.40}
\end{eqnarray}
Now in order to write the above equation in terms of the noncommutative horizon area ($A_\theta$), we use (\ref{3.26}) to obtain,
\begin{eqnarray}
A_\theta=4\pi r_H^2=16\pi M^2 - 64 \sqrt{\frac{\pi}{\theta}} M^3 e^{-\frac{M^2}{\theta}}+{\cal{O}}(\sqrt\theta e^{-\frac{M^2}{\theta}}).
\label{3.41}
\end{eqnarray}
Comparing equations (\ref{3.40}) and (\ref{3.41}) we see that at the leading order the noncommutative black hole entropy satisfies the area law
\begin{eqnarray}
S_{bh}=\frac{A_\theta}{4\hbar}.
\label{3.42}
\end{eqnarray}
This is functionally identical to the Bekenstein-Hawking area law in the commutative space.

   Considering $\theta\rightarrow 0$ limit in (\ref{3.39}) we have the corrected form of Bekenstein-Hawking entropy for commutative Schwarzschild black hole with back reaction effect \cite{Fursaev:1994te,Banerjee:2008ry}. The well known logarithmic correction \cite{Page:2004xp} is reproduced (see equation (\ref{3.13})).

     Now using the second law of thermodynamics (\ref{2.53}) we can find the corrected form of the Hawking temperature $T_H$ due to back reaction. This is obtained from (\ref{3.39}) as,
\begin{eqnarray}
\frac{1}{T_H}=\frac{dS_{bh}}{dM}=\frac{8\pi M^3}{\hbar(M^2+\alpha)}+\frac{32}{\hbar}\frac{\sqrt{\pi}}{\theta^{\frac{3}{2}}}M^4 e^{-\frac{M^2}{\theta}}+{\cal{O}}(\frac{1}{\sqrt\theta}e^{-\frac{M^2}{\theta}}).
\label{3.43}
\end{eqnarray}
Therefore the back reaction corrected noncommutative Hawking temperature is given by
\begin{eqnarray}
T_H &=& \frac{\hbar(M^2+\alpha)}{8\pi M^3} -\frac{\hbar M^2}{2(\pi\theta)^{\frac{3}{2}}}e^{-\frac{M^2}{\theta}}+ {\cal{O}}(\frac{1}{\sqrt\theta}e^{-\frac{M^2}{\theta}}).
\label{3.44}
\end{eqnarray}

   We now provide a simple consistency check on the relation (\ref{3.34}). The Hawking temperature is recalculated using this relation and showing that it reproduces (\ref{3.44}). For the large radius limit, (\ref{3.34}) takes the value,
\begin{eqnarray}
T_H\simeq\frac{\hbar}{4\pi}\big[\frac{1}{r_H}-\frac{r_H^2}{2\sqrt{\pi}\theta^{3/2}}e^{-\frac{r_H^2}{4\theta}}\big]\big(1+\frac{4\alpha}{r_H^2}\big).
\label{3.45}
\end{eqnarray}
Now the approximated form of $r_H$ in terms of $M$ (\ref{3.26}) is substituted in (\ref{3.45}) to get the relation (\ref{3.44}) upto the leading order in the noncommutative parameter. This shows the self consistency of our calculation.

   For $\alpha=\theta=0$, the expression (\ref{3.44}) reduces to the usual Hawking temperature $T_H=\frac{\hbar}{8\pi M}$ for a Schwarzschild black hole. Also, keeping the back reaction ($\alpha$) but taking $\theta\rightarrow 0$ limit, we reproduce the commutative Hawking temperature (\ref{3.14}) \cite{Lousto:1988sp,Fursaev:1994te,Banerjee:2008ry}. 
%%%%%%%%%%%%%%%%%%%%%%%%%%%%%%%%%%%%%%%%%%%%%%%%%%%%%%%%
\section{Discussions}
       We have considered self-gravitation and (one loop) back reaction effects in tunneling formalism for Hawking radiation. The modified tunneling rate was computed. From this modification, corrections to the semiclassical expressions for  entropy and Hawking temperature were obtained. Also, the logarithmic correction to the semiclassical Bekenstein-Hawking area law was reproduced.

      The other significant part of this chapter was the application of our formulation to a noncommutative Schwarzschild metric, keeping in mind the consequence of back reaction. Several thermodynamic entities like the temperature and entropy were computed. The tunneling rate was also derived. The temperature, in particular, was obtained in a closed form. This result was analyzed in detail using two graphical representations. We gave particular attention to the small scale behaviour of black hole temperature where the effects of both noncommutativity and back reaction are highly nontrivial. The graphs presented here are naturally more general than \cite{Nicolini:2005vd,Lousto:1988sp}, because in \cite{Nicolini:2005vd} the effect of back reaction was not included and in \cite{Lousto:1988sp} space time was taken to be commutative in nature. Expectedly in suitable limits, the results of our paper reduced to that of \cite{Nicolini:2005vd,Lousto:1988sp}, but the combination of noncommutativity and back reaction, as shown here, gave new results at small scale. In particular, it was shown that in the presence of both noncommutativity and back reaction, the back reaction parameter $\alpha$ cannot be negative. Interestingly, even for the commutative case, arguments based on quantum geometry \cite{Banerjee:2008ry,Page:2004xp,Kaul:2000kf} fix a positive value for $\alpha$.

   In the noncommutative analysis, with positive $\alpha$, (Fig \ref{fig1}), the maximum Hawking temperature got enhanced in the presence of back reaction. However, the Hawking paradox was avoided whether or not the back reaction is included.

   Apart from the temperature, other variables like the tunneling rate and entropy were given upto the leading order in the noncommutative parameter. The entropy was expressed in terms of the area. The result was a noncommutative deformation of the Bekenstein-Hawking area law, preserving the usual functional form. Since both $T_H=\frac{\hbar\kappa}{2\pi}$ and the area law retained their standard forms it suggests that the laws of noncommutative black hole thermodynamics are a simple noncommutative deformation of the usual laws. However, it must be remembered this result was obtained only in the leading order approximation. For $r \sim \sqrt{\theta}$ this approximation is expected to be significant.

   As a final remark we mention that although our results are presented for the  Schwarzschild metric, the formulation is resilient enough to discuss both back reaction and noncommutativity in other types of black holes.

%%%%%%%%%%%%%%%%%%%%%%%%%%%%%%%%%%%%%%%%%%%
\begin{subappendices}
\chapter*{Appendix}
\section{\label{appendix1A}Incomplete gamma function}
\renewcommand{\theequation}{3A.\arabic{equation}}
\setcounter{equation}{0}  % reset counter
The lower incomplete gamma function is given by
\begin{eqnarray}
\gamma(a,x)=\int_0^x t^{a-1}e^{-t}dt
\label{app1}
\end{eqnarray}
whereas the upper incomplete gamma function is
\begin{eqnarray}
\Gamma(a,x)=\int_x^\infty t^{a-1}e^{-t}dt
\label{app2}
\end{eqnarray}
and they are related to the total gamma function through the following relation
\begin{eqnarray}
\Gamma_{{\textrm{total}}}(a)=\gamma(a,x)+\Gamma(a,x)=\int_0^\infty t^{a-1}e^{-t}dt.
\label{app3}
\end{eqnarray}
Furthermore, for large $x$, i.e. $x>>1$, the asymptotic expansion of the lower incomplete gamma function is given by
\begin{eqnarray}
\gamma(\frac{3}{2},x)&=&\Gamma_{{\textrm{total}}}(\frac{3}{2})-\Gamma(\frac{3}{2},x)
\nonumber
\\
&\simeq&\frac{\sqrt{\pi}}{2}\Big[1-e^{-x}\sum_{p=0}^{\infty}\frac{x^{\frac{1-2p}{2}}}{\Gamma_{{\textrm{total}}}(\frac{3}{2}-p)}\Big].
\label{app4}
\end{eqnarray}
Using the definition (\ref{app1}) and then integrating by parts we have
\begin{eqnarray}
\gamma(a+1,x)=\int_0^x t^ae^{-t}dt&=&-t^ae^{-t}|_0^x+a\int_o^x t^{a-1}e^{-t}dt
\nonumber
\\
&=&-x^ae^{-x}+a\gamma(a,x).
\label{app7}
\end{eqnarray}
Similarly by the definition (\ref{app2}) one can show
\begin{eqnarray}
\Gamma(a+1,x)=x^ae^{-x}+a\Gamma(a,x).
\label{app8}
\end{eqnarray}
\section{\label{appendix1B}Some useful formulas}
\renewcommand{\theequation}{3B.\arabic{equation}}
\setcounter{equation}{0}  % reset counter
\begin{eqnarray}
I_1=\int_a^b e^{-\alpha x^2}dx=\frac{1}{2{\alpha}^{\frac{1}{2}}}\Big[\sqrt{\pi}-\gamma(\frac{1}{2},\alpha a^2)-\Gamma(\frac{1}{2},\alpha b^2)\Big]
\label{app51}
\end{eqnarray}
\begin{eqnarray}
I_2=\int_a^b x^2e^{-\alpha x^2}dx=\frac{1}{2{\alpha}^{\frac{3}{2}}}\Big[\frac{\sqrt{\pi}}{2}-\gamma(\frac{3}{2},\alpha a^2)-\Gamma(\frac{3}{2},\alpha b^2)\Big]
\label{app5}
\end{eqnarray}
\begin{eqnarray}
I_3=\int_a^b x^4e^{-\alpha x^2}dx=\frac{1}{2{\alpha}^{\frac{5}{2}}}\Big[\frac{3\sqrt{\pi}}{4}-\gamma(\frac{5}{2},\alpha a^2)-\Gamma(\frac{5}{2},\alpha b^2)\Big]
\label{app6}
\end{eqnarray} 
\end{subappendices}

%%%%%%%%%%%%%%%%%%%%%%%%%%%%%%%%%%%%%%%%%%%%%%
\chapter{\label{chap:anomaly}Tunneling mechanism and anomaly}
%%%%%%%%%%%%%%%%%%%%%%%%%%%%%%%%%%%%%%%%%%%%%%%%%%%%%%%%%%%%%%%%%%
    Ever since Hawking's original observation \cite{Hawking:1974rv,Hawking:1974sw} that black holes radiate, there have been several derivations \cite{Gibbons:1976ue,Christensen:1977jc,Paddy,Wilczek,Robinson:2005pd,Iso:2006wa,Banerjee:2007qs,Banerjee:2008az,Banerjee:2007uc} of this effect. A common feature in these derivations is the universality of the phenomenon; the Hawking radiation is determined by the horizon properties of the black hole leading to the same answer. This, in the absence of direct experimental evidence, definitely reinforces Hawking's original conclusion. Moreover, it strongly suggests that there is some fundamental mechanism which could, in some sense, unify the various approaches.

    In this chapter we show that chirality is the common property which connects the tunneling formalism \cite{Paddy,Wilczek} and the anomaly method \cite{Christensen:1977jc,Robinson:2005pd,Iso:2006wa,Banerjee:2007qs,Banerjee:2008az,Banerjee:2007uc,Banerjee:2008wq,Bonora:2008he,Bonora:2008nk,Ghosh:2008tg,Iso:2007nf} in studying Hawking effect. Apart from being among the most widely used approaches, interest in both the anomaly and tunneling methods has been revived recently leading to different variations and refinements in them \cite{Banerjee:2007qs,Banerjee:2007uc,Bonora:2008he,Bonora:2008nk,Ghosh:2008tg,Iso:2007nf,Jiang,Chen,Medved1,Akhmedov,Banerjee:2008cf,Morita:2008qn}. The calculation will be performed using a family of metrics that includes a subset of the stationary, spherically symmetric space-times which are asymptotically flat. Also, the results are derived using mostly physical reasoning and do not require any specific technical skill.

    Before commencing on our analysis we briefly recapitulate the basic tenets of the tunneling and anomaly methods. The idea of a tunneling description, quite akin to what we know in usual quantum mechanics where classically forbidden processes might be allowed through quantum tunneling, dates back to 1976 \cite{Damour:1976jd}. Present day computations generally follow either the null geodesic method \cite{Wilczek} or the Hamilton-Jacobi method \cite{Paddy}, both of which rely on the semi-classical WKB approximation yielding equivalent results. The essential idea, as explained earlier, is that a particle-antiparticle pair forms close to the event horizon. The ingoing negative energy mode is trapped inside the horizon while the outgoing positive energy mode is observed at infinity as the Hawking flux.

   Although the notion of an anomaly, which represents the breakdown of some classical symmetry upon quantisation, is quite old, its implications for Hawking effect were first studied in \cite{Christensen:1977jc}. It was based on the conformal (trace) anomaly but the findings were confined only to two dimensions. However it is possible to apply this method to general dimensions. Recently a new method was put forward in \cite{Robinson:2005pd,Iso:2006wa} where a general (any dimensions) derivation was given. It was based on the well known fact that the effective theory near the event horizon is a two dimensional conformal theory. The ingoing modes are trapped within the horizon and cannot contribute to the effective theory near the horizon. Thus the near horizon theory becomes a two dimensional chiral theory. Such a chiral theory suffers from a general coordinate (diffeomorphism) anomaly manifested by a nonconservation of the stress tensor. Using this gravitational anomaly and a suitable boundary condition the Hawking flux was obtained. A covariant version of this method, that was also technically simpler, was given in \cite{Banerjee:2007qs}. This was followed by another, new, effective action based approach in \cite{Banerjee:2007uc,Banerjee:2008az}.

    The first step in our procedure is to derive the two dimensional gravitational anomaly using the notion of chirality. This is a new method of obtaining the gravitational anomaly. Once this anomaly is obtained, the flux is easily deduced. Exploiting the same notion of chirality the probability of the outgoing mode in the tunneling approach will be computed. The Hawking temperature then follows from this probability. At an intermediate stage of this computation we further show that the chiral modes obtained in the tunneling formalism reproduce the gravitational anomaly thereby completing the circle of arguments regarding the connection of the two approaches. 

\section{Metric and null coordinates}   
   Consider a black hole characterised by a spherically symmetric, static space-time  and asymptotically flat metric of the form (\ref{2.01}). For simplicity we consider here $f(r)=g(r)=F(r)$ and hence the
event horizon $r=r_H$ is defined by $F(r_H)=0$. 
Now it is well known \cite{Robinson:2005pd,Iso:2006wa,Solodukhin:1998tc,Carlip:1998wz} that near the event horizon the effective theory reduces to a two dimensional conformal theory whose metric is given by the ($r-t$) sector of the original metric (\ref{2.01}).

     It is convenient to express (\ref{2.01}) in the null tortoise coordinates which are defined as,
\begin{eqnarray}
u=t-r^*,\,\,\, v=t+r^*;
\label{4.01}
\end{eqnarray}
where $r^*$ is defined by the relation (\ref{2.21}).
Under these set of coordinates the relevant ($r-t$)-sector of the metric (\ref{2.01}) takes the form,
\begin{eqnarray}
ds^2=-\frac{F(r)}{2}(du~dv+dv~du)
\label{4.02}
\end{eqnarray}
Chiral conditions, to be discussed in the next section, are most appropriately described in these coordinates.
%%%%%%%%%%%%%%%%%%%%%%%%%%%%%%%%%%%%%%%%%%%%%%%%%%%%%%%%%%%%%%%%%%%%%%%%
\section{Chirality conditions}
     Consider the Klein-Gordon (KG) equation (\ref{2.02}) for a massless scalar particle governed by the metric (\ref{4.02}). Then the KG equation reduces to the following form:
\begin{eqnarray}
2\partial_u\partial_v\phi(u,v)=0.
\label{4.03}
\end{eqnarray}
The general solution of this can be taken as
$\phi(u,v)=\phi^{(R)}(u)+\phi^{(L)}(v)$
where $\phi^{(R)}(u)$ and $\phi^{(L)}(v)$ are the right (outgoing) and left (ingoing) modes (see Appendix \ref{appendixmode}) satisfying
\begin{eqnarray}
\nabla_v\phi^{(R)}=0,\,\, \nabla_u\phi^{(R)}\neq0;\,\,\,\
\nabla_u\phi^{(L)}=0,\,\,\, \nabla_v\phi^{(L)}\neq0.
\label{4.04}
\end{eqnarray}
These equations are expressed simultaneously as,
\begin{eqnarray} 
\nabla_\mu\phi=\pm\bar{\epsilon}_{\mu\nu}\nabla^\nu\phi
=\pm \sqrt{-g}\epsilon_{\mu\nu}\nabla^\nu\phi
\label{4.05}
\end{eqnarray}
where $+(-)$ stand for left (right) mode and $\epsilon_{\mu\nu}$ is the numerical antisymmetric tensor with $\epsilon_{uv}=\epsilon_{tr}=-1$.
This is the chirality condition {\footnote{In analogy with studies in 2d CFT this condition is usually referred as holomorphy condition and the chiral modes $\phi^{(L,R)}$ are called the holomorphic modes.}}. In fact the condition (\ref{4.05}) holds for any chiral vector $J_\mu$ in which case $J_\mu=\pm{\bar{\epsilon}}_{\mu\nu}J^\nu$. Likewise, the chirality condition for the energy-momentum tensor is \cite{Banerjee:2008wq},
\begin{eqnarray}
T_{\mu\nu}=\pm\frac{1}{2}(\bar{\epsilon}_{\mu\sigma}T^\sigma_\nu+\bar{\epsilon}_{\nu\sigma}T^\sigma_\mu)+\frac{1}{2}g_{\mu\nu}T^\alpha_\alpha
\label{4.06}
\end{eqnarray}
The $+$ ($-$) sign corresponds to the left (right) mode satisfying,
\begin{eqnarray}
T^{(R)}_{vv}=0,\,\,\,\,  T^{(R)}_{uu}\neq0
\label{4.07}
\\
T^{(L)}_{uu}=0,\,\,\,\,  T^{(L)}_{vv}\neq0~.
\label{4.06n1}
\end{eqnarray}   
These are the analogous of (\ref{4.04}).
%which is analogous to the first equation of (\ref{4.04}). 
%The other sign of (\ref{4.06}) gives the energy-momentum tensor for the left mode. This essentially corresponds to the 
They manifest the symmetry under the interchange $u\leftrightarrow v$  and $L \leftrightarrow R$. In the next section, using these chirality conditions we will derive the explicit form for the gravitational anomaly that reproduces the Hawking flux.     
%%%%%%%%%%%%%%%%%%%%%%%%%%%%%%%%%%%%%%%%%%%%%%%%%%%%%%%%%%%%%%%%%
\section{Chirality, gravitational anomaly and Hawking flux}
      It is well known that for a non-chiral (vector like) theory it is not possible to simultaneously preserve, at the quantum level, general coordinate invariance as well as conformal invariance. Since the former invariance is fundamental in general relativity, conformal invariance is sacrificed leading to a nonvanishing trace of the stress tensor, called the trace anomaly. Using this trace anomaly and the chirality condition we will derive an expression for the chiral gravitational (diffeomophism) anomaly from which the Hawking flux is computed.

   The energy-momentum tensor near an evaporating black hole is split into a traceful and traceless part by \cite{Fulling1},
\begin{eqnarray}
T_{\mu\nu}=\frac{R}{48\pi}g_{\mu\nu}+\theta_{\mu\nu}
\label{4.08}
\end{eqnarray}
where $\theta_{\mu\nu}$ is symmetric (i.e. $\theta_{\mu\nu}=\theta_{\nu\mu}$), so that it preserves the symmetricity of $T_{\mu\nu}$, and traceless (i.e. $\theta_\mu^\mu=0$ so that in $u,v$ coordinates $\theta_{uv}=0$). The traceful part is contained in the first piece leading to the trace anomaly,
$T^\mu_\mu=\frac{R}{24\pi}$. 
Also, since general coordinate invariance is preserved, $\nabla^\mu T_{\mu\nu}=0$, from which it follows that the solutions of $\theta_{\mu\nu}$ satisfy,
\begin{eqnarray}
\nabla^\mu\theta_{\mu\nu}=-\frac{1}{48\pi}\nabla_\nu R
\label{4.09}
\end{eqnarray}
   
      Now the energy-momentum tensor (\ref{4.08}) can be regarded as the sum of the contributions from the right and left moving modes. Symmetry principle tells that the contribution from one mode is exactly equal to that from the other mode, only that $u,v$ have to be interchanged. Since $T_{\mu\nu}$ is symmetric we have
$T_{\mu\nu}=T_{\mu\nu}^{(R)}+T_{\mu\nu}^{(L)}$ 
with
\begin{eqnarray}
T_{\mu\nu}^{(R/L)}=\frac{R}{96\pi}g_{\mu\nu}+\theta_{\mu\nu}^{(R/L)}
\label{4.10}
\end{eqnarray}
where $\theta_{\mu\nu}=\theta_{\mu\nu}^{(R)}+\theta_{\mu\nu}^{(L)}$ (in analogy with $T_{\mu\nu}$). Therefore the chirality condition (\ref{4.07}) and the traceless condition of $\theta_{\mu\nu}$ immediately show
\begin{eqnarray}
&&\theta_{uv}^{(R)}=0,\,\,\, \theta_{vv}^{(R)}=0,\,\,\,\, \theta_{uu}^{(R)}\neq 0;\,\,\, \theta_{uv}^{(L)}=0,\,\,\, \theta_{uu}^{(L)}=0,\,\,\,\, \theta_{vv}^{(L)}\neq 0~.
\label{4.11}
\end{eqnarray}
The trace anomaly for the chiral components follows from (\ref{4.10}) and (\ref{4.11}),
\begin{eqnarray}
T{^\mu_\mu}{^{(R)}}= T{^\mu_\mu}{^{(L)}}=\frac{1}{2}T^\mu_\mu=\frac{R}{48\pi}~.
\label{4.12}
\end{eqnarray}

     To find out the diffeomorphism anomaly for the chiral components we will use (\ref{4.10}). Considering only the right mode, for example, we have
\begin{eqnarray}
\nabla^\mu T_{\mu\nu}^{(R)}=\frac{1}{96\pi}\nabla_\nu R+\nabla^\mu\theta_{\mu\nu}^{(R)}~.
\label{4.13}
\end{eqnarray}
Next, using (\ref{4.09}) and (\ref{4.11}) for the right mode we obtain,
\begin{eqnarray}
\nabla^\mu \theta_{\mu u}^{(R)}=-\frac{1}{48\pi}\nabla_u R;\,\,\, \nabla^\mu\theta_{\mu v}^{(R)}=0
\label{4.14}
\end{eqnarray}
Substituting these in (\ref{4.13}) we get, once for $\nu=u$ and then $\nu=v$,
\begin{eqnarray}
\nabla^\mu T_{\mu u}^{(R)}=-\frac{1}{96\pi}\nabla_u R;\,\,\,\,\, \nabla^\mu T_{\mu v}^{(R)}=\frac{1}{96\pi}\nabla_v R.
\label{4.15}
\end{eqnarray}
Therefore, combining both the above results yields
\begin{eqnarray}
\nabla^{\mu}T_{\mu\nu}^{(R)}=\frac{1}{96\pi}\bar{\epsilon}_{\nu\lambda}\nabla^\lambda R
\label{4.16}
\end{eqnarray}
which is the chiral (gravitational) anomaly for the right mode. Similarly the chiral anomaly for left mode can also be obtained which has a similar form except for a minus sign on the right side of (\ref{4.16}). This anomaly is in covariant form and so it is also called the covariant gravitational anomaly. The structure, including the normalization, agrees with that found by using explicit regularization of the chiral stress tensor \cite{Fulling:1986rk,Bat}.

    From (\ref{4.16}) and (\ref{4.12}) a simple relation follows between the gravitational anomaly (${\cal{A}}_\nu$) and the trace anomaly ($T$),
\begin{eqnarray}
{\cal{A}}_\nu=\frac{1}{2}\bar{\epsilon}_{\nu\lambda}\nabla^\lambda T.
\label{4.17}
\end{eqnarray}
Such a relation is not totally unexpected since covariant expressions must involve the Ricci scalar. However (\ref{4.17}) should not be interpreted as a Wess-Zumino consistency condition which involves only `consistent' expressions \cite{Bat}. Here, on the contrary, we are dealing with covariant expressions.

    The covariant anomaly (\ref{4.16}) is now used to obtain the Hawking flux. As was mentioned earlier the effective two dimensional theory near the horizon becomes chiral. The chiral theory has the anomaly (\ref{4.16}). Taking its $\nu=u$ component we obtain,
\begin{eqnarray}
\partial_rT_{uu}^{(R)}=\frac{F}{96\pi}\partial_r R=\frac{F}{96\pi}\partial_r(F'')=\frac{1}{96\pi}\partial_r(FF''-\frac{F'^2}{2})
\label{4.18}
\end{eqnarray}
which yields,
\begin{eqnarray}
T_{uu}^{(R)}=\frac{1}{96\pi}\Big(FF^{''}-\frac{F'^{2}}{2}\Big)+C_{uu}
\label{4.19}
\end{eqnarray}
where $C_{uu}$ is an integration constant.

  Now, in the coordinates $U = -\kappa e^{-\kappa u}$ and $V=\kappa e^{\kappa v}$, we have the following relations for components of the energy-momentum tensor:
\begin{eqnarray}
T_{UU}^{(R)} = \frac{T_{uu}^{(R)}}{(\kappa U)^2}
\label{4.19n1}
\\
T_{VV}^{(R)} = \frac{T_{vv}^{(R)}}{(\kappa V)^2}~.
\label{4.19n2}
\end{eqnarray}
According to the definition of Unruh vacuum (proper vacuum for studying Hawking effect) for outgoing mode $T_{UU}$ must be finite at future horizon ($U\rightarrow 0$), implying that a freely falling observer sees a finite amount of flux at the outer horizon.
This requires $T_{uu}^{(R)}(r\rightarrow r_H)=0$, leads to $C_{uu}=\frac{F'^2(r_H)}{192\pi}$. The corresponding condition on the ingoing mode for the Unruh vacuum - $T_{VV}$ is finite at infinity -  is satisfied by default since, due to chirality, these are absent ($T_{vv}^{(R)}=0$). This choice of the Unruh vacuum is similar to imposing the covariant boundary condition \cite{Banerjee:2008wq}.  Note, however, that the Unruh condition on the ingoing modes $T_{vv}^{(R)}(r\rightarrow\infty)=0$ is applied at asymptotic infinity where the theory is non-chiral. This does not affect our interpretation since, asymptotically, the anomaly (\ref{4.16}) vanishes. Hence the results from the chiral expressions will agree with the non-chiral ones at asymptotic infinity. Indeed, the Hawking flux, obtained by taking the asymptotic infinity limit ($r\rightarrow \infty$) of (\ref{4.19}),  
\begin{eqnarray}
T_{uu}^{(R)}(r\rightarrow\infty)=C_{uu}=\frac{F'^2(r_H)}{192\pi}=\frac{\kappa^2}{48\pi}
\label{4.20}
\end{eqnarray}
where $\kappa$ is the surface gravity of the black hole given by (\ref{2.20}), reproduces the known result corresponding to the Hawking temperature (\ref{2.27}) in $\hbar=1$ unit \cite{Robinson:2005pd,Iso:2006wa,Iso:2006ut,Banerjee:2007qs,Banerjee:2008az,Banerjee:2007uc,Banerjee:2008wq,Bonora:2008he,Bonora:2008nk,Iso:2007nf,Morita:2008qn}. The other terms in (\ref{4.19}) drop out due to asymptotic flatness.  
%%%%%%%%%%%%%%%%%%%%%%%%%%%%%%%%%%%%%%%%%%%%%%%%%%%%%%%%%%  
\section{Chirality, quantum tunneling and Hawking temperature}
    Here, using the chirality condition (\ref{4.05}), we will derive the tunneling probability, which will eventually yield the Hawking temperature. Under the ($t-r$) sector of the metric (\ref{2.01}), this condition corresponds to,
\begin{eqnarray}
\partial_t\phi(r,t)=\pm F(r)\partial_r\phi(r,t)
\label{4.21}
\end{eqnarray}
As before $+(-)$ stand for left (right) mode. Putting the standard WKB ansatz (\ref{2.04}) and the expansion for $S(r,t)$ (\ref{2.06})
in (\ref{4.21}), we get in the $\hbar\rightarrow 0$ limit the familiar semiclassical Hamilton-Jacobi equation (\ref{2.07}),
which is the basic equation in the tunneling mechanism for studying Hawking radiation. This has been derived earlier from the Klein-Gordon equation with the background metric (\ref{2.01}) and the ansatz (\ref{2.04}) \cite{Paddy,Banerjee:2008cf}.

      Now proceeding in the similar way as earlier, 
we obtain the solution for $S_0(r,t)$ as,
\begin{eqnarray}
S_0(r,t)=\omega t\pm\omega\int\frac{dr}{F(r)}
\label{4.22}
\end{eqnarray}
which is nothing but (\ref{2.10}) for $f(r)=g(r)=F(r)$.
Expressing (\ref{4.22}) in the null tortoise coordinates (see (\ref{4.01})), defined inside and outside of the event horizon,  
we obtain
\begin{eqnarray}
\Big(S_0^{(R)}(r,t)\Big)_{in} = \omega(t_{in}-r^*_{in}) = \omega u_{in};
\\
\Big(S_0^{(L)} (r,t)\Big)_{in} = \omega (t_{in}+r^*_{in}) = \omega v_{in}
\label{4.22n1}
\\
\Big(S_0^{(R)}(r,t)\Big)_{out} = \omega(t_{out}-r^*_{out}) = \omega u_{out}; 
\\
\Big(S_0^{(L)} (r,t)\Big)_{out} = \omega (t_{out}+r^*_{out}) = \omega v_{out}~.
\label{4.22n2}
\end{eqnarray}
Substituting these in (\ref{2.04}) one can obtain the right and left modes for both sectors:
\begin{eqnarray}
&&\Big(\phi^{(R)}\Big)_{in}=e^{-\frac{i}{\hbar}\omega u_{in}};\,\,\, \Big(\phi^{(L)}\Big)_{in}=e^{-\frac{i}{\hbar}\omega v_{in}}
\label{4.23n}
\\
&&\Big(\phi^{(R)}\Big)_{out}=e^{-\frac{i}{\hbar}\omega u_{out}};\,\,\, \Big(\phi^{(L)}\Big)_{out}=e^{-\frac{i}{\hbar}\omega v_{out}}
\label{4.23}
\end{eqnarray}
which satisfy the condition (\ref{4.04}). Precisely these modes were  used previously to find the trace anomaly \cite{Fulling1} as well as the chiral (gravitational) anomaly \cite{Fulling:1986rk} by the point splitting regularization technique. In our formulation these modes (\ref{4.23}) are a natural consequence of chirality.

     Now in the tunneling formalism, as stated earlier, a virtual pair of particles is produced in the black hole. One of this pair can quantum mechanically tunnel through the horizon. This particle is observed at infinity while the other goes towards the center of the black hole. While crossing the horizon the nature of the coordinates changes. This can be explained in the following way. The Kruskal time ($T$) and space ($X$) coordinates inside and outside the horizon are defined by (\ref{2.18}) and (\ref{2.19}) respectively.
In section 2.1.1 of chapter 2 it has been shown that these two sets of coordinates are connected by the relations (\ref{2.22}) and (\ref{2.23}),
so that the Kruskal coordinates get identified as $T_{in} = T_{out}$ and $X_{in} = X_{out}$.
In particular, for the Schwarzschild metric, the surface gravity is $\kappa=\frac{1}{4M}$
and thus the extra term connecting $t_{in}$ and $t_{out}$ is given by ($-2\pi iM$).
Such a result (for the Schwarzschild case) was earlier discussed in \cite{Akhmedov}.
It should be mentioned that instead of Kruskal coordinates one can do the analysis employing
the Painleve coordinates \cite{Painleve} since in these coordinates the apparent singularity at
the horizon is also removed. Nevertheless it is noteworthy that the coordinate transformation from
the Schwarzschild-like to the Painleve coordinates contains a singularity at the horizon while transformations
(\ref{2.18}) and (\ref{2.19}) do not have such singularity. Therefore,
Painleve coordinates are not suitable for the present analysis. In addition, there is an
arbitrariness in the mapping $T_{in} =  T_{out}$ and $X_{in} = X_{out}$ because they can also be obtained
if, instead of (\ref{2.22}) and (\ref{2.23}), we use the following relations
\begin{eqnarray}
t_{in} = t_{out}+i\frac{\pi}{2\kappa};\,\,\,\, r^*_{in} = r^*_{out}-i\frac{\pi}{2\kappa}~.
\label{4.24}
\end{eqnarray}
However, this set of coordinates gives unphysical results. This issue will be clarified in the subsequent analysis.
Therefore, we can exclude the set of coordinates given by equation (\ref{4.24}).

   Employing equations (\ref{2.22}) and (\ref{2.23}) in equation (\ref{4.01}),
we can obtain the relations that connect the null coordinates defined inside
and outside the black hole event horizon
\begin{eqnarray}
&&u_{in}=t_{in} -  r^{*}_{in} = u_{out}-i\frac{\pi}{\kappa}
\label{4.25}
\\
&&v_{in}=t_{in}+ r^{*}_{in} = v_{out}~.
\label{4.26}
\end{eqnarray}
Under these transformations the modes in equations (\ref{4.23n}) and (\ref{4.23}) which
are travelling in the ``$in$'' and ``$out$'' sectors of the black hole horizon are connected through the expressions
\begin{eqnarray}
&&\phi^{(R)}_{in} = e^{-\frac{\pi\omega}{\hbar \kappa}} \phi^{(R)}_{out}
\label{4.27}
\\
&&\phi^{(L)}_{in} =  \phi^{(L)}_{out}~.
\label{4.28}
\end{eqnarray}
Since the left moving mode travels towards the center of the black hole, its probability to go inside,
as measured by an external observer, is expected to be unity. This is easily verified by computing
\begin{eqnarray}
P^{(L)}=|\phi^{(L)}_{in}|^2 = |\phi^{(L)}_{out}|^2=1
\label{4.29}
\end{eqnarray}
where we have used (\ref{4.28}) to recast $\phi^{(L)}_{in}$ in terms of $\phi^{(L)}_{out}$
since measurements are done by an outside observer.
This shows that the left moving (ingoing) mode is trapped inside the black hole, as expected.

   On the other hand the right moving mode, i.e. $\phi^{(R)}_{in}$, tunnels through the event horizon.
So to calculate the tunneling probability as seen by an external observer one has to use the transformation
(\ref{4.27}) to recast $\phi^{(R)}_{in}$ in terms of $\phi^{(R)}_{out}$. Then we find
\begin{eqnarray}
P^{(R)}=|\phi^{(R)}_{in}|^2 = |e^{-\frac{\pi\omega}{\hbar \kappa}}\phi^{(R)}_{out}|^2
=e^{-\frac{2\pi\omega}{\hbar \kappa}}~.
\label{4.30}
\end{eqnarray}
Finally, using the principle of ``detailed balance'' \cite{Paddy}, i.e.
$P^{(R)}=e^{-\frac{\omega}{T_H}}P^{(L)}=e^{-\frac{\omega}{T_H}}$,  and making comparison with
equation (\ref{4.30}), one immediately reproduces the Hawking temperature (\ref{2.27}). This is the standard expression corresponding to the flux (\ref{4.20}) in units of $\hbar=1$.

   It should be pointed out that the tunneling probability given by equation (\ref{4.30})
goes to zero in the classical limit ($\hbar\rightarrow 0$), which is expected since classically
a black hole cannot radiate. On the other hand, if the above analysis is repeated by utilizing the set of
coordinates given in equation (\ref{4.24}), then $P^{(R)}=e^{\frac{2\pi\omega}{\hbar \kappa}}$.
This probability diverges in the classical limit which is unphysical.
Therefore, the set of coordinates presented in equation (\ref{4.24}) are not appropriate
for our study.

     As we observe the ingoing modes are trapped and do not play any role in the computation of the Hawking temperature. A similar feature occurs in the anomaly approach where the ingoing modes are neglected leading to a chiral theory that eventually yields the flux. These observations provide a physical picture of chirality connecting the tunneling and anomaly methods. 
%%%%%%%%%%%%%%%%%%%%%%%%%%%%%%%%%%%%%%%%%%%%%%%%%%%%%%%%%%%%%%     

\section{Discussions}
     We have shown that the notion of chirality pervades the anomaly and tunneling formalisms thereby providing a close connection between them. This is true both from a physical as well as algebraic perspective. The chiral restrictions play a pivotal role in the abstraction of the anomaly from which the flux is computed. The same restrictions, in the tunneling formalism, lead to the Hawking temperature corresponding to that flux.

    A dimensional reduction is known to reduce the theory effectively to a two dimensional conformal theory near the event horizon. The ingoing (left moving) modes are lost inside the horizon. They cannot contribute to the near horizon theory thereby rendering it chiral and, hence, anomalous. Using the restrictions imposed by chirality we obtained a form for this (gravitational) anomaly, manifested by a nonconservation of the stress tensor, by starting from the familiar form of the trace anomaly. From a knowledge of the gravitational anomaly we were able to obtain the flux.

   The chirality constraints were then exploited to obtain the equations for the ingoing and outgoing modes in the tunneling formalism, following the standard geometrical (WKB) approximation. We reformulated the tunneling mechanism to highlight the role of coordinate systems in the chiral framework. A specific feature of this reformulation is that explicit treatment of the singularity in (\ref{4.22}) is not required since we do not carry out the integration. Only the modes inside ($\phi_{in}$) and outside ($\phi_{out}$) the horizon, both of which are well defined, are required. The singularity now manifests in the complex transformations (\ref{2.22}) and (\ref{2.23}) that connect these modes across the horizon. In this way our formalism, contrary to the traditional approaches \cite{Paddy,Wilczek} avoids explicit complex path analysis. It is implicit only in the expression for $S_0(r,t)$ (\ref{4.22}). 
The probability for finding the ingoing modes was shown to be unity. These modes do not play any role in the tunneling approach which is the exact analogue of omitting them when considering the effective near horizon theory in the anomaly method.

   It is useful to observe that the crucial role of chirality in both approaches is manifested in the near horizon regime. This reaffirms the universality of the Hawking effect being governed by the properties of the event horizon.

%%%%%%%%%%%%%%%%%%%%%%%%%%%%%%%%%%%%%%%%%%%%%%%%%%%%%%%%%%%%%%%%%%%%%%%%%%%%%%
\chapter{\label{chap:spectrum}Black body spectrum from tunneling mechanism}
%%%%%%%%%%%%%%%%%%%%%%%%%%%%%%%%%%%%%%%%%%%%%%%%%%%%%%%%%%%%%%%%%%%%%%%%%%%%%%%
      So far we have discussed the Hawking effect by the tunneling mechanism.  However, the analysis was confined to obtaining the Hawking temperature only by comparing the tunneling probability of an outgoing particle with the Boltzmann factor. There was no discussion of the spectrum. Hence it is not clear whether this temperature really corresponds to the temperature of a black body spectrum associated with black holes. One has to take recourse to other results to really justify the fact that the temperature found in the tunneling approach is indeed the Hawking black body temperature. Indeed, as far as we are aware, there is no discussion of the spectrum in the different variants of the tunneling formalism. In this sense the tunneling method, presented so far, is incomplete.

    In this chapter we rectify this shortcoming. Using density matrix techniques we will directly find the spectrum from a reformulation of the tunneling mechanism discussed in the previous chapter. For both bosons and fermions we obtain a black body spectrum with a temperature that corresponds to the familiar semi-classical Hawking expression. Our results are valid for black holes with spherically symmetric geometry.  
%%%%%%%%%%%%%%%%%%%%%%%%%%%%%%%%%%%%%%%%%%%%%%%%%%%%%%%%% 
\section{Black body spectrum and Hawking flux}
     Here the emission spectrum of the black hole will be calculated by the density matrix technique. It has been shown in chapter \ref{chap:anomaly} that a pair created inside the black hole is represented by the modes (\ref{4.23n}). Since the Hawking effect is observed from outside the black hole, one must recast these modes in terms of the outside coordinates. This will yield the relations between the ``$in$'' and ``$out$'' modes. These are given by (\ref{4.27}) and (\ref{4.28}). These transformations are the essential ingredients of constructing all the physical observables regarding the Hawking effect, because the observer is situated outside the event horizon of the black hole.

     Now to find the black body spectrum and Hawking flux, we first consider $n$ number of non-interacting virtual pairs that are created inside the black hole. Each of these pairs is represented by the modes defined by (\ref{4.23n}). Then the physical state of the system, observed from outside, is given by,
\begin{eqnarray}
|\Psi> = N \sum_n |n^{(L)}_{in}>\otimes|n^{(R)}_{in}>
 = N \sum_n e^{-\frac{\pi n\omega}{\hbar \kappa}}|n^{(L)}_{out}>\otimes|n^{(R)}_{out}>
\label{5.01}
\end{eqnarray}
where  use has been made of the transformations (\ref{4.27}) and (\ref{4.28}). Here $|n^{(L)}_{out}>$ corresponds to $n$ number of left going modes and so on while $N$ is a normalization constant which can be determined by using the normalization condition $<\Psi|\Psi>=1$.  This immediately yields, 
\begin{eqnarray}
N=\frac{1}{\Big(\displaystyle\sum_n e^{-\frac{2\pi n\omega}{\hbar \kappa}}\Big)^{\frac{1}{2}}}~.
\label{5.02}
\end{eqnarray} 
The above sum will be calculated for both bosons and fermions. For bosons $n=0,1,2,3,....$ whereas for fermions $n=0,1$. With these values of $n$ we obtain the normalization constant (\ref{5.02}) as
\begin{eqnarray}
N_{(\textrm {boson})}=\Big(1-e^{-\frac{2\pi\omega}{\hbar \kappa}}\Big)^{\frac{1}{2}}
\label{5.03}
\\
N_{(\textrm {fermion})}=\Big(1+e^{-\frac{2\pi\omega}{\hbar \kappa}}\Big)^{-\frac{1}{2}}~.
\label{5.04}
\end{eqnarray}
Therefore the normalized physical states of the system for bosons and fermions are, respectively,
\begin{eqnarray}
|\Psi>_{(\textrm{boson})}= \Big(1-e^{-\frac{2\pi\omega}{\hbar \kappa}}\Big)^{\frac{1}{2}} \sum_n e^{-\frac{\pi n\omega}{\hbar \kappa}}|n^{(L)}_{out}>\otimes|n^{(R)}_{out}>,
\label{5.05}
\\
|\Psi>_{(\textrm{fermion})}= \Big(1+e^{-\frac{2\pi\omega}{\hbar \kappa}}\Big)^{-\frac{1}{2}} \sum_n e^{-\frac{\pi n\omega}{\hbar \kappa}}|n^{(L)}_{out}>\otimes|n^{(R)}_{out}>~.
\label{5.06}
\end{eqnarray}
From here on our analysis will be only for bosons since for fermions the analysis is identical. For bosons the density matrix operator of the system is given by,
\begin{eqnarray}
{\hat\rho}_{(\textrm{boson})}&=&|\Psi>_{(\textrm{boson})}<\Psi|_{(\textrm{boson})}
\nonumber
\\
&=&\Big(1-e^{-\frac{2\pi\omega}{\hbar \kappa}}\Big) \sum_{n,m} e^{-\frac{\pi n\omega}{\hbar \kappa}} e^{-\frac{\pi m\omega}{\hbar \kappa}} |n^{(L)}_{out}>\otimes|n^{(R)}_{out}>  <m^{(R)}_{out}|\otimes<m^{(L)}_{out}|~.
\label{5.07}
\end{eqnarray}
Now since, as explained in the previous chapter, the ingoing ($L$) modes are completely trapped, they do not contribute to the emission spectrum from the black hole event horizon. Hence tracing out the ingoing (left) modes we obtain the density matrix for the outgoing modes,
\begin{eqnarray}
{\hat{\rho}}^{(R)}_{(\textrm{boson})}= \Big(1-e^{-\frac{2\pi\omega}{\hbar \kappa}}\Big) \sum_{n} e^{-\frac{2\pi n\omega}{\hbar \kappa}}|n^{(R)}_{out}>  <n^{(R)}_{out}|~.
\label{5.08}
\end{eqnarray}
Therefore the average number of particles detected at asymptotic infinity is given by,
\begin{eqnarray}
<n>_{(\textrm{boson})}={\textrm{trace}}({\hat{n}} {\hat{\rho}}^{(R)}_{(\textrm{boson})})&=& \Big(1-e^{-\frac{2\pi\omega}{\hbar \kappa}}\Big) \sum_{n} n e^{-\frac{2\pi n\omega}{\hbar \kappa}}
\nonumber
\\
&=&\Big(1-e^{-\frac{2\pi\omega}{\hbar \kappa}}\Big) (-\frac{\hbar \kappa}{2\pi})\frac{\partial}{\partial\omega}\Big(\sum_{n}  e^{-\frac{2\pi n\omega}{\hbar \kappa}}\Big)
\nonumber
\\
&=&\Big(1-e^{-\frac{2\pi\omega}{\hbar \kappa}}\Big) (-\frac{\hbar \kappa}{2\pi})\frac{\partial}{\partial\omega}\Big(\frac{1}{1-e^{-\frac{2\pi\omega}{\hbar \kappa}}}\Big)
\nonumber
\\
&=&\frac{1}{e^{\frac{2\pi\omega}{\hbar \kappa}}-1}
\label{5.09}
\end{eqnarray}
where the trace is taken over all $|n^{(R)}_{out}>$ eigenstates. This is the Bose distribution. Similar analysis for fermions leads to the Fermi distribution:
\begin{eqnarray}
<n>_{(\textrm{fermion})}=\frac{1}{e^{\frac{2\pi\omega}{\hbar \kappa}}+1}~.
\label{5.10} 
\end{eqnarray}
Note that both these distributions correspond to a black body spectrum with a temperature given by the Hawking expression (\ref{2.27}). 
Correspondingly, the Hawking flux can be obtained by integrating the above distribution functions over all $\omega$'s. For fermions it is given by,
\begin{eqnarray}
{\textrm{Flux}}=\frac{1}{\pi}\int_0^\infty \frac{\omega ~d\omega}{e^{\frac{2\pi\omega}{\hbar K}}+1} =\frac{\hbar^2 \kappa^2}{48\pi}
\label{5.11}
\end{eqnarray}
Similarly, the Hawking flux for bosons can be calculated, leading to the same answer.
%%%%%%%%%%%%%%%%%%%%%%%%%%%%%%%%%%%%%%%%%%%%%%%%%%%%%%%%%%%%%%%%%%

\section{Discussions}
    We have adopted a novel formulation of the tunneling mechanism which was elaborated in the previous chapter to find the emission spectrum from the black hole event horizon. 
%This reformulation highlights the role of coordinate systems. A particular feature of this reformulation, as seen from the analysis of the earlier chapter, is that explicit treatment of the singularity in (\ref{4.22}) is not required since we do not carry out the complex path integration. Of course, the singularity at the event horizon is manifested in the transformations (\ref{2.22}) and (\ref{2.23}). In this way our formalism, contrary to the traditional approaches \cite{Paddy,Wilczek} avoids explicit complex path analysis. It is implicit only in the expression for $S_0(r,t)$ (\ref{4.22}). 
Here the computations were done in terms of the basic modes obtained earlier in Chapter \ref{chap:anomaly}. From the density matrix constructed from these modes we were able to directly reproduce the black body spectrum, for either bosons or fermions, from a black hole with a temperature corresponding to the standard Hawking expression. We feel that the lack of such an analysis was a gap in the existing tunneling formulations \cite{Paddy,Wilczek,Jiang,Chen,Medved1,Medved2,Singleton} which yield only the temperature rather that the actual black body spectrum. Finally, although our analysis was done for a static spherically symmetric space-time in Einstein gravity, this can be applied as well for a stationary black hole, for example Kerr-Newman metric \cite{Umetsu:2009ra} and also for black holes in other gravity theory like Ho\v{r}ava-Lifshit theory \cite{Majhi:2009xh}.

%%%%%%%%%%%%%%%%%%%%%%%%%%%%%%%%%%%%%%%%%%%%%%
\chapter{\label{chap:GEMS}Global embedding and Hawking-Unruh effect}
%%%%%%%%%%%%%%%%%%%%%%%%%%%%%%%%%%%%%%%%%%%%%%%%%%%%%%%%%%%%%%%%%%
 After Hawking's famous work \cite{Hawking:1974rv} - radiation of black holes - known as {\it{Hawking effect}}, it is now well understood that this is related to the event horizon of a black hole. A closely related effect is the {\it{Unruh effect}} \cite{Unruh:1976db}, where a similar type of horizon is experienced by a uniformly accelerated observer on the Minkowski space-time. A unified description of them was first put forwarded by Deser and Levin \cite{Deser:1997ri,Deser:1998xb} which was a sequel to an earlier attempt \cite{Narnhofer:1996zk}. This is called the global embedding Minkowskian space (GEMS) approach. In this approach, the relevant detector in curved space-time (namely Hawking detector) and its event horizon map to the Rindler detector in the corresponding flat higher dimensional embedding space \cite{Goenner,Rosen} and its event horizon. 
Then identifying the acceleration of the Unruh detector, the Unruh temperature can be calculated. Finally, use of the Tolman relation \cite{Tolman} yields the Hawking temperature. In this picture the Unruh temperature is interpreted as a local Hawking temperature. Subsequently, this unified approach to determine the Hawking temperature using the Unruh effect was applied for several black hole space-times \cite{Kim:2000ct,Tian:2005yj,Brynjolfsson:2008uc,Hong:2003xz}. However the results were confined to four dimensions and the calculations were done case by case, taking specific black hole metrics. It was not clear whether the technique was applicable to complicated examples like the Kerr-Newman metric which lacks spherical symmetry.

      The motivation of this chapter is to give a modified presentation of the GEMS approach that naturally admits generalization. Higher dimensional black holes with different metrics, including Kerr-Newman, are considered. 
Using this new embedding, the local Hawking temperature (Unruh temperature) will be derived. Then the Tolman formula leads to the Hawking temperature.

      We shall first introduce a new global embedding which embeds only the ($t-r$)-sector of the curved metric into a flat space. It will be shown that this embedding is enough to derive the Hawking result using the Deser-Levin approach \cite{Deser:1997ri,Deser:1998xb}, instead of the full embedding of the curved space-time. Hence we might as well call this the reduced global embedding. This is actually motivated from the fact that an $N$-dimensional black hole metric effectively reduces to a $2$ -dimensional metric (only the ($t-r$)-sector) near the event horizon by the dimensional reduction technique \cite{Robinson:2005pd,Solodukhin:1998tc,Carlip:1998wz,Iso:2006ut,Umetsu:2009ra} (for examples see Appendix \ref{appendixdim1}). Furthermore, this $2$-dimensional metric is enough to find the Hawking quantities if the back scattering effect is ignored. Several spherically symmetric static metrics will be exemplified. Also, to show the utility of this reduced global embedding, we shall discuss the most general solution of the Einstein gravity - Kerr-Newman space-time, whose full global embedding is difficult to find. Since the reduced embedding involves just the two dimensional ($t-r$)-sector, black holes in arbitrary dimensions can be treated. 
In this sense our approach is valid for any higher dimensional black hole.

     The organization of the chapter is as follows. In section \ref{reduce1} we shall find the reduced global embedding of several black hole space-times which are spherically symmetric. In the next section the power of this approach will be exploited to find the Unruh/Hawking temperature for the Kerr-Newman black hole. Finally, we shall give our concluding remarks. One appendix, briefly reviewing dimensional reduction, is also included.

\section{\label{reduce1}Reduced global embedding}
    A unified picture of Hawking effect \cite{Hawking:1974rv} and Unruh effect \cite{Unruh:1976db} was established by the global embedding of a curved space-time into a higher dimensional flat space \cite{Deser:1998xb}. Subsequently, this unified approach to determine the Hawking temperature using the Unruh effect was applied for several black hole space-times \cite{Kim:2000ct,Tian:2005yj}, but usually these are spherically symmetric. For instance, no discussion on the Kerr-Newman black hole has been given, because it is difficult to find the full global embedding.

     Since the Hawking effect is governed solely by properties of the event horizon, it is enough to consider the near horizon theory. As already stated, this is a two dimensional theory obtained by dimensional reduction of the full theory. Its metric is just the ($t-r$)-sector of the original metric.

   In the following sub-sections we shall find the global embedding of the near horizon effective $2$-dimensional theory. Then the usual local Hawking temperature will be calculated. Technicalities are considerably simplified and our method is general enough to include different black hole metrics.

\subsection{Schwarzschild metric}
   Near the event horizon the physics is given by just the two dimensional ($t-r$) -sector of the full Schwarzschild metric \cite{Robinson:2005pd} (see also Appendix \ref{appendixdim1}):
\begin{eqnarray}
ds^2 = g_{tt}dt^2 + g_{rr}dr^2 = \Big(1-\frac{2M}{r}\Big)dt^2 -\frac{dr^2}{1-\frac{2M}{r}}.
\label{12.01}
\end{eqnarray} 
It is interesting to see that this can be globally embedded in a flat $D=3$ space as,
\begin{eqnarray}
ds^2 = (dz^0)^2 - (dz^1)^2 - (dz^2)^2
\label{12.02}
\end{eqnarray}
by the following relations among the flat and curved coordinates:
\begin{eqnarray}
&&z^0_{out} = \kappa^{-1} \Big(1-\frac{2M}{r}\Big)^{1/2} \textrm{sinh}(\kappa t),\,\,\,\
z^1_{out} = \kappa^{-1} \Big(1-\frac{2M}{r}\Big)^{1/2} \textrm{cosh}(\kappa t),
\nonumber
\\
&&z^0_{in} = \kappa^{-1} \Big(\frac{2M}{r} - 1\Big)^{1/2} \textrm{cosh}(\kappa t),\,\,\,\
z^1_{in} = \kappa^{-1} \Big(\frac{2M}{r} - 1\Big)^{1/2} \textrm{sinh}(\kappa t),
\nonumber
\\
&&z^2 = \int dr \Big(1+\frac{r_Hr^2 + r_H^2r + r_H^3}{r^3}\Big)^{1/2},
\label{12.03}
\end{eqnarray}
where the surface gravity $\kappa=\frac{1}{4M}$ and the event horizon is located at $r_H=2M$. The suffix ``$in$'' (``$out$'') refer to the inside (outside) of the event horizon while variables without any suffix (like $z^2$) imply that these are valid on both sides of the horizon. We shall follow these notations throughout the chapter.

      Now if a detector moves according to constant $r$ (Hawking detector) outside the horizon in the curved space, then the detector corresponding to the $z$ coordinates, moves on the constant $z^2$ plane and it will follow the hyperbolic trajectory
\begin{eqnarray}
\Big(z^1_{out}\Big)^2 - \Big(z^0_{out}\Big)^2 = 16 M^2 \Big(1 - \frac{2M}{r}\Big) 
%= \frac{1}{{\tilde{a}}^2}.
\label{12.04}
\end{eqnarray}
Such a detector is usually called as the Unruh detector, since the metric corresponding to $z^2$ constant plane:
\begin{eqnarray}
ds^2_{(z^0,z^1)} &=& (dz^0_{out})^2 - (dz^1_{out})^2 
\nonumber
\\
&=& \Big(1-\frac{2M}{r}\Big) dt^2 - \frac{16M^4}{r^4} \Big(1-\frac{2M}{r}\Big)^{-1} dr^2
\label{12.04n1}
\end{eqnarray}
is in generalized Rindler form,
\begin{equation}
ds^2_{Rind} = \alpha^2 H(r)^2 dt^2 - H'(r)^2 dr^2
\label{12.04n2}
\end{equation}
with
\begin{eqnarray}
H(r) = \kappa^{-1} \Big(1-\frac{2M}{r}\Big)^{1/2};\,\,\,\ \alpha = \kappa~.
\label{12.04n3}
\end{eqnarray}
For the generalized Rindler metric (\ref{12.04n2}) the acceleration of the Unruh detector is given by \cite{Carrol},
\begin{eqnarray}
{\tilde{a}} = \frac{1}{H(r)}
\label{12.04n4}
\end{eqnarray}
and according to Unruh \cite{Unruh:1976db}, the accelerated detector will see a thermal spectrum in the Minkowski vacuum with the local Hawking (Unruh) temperature given by (\ref{1.08}). 
This shows that the Unruh detector is moving in the ($z^0_{out}, z^1_{out}$) flat plane with a uniform acceleration ${\tilde{a}}= \frac{1}{4M}\Big(1 - \frac{2M}{r}\Big)^{-1/2}$ and it will see a thermal spectrum in the Minkowski vacuum with local Hawking temperature given by,
\begin{eqnarray}
T = \frac{\hbar {\tilde{a}}}{2\pi} = \frac{\hbar}{8\pi M} \Big(1 - \frac{2M}{r}\Big)^{-1/2}.
\label{12.05}
\end{eqnarray}
So we see that with the help of the reduced global embedding the local Hawking temperature near the horizon can easily be obtained. 
The same analysis can also be done in the upcoming discussions, although we shall not mention explicitly. We shall only read off the acceleration of the Unruh detector by finding the appropriate hyperbolic trajectory and thereby the local Hawking (Unruh) temperature will be derived.

    Now the temperature measured by any observer away from the horizon can be obtained by using the Tolman formula \cite{Tolman} which ensures constancy between the product of temperatures and corresponding Tolman factors measured at two different points in space-time. This formula is given by \cite{Tolman}:
\begin{eqnarray}
\sqrt{g_{tt}}~ T = \sqrt{g_{0_{tt}}}~ T_0
\label{12.06}
\end{eqnarray}
where, in this case, the quantities on the left hand side are measured near the horizon whereas those on the right hand side are measured away from the horizon (say at $r_0$). Since away from the horizon the space-time is given by the full metric, $g_{0_{tt}}$ must correspond to the $dt^2$ coefficient of the full (four dimensional) metric.

   For the case of Schwarzschild metric $g_{tt} = 1-2M/r$, $g_{0_{tt}} = 1-2M/r_0$. Now the Hawking effect is observed at infinity ($r_0 = \infty$), where $g_{0_{tt}} = 1$. Hence, use of the Tolman formula (\ref{12.06}) immediately yields the Hawking temperature:
\begin{eqnarray}
T_H \equiv T_0 = {\sqrt{g_{tt}}}~ T =  \frac{\hbar}{8\pi M }.
\label{12.07}
\end{eqnarray}
Thus, use of the reduced embedding instead of the embedding of the full metric is sufficient to get the answer.

\subsection {Reissner-Nordstr$\ddot{\textrm{o}}$m metric}
   In this case, the effective metric near the event horizon is given by \cite{Robinson:2005pd} (see also appendix \ref{appendixdim1}),
\begin{eqnarray}
ds^2 = \Big(1 - \frac{2M}{r} + \frac{Q^2}{r^2}\Big)dt^2 - \frac{dr^2}{1 - \frac{2M}{r} + \frac{Q^2}{r^2}}.
\label{12.08}
\end{eqnarray}
This metric can be globally embedded into the $D=4$ dimensional flat metric as,
\begin{eqnarray}
ds^2 = (dz^0)^2 - (dz^1)^2 - (dz^2)^2 + (dz^3)^2
\label{12.09}
\end{eqnarray}
where the coordinate transformations are:
\begin{eqnarray}
&&z^0_{out} = \kappa^{-1} \Big(1-\frac{2M}{r} + \frac{Q^2}{r^2}\Big)^{1/2} \textrm{sinh}(\kappa t),\,\,\,\
z^1_{out} = \kappa^{-1} \Big(1-\frac{2M}{r} + \frac{Q^2}{r^2}\Big)^{1/2} \textrm{cosh}(\kappa t),
\nonumber
\\
&&z^0_{in} = \kappa^{-1} \Big(\frac{2M}{r} - \frac{Q^2}{r^2} - 1\Big)^{1/2} \textrm{cosh}(\kappa t),\,\,\,\
z^1_{in} = \kappa^{-1} \Big(\frac{2M}{r} - \frac{Q^2}{r^2} - 1\Big)^{1/2} \textrm{sinh}(\kappa t),
\nonumber
\\
&&z^2 = \int dr \Big[1+\frac{r^2(r_+ + r_-) + r_+^2(r + r_+)}{r^2(r-r_-)}\Big]^{1/2},
\nonumber
\\
&&z^3 = \int dr \Big[\frac{4r_+^5r_-}{r^4(r_+ - r_-)^2}\Big]^{1/2}.
\label{12.10}
\end{eqnarray}
Here in this case the surface gravity $\kappa = \frac{r_+ - r_-}{2r_+^2}$ and $r_{\pm}=M\pm\sqrt{M^2-Q^2}$. The black hole event horizon is given by $r_H=r_+$. Note that for $Q=0$, the above transformations reduce to the Schwarzschild case (\ref{12.03}). 
The Hawking detector moving in the curved space outside the horizon, following a constant $r$ trajectory, maps to the Unruh detector on the constant ($z^2,z^3$) surface. The trajectory of the Unruh detector is given by
\begin{eqnarray}
\Big(z^1_{out}\Big)^2 - \Big(z^0_{out}\Big)^2 = \Big(\frac{r_+ - r_-}{2r_+^2}\Big)^{-2} \Big(1-\frac{2M}{r} + \frac{Q^2}{r^2}\Big)=\frac{1}{{\tilde{a}}^2}.
\label{12.11}
\end{eqnarray}
This, according to Unruh \cite{Unruh:1976db}, immediately leads to the local Hawking temperature 
\begin{eqnarray}
T=\frac{\hbar {\tilde{a}}}{2\pi}=\frac{\hbar(r_+ - r_-)}{4\pi r_+^2\sqrt{1-2M/r+Q^2/r^2}}
\label{12.11n1}
\end{eqnarray}
which was also obtained from the full global embedding \cite{Deser:1998xb}. Again, since in this case $g_{0_{tt}} = 1-2M/r_0 + Q^2/r_0^2$ which reduces to unity at $r_0=\infty$ and $g_{tt} = 1-2M/r+Q^2/r^2$, use of Tolman formula (\ref{12.06}) leads to the standard Hawking temperature 
\begin{eqnarray}
T_H \equiv T_0=\sqrt{g_{tt}}~ T=\frac{\hbar(r_+ - r_-)}{4\pi r_+^2}~.
\label{12.11n2}
\end{eqnarray}

\subsection{Schwarzschild-AdS metric}
   Near the event horizon the relevant effective metric is \cite{Robinson:2005pd} (see also Appendix \ref{appendixdim1}),
\begin{eqnarray}
ds^2 = \Big(1-\frac{2M}{r}+\frac{r^2}{R^2}\Big)dt^2 - \frac{dr^2}{\Big(1-\frac{2M}{r}+\frac{r^2}{R^2}\Big)},
\label{12.12}
\end{eqnarray}
where $R$ is related to the cosmological constant $\Lambda= -1/R^2$.
This metric can be globally embedded in the flat space (\ref{12.09}) with the following coordinate transformations:
\begin{eqnarray}
&&z^0_{out} = \kappa^{-1} \Big(1-\frac{2M}{r} + \frac{r^2}{R^2}\Big)^{1/2} \textrm{sinh}(\kappa t),\,\,\
z^1_{out} = \kappa^{-1} \Big(1-\frac{2M}{r} + \frac{r^2}{R^2}\Big)^{1/2} \textrm{cosh}(\kappa t),
\nonumber
\\
&&z^0_{in} = \kappa^{-1} \Big(\frac{2M}{r} - \frac{r^2}{R^2} - 1\Big)^{1/2} \textrm{cosh}(\kappa t),\,\,\,\
z^1_{in} = \kappa^{-1} \Big(\frac{2M}{r} - \frac{r^2}{R^2} - 1\Big)^{1/2} \textrm{sinh}(\kappa t),
\nonumber
\\
&&z^2 = \int dr \Big[1+\Big(\frac{R^3 + R r_H^2}{R^2 + 3r_H^2}\Big)^2 \frac{r^2r_H+rr_H^2 + r_H^3}{r^3(r^2+rr_H+r_H^2+R^2)}\Big]^{1/2},
\nonumber
\\
&&z^3 = \int dr \Big[\frac{(R^4 + 10R^2r_H^2+9r_H^4)(r^2+rr_H+r_H^2)}{(r^2+rr_H+r_H^2+R^2)(R^2+3r_H^2)^2}\Big]^{1/2}
\label{12.13}
\end{eqnarray}
where the surface gravity $\kappa=\frac{R^2+3r_H^2}{2r_HR^2}$ and the event horizon $r_H$ is given by the largest root of the equation $1-\frac{2M}{r_H} + \frac{r^2_H}{R^2}=0$.
Note that in the $R\rightarrow\infty$ limit these transformations reduce to those for the Schwarzschild case (\ref{12.03}). We observe that the Unruh detector on the ($z^2,z^3$) surface (i.e. the Hawking detector moving outside the event horizon on a constant $r$ surface) follows the hyperbolic trajectory:
\begin{eqnarray}
\Big(z^1_{out}\Big)^2 - \Big(z^0_{out}\Big)^2 = \Big(\frac{R^2+3r_H^2}{2r_HR^2}\Big)^{-2}\Big(1-\frac{2M}{r} + \frac{r^2}{R^2}\Big)=\frac{1}{{\tilde{a}}^2}
\label{12.14}
\end{eqnarray} 
leading to the local Hawking temperature 
\begin{eqnarray}
T=\frac{\hbar {\tilde{a}}}{2\pi}=\frac{\hbar\kappa}{2\pi\Big(1-\frac{2M}{r}+\frac{r^2}{R^2}\Big)^{1/2}}.
\label{12.14n1}
\end{eqnarray}
This result was obtained earlier \cite{Deser:1998xb}, but with more technical complexities,  from the embedding of the full metric.

      It may be pointed out that for the present case, the observer must be at a finite distance away from the event horizon, since the space-time is asymptotically AdS. Therefore, if the observer is far away from the horizon ($r_0>>r$) where $g_{0_{tt}}=1-2M/r_0+r_0^2/R^2$, then use of (\ref{12.06}) immediately leads to the temperature measured at $r_0$:
\begin{eqnarray}
T_0 = \frac{\hbar\kappa}{2\pi\sqrt{1-2M/r_0 + r_0^2/R^2}}.
\label{12.15}
\end{eqnarray}     
Now, this shows that $T_0\rightarrow 0$ as $r_0\rightarrow \infty$; i.e. no Hawking particles are present far from horizon.

\section{Kerr-Newman metric}
           So far we have discussed a unified picture of Unruh and Hawking effects using our reduced global embedding approach for spherically symmetric metrics, reproducing standard results. However, our approach was technically simpler since it involved the embedding of just the two dimensional near horizon metric. Now we shall explore the real power of this new embedding.

           The utility of the reduced embedding approach comes to the fore for the Kerr-Newman black hole which is not spherically symmetric. The embedding for the full metric, as far as we are aware, is not done in the literature.

   The effective $2$-dimensional metric near the event horizon is given by (\ref{app1.14}) \cite{Iso:2006ut,Umetsu:2009ra} (see Appendix \ref{appendixdim1}),
%\begin{eqnarray}
%ds^2 = \frac{\Delta}{r^2+a^2}dt^2 - \frac{r^2+a^2}{\Delta}dr^2,
%\label{12.16}
%\end{eqnarray} 
%where 
%\begin{eqnarray}
%&&\Delta = r^2-2Mr+a^2+Q^2 = (r-r_+)(r-r_-);\,\,\,\ a=\frac{J}{M};
%\nonumber
%\\
%&&r_\pm = M\pm\sqrt{M^2-a^2-Q^2}.
%\label{12.17}
%\end{eqnarray}
%The event horizon is located at $r=r_+$. 
This metric can be embedded in the following $D=5$-dimensional flat space:
\begin{eqnarray}
ds^2=\Big(dz^0\Big)^2 -\Big(dz^1\Big)^2-\Big(dz^2\Big)^2 + \Big(dz^3\Big)^2 + \Big(dz^4\Big)^2,
\label{12.18}
\end{eqnarray}
where the coordinate transformations are
\begin{eqnarray}
&&z^0_{out} = \kappa^{-1} \Big(1-\frac{2Mr}{r^2+a^2} + \frac{Q^2}{r^2+a^2}\Big)^{1/2} \textrm{sinh}(\kappa t),
\nonumber
\\
&&z^1_{out} = \kappa^{-1} \Big(1-\frac{2Mr}{r^2+a^2} + \frac{Q^2}{r^2+a^2}\Big)^{1/2} \textrm{cosh}(\kappa t),
\nonumber
\\
&&z^0_{in} = \kappa^{-1} \Big(\frac{2Mr}{r^2+a^2} - \frac{Q^2}{r^2+a^2} - 1\Big)^{1/2} \textrm{cosh}(\kappa t),
\nonumber
\\
&&z^1_{in} = \kappa^{-1} \Big(\frac{2Mr}{r^2+a^2} - \frac{Q^2}{r^2+a^2} - 1\Big)^{1/2} \textrm{sinh}(\kappa t),
\nonumber
\\
&&z^2 = \int dr \Big[1+\frac{(r^2+a^2)(r_+ + r_-) + r_+^2(r + r_+)}{(r^2+a^2)(r-r_-)}\Big]^{1/2},
\nonumber
\\
&&z^3 = \int dr \Big[\frac{4r_+^5r_-}{(r^2+a^2)^2(r_+ - r_-)^2}\Big]^{1/2},
\nonumber
\\
&&z^4 = \int dr a\Big[\frac{r_+ + r_-}{(a^2+r_-^2)(r_- - r)} + \frac{4(a^2 + r_+^2)(a^2-r_+r_- + (r_+ +r_-)r)}{(r_+ - r_-)^2 (a^2 + r^2)^3} 
\nonumber
\\
&&+ \frac{4r_+r_-(a^2+2r_+^2)}{(r_+ - r_-)^2(a^2+r^2)^2} + \frac{rr_- - a^2 + r_+(r+r_-)}{(a^2+r_-^2)(a^2+r^2)}\Big]^{1/2}.
\label{12.19}
\end{eqnarray}
Here the surface gravity $\kappa = \frac{r_+-r_-}{2(r_+^2 + a^2)}$.
For $Q=0, a=0$, as expected, the above transformations reduce to the Schwarzschild case (\ref{12.03}) while only for $a=0$ these reduce to the Reissner-Nordstr$\ddot{\textrm{o}}$m case (\ref{12.10}).

    As before, the trajectory adopted by the Unruh detector on the constant ($z^2,z^3,z^4$) surface corresponding to the Hawking detector on the constant $r$ surface is given by the hyperbolic form,
\begin{eqnarray}
\Big(z^1_{out}\Big)^2 - \Big(z^0_{out}\Big)^2 = \kappa^{-2}\Big(1-\frac{2Mr}{r^2+a^2} + \frac{Q^2}{r^2+a^2}\Big)=\frac{1}{{\tilde{a}}^2}.
\label{12.20}
\end{eqnarray}
Hence the Unruh or local Hawking temperature is
\begin{eqnarray}
T=\frac{\hbar {\tilde{a}}}{2\pi}=\frac{\hbar\kappa}{2\pi\sqrt{\Big(1-\frac{2Mr}{r^2+a^2} + \frac{Q^2}{r^2+a^2}\Big)}}.
\label{12.21}
\end{eqnarray} 
Finally, since $g_{tt} = 1-\frac{2Mr}{r^2+a^2} + \frac{Q^2}{r^2+a^2}$ (corresponding to the near horizon reduced two dimensional metric) and $g_{0_{tt}}=\frac{r_0^2-2Mr_0+a^2+Q^2-a^2 {\textrm{sin}}^2\theta}{r_0^2+a^2{\textrm{cos}}^2\theta}$ (corresponding to the full four dimensional metric), use of the Tolman relation (\ref{12.06}) leads to the Hawking temperature
\begin{eqnarray}
T_H \equiv T_0=\frac{\sqrt{g_{tt}}}{\sqrt{(g_0{_{tt}})_{r_0\rightarrow\infty}}}~T = \frac{\hbar\kappa}{2\pi} = \frac{\hbar(r_+-r_-)}{4\pi(r_+^2 + a^2)},
\label{12.22}
\end{eqnarray}
which is the well known result \cite{Iso:2006ut}.

\section{Conclusion}
      We provided a new approach to the study of Hawking/Unruh effects including their unification, initiated in \cite{Deser:1997ri,Deser:1998xb,Narnhofer:1996zk}, popularly known as global embedding Minkowskian space-time (GEMS). Contrary to the usual formulation \cite{Deser:1997ri,Deser:1998xb,Narnhofer:1996zk,Kim:2000ct,Tian:2005yj,Brynjolfsson:2008uc}, the full embedding was avoided. Rather, we required the embedding of just the two dimensional ($t-r$)-sector of the theory. This was a consequence of the fact that the effective near horizon theory is basically two dimensional. Only near horizon theory is significant since Hawking/Unruh effects are governed solely by properties of the event horizon.

    This two dimensional embedding ensued remarkable technical simplifications whereby the treatment of more general black holes (e.g. those lacking spherical symmetry like the Kerr-Newman) was feasible. Also, black holes in any dimensions were automatically considered since the embedding just required the ($t-r$)-sector.

%%%%%%%%%%%%%%%%%%%%%%%%%%%%%%%%%%%%%%%%%%%%%%%%%
\begin{subappendices}
\chapter*{Appendix}
\section{\label{appendixdim1}Dimensional reduction technique}
\renewcommand{\theequation}{6A.\arabic{equation}}
\setcounter{equation}{0}  % reset counter
       Dimensional reduction has been discussed in various contexts in the literature \cite{Robinson:2005pd,Carlip:1998wz,Iso:2006ut,Umetsu:2009ra}. Here we briefly summarise the technique and findings relevant for our study. Two specific examples are considered.
\vskip 5mm
{\bf{Spherically symmetric static metric:}}\\
     Let us consider a spherically symmetric static metric
\begin{eqnarray}
ds^2 = f(r)dt^2 - \frac{dr^2}{f(r)}-r^2(d\theta^2 + \textrm{sin}^2\theta d\phi^2)
\label{app1.01}
\end{eqnarray}
whose event horizon is given by $f(r=r_H)=0$. Now in terms of the tortoise coordinate (\ref{2.21})
the above metric takes the following form
\begin{eqnarray}
ds^2=f(r(r^*))\Big(dt^2 -dr^{*^2}\Big) - r^2(r^*)(d\theta^2+\textrm{sin}^2\theta d\phi^2)
\label{app1.02}
\end{eqnarray}
Then the free action for massless scalar field under this background is given by
\begin{eqnarray}
A &=& -\int d^4x~\sqrt{-g}~\Phi\nabla_\mu\nabla^\mu\Phi
\nonumber
\\ 
&=&  -\int dtdr^*d\theta d\phi~\textrm{sin}\theta~\Phi\Big[r^2(r^*)(\partial^2_t - \partial^2_{r^*}) - 2r(r^*)f(r(r^*))\partial_{r^*}\Big]\Phi
\nonumber
\\
&-& \int dtdr^*d\theta d\phi ~ f(r(r^*))\textrm{sin}\theta~\Phi L^2\Phi,
\label{app1.03}
\end{eqnarray}
where
\begin{eqnarray}
L^2 = -\frac{1}{\textrm{sin}^2\theta}\partial^2_\phi - \textrm{cot}\theta \partial_\theta - \partial^2_\theta.
\label{app1.04}
\end{eqnarray} 
Substituting the partial wave decomposition for $\Phi$ 
\begin{eqnarray}
\Phi(t,r^*,\theta,\phi) = \displaystyle\sum_{l,n}\phi_{ln}(t,r^*)Y_{ln}(\theta,\phi)
\label{app1.05}
\end{eqnarray}
in (\ref{app1.03}) and using the eigenvalue equation $L^2Y_{ln}(\theta,\phi) = l(l+1)Y_{ln}(\theta,\phi)$ followed by the orthonormality condition, $\int d\theta d\phi~{\textrm{sin}}\theta~Y_{l'n'}Y_{ln} = \delta_{l'l}\delta_{n'n}$, we obtain,
\begin{eqnarray}
A &=& -\displaystyle\sum_{l,n}\int dtdr^* r^2(r^*)\phi_{ln}\Big[\partial^2_t -\partial^2_{r^*}\Big]\phi_{ln}
\nonumber
\\
&+& \displaystyle\sum_{l,n}\int dtdr^*r^2(r^*)\phi_{ln} f(r(r^*))\Big[\frac{l(l+1)}{r^2(r^*)}+\frac{1}{r(r^*)}\partial_{r}f(r)\Big]\phi_{ln}.
\label{app1.06}
\end{eqnarray}
Now near the horizon ($r\rightarrow r_H$), $f(r)\rightarrow 0$, and hence the above action reduces to the following form:
\begin{eqnarray}
A \simeq -\displaystyle\sum_{l,n}\int dtdr^* r_H^2(r^*)\phi_{ln}\Big[\partial^2_t -\partial^2_{r^*}\Big]\phi_{ln}.
\label{app1.07}
\end{eqnarray}
Transforming back to the original coordinates ($t,r$), yields
\begin{eqnarray}
A \simeq -\displaystyle\sum_{l,n}\int dtdr r_H^2\phi_{ln}\Big[\frac{1}{f(r)}\partial^2_t -\partial_r(f\partial_{r})\Big]\phi_{ln}.
\label{app1.08}
\end{eqnarray}
It must be noted that the above action is the original action for the infinite collection of free scalar fields under the metric \cite{Iso:2006ut},
%The factor $r^2$ in the above action can be interpreted as a dilaton background coupled to the scalar field. Thus, physics near the horizon can be effectively described by an infinite collection of free massless scalar fields, each propagating in a ($1+1$) dimensional space-time given by the ($t-r$)-sector of the ($3+1$)-dimensional metric (\ref{app1}), that is
\begin{eqnarray}
ds^2 = f(r)dt^2 - \frac{dr^2}{f(r)},
\label{app1.09}
\end{eqnarray} 
which is just the ($t-r$)-sector of the ($3+1$)-dimensional metric (\ref{app1.01}). It is simple to extend this analysis for arbitrary dimensions \cite{RN1}. The effective theory is again given by the metric (\ref{app1.09}).

{\bf{Kerr-Newman metric:}}\\
     The most general black hole in four dimensional Einstein theory is given by the Kerr-Newman metric,
\begin{eqnarray}
ds^2&=&\frac{\Delta -a^2 \sin^2\theta}{\Sigma}dt^2+\frac{2a\sin^2\theta}{\Sigma}(r^2+a^2-\Delta)dtd\varphi
\nonumber
\\
&-&\frac{a^2\Delta \sin^2\theta-(r^2+a^2)^2}{\Sigma}\sin^2\theta d\varphi^2-\frac{\Sigma}{\Delta}dr^2-\Sigma d\theta^2
\label{app1.10}
\end{eqnarray}
where
\begin{eqnarray}
&&a \equiv  \frac{J}{M};\,\,\ 
\Sigma \equiv  r^2+a^2\cos^2 \theta;\,\,\
\Delta \equiv  r^2-2Mr+a^2+Q^2=(r-r_+)(r-r_-),
\nonumber
\\
&&r_\pm  = M\pm \sqrt{M^2-a^2-Q^2},
\label{app1.11}
\end{eqnarray}
while $M, J, Q$ and $r_{+(-)}$ are the mass, angular momentum, electrical charge and
the outer (inner) horizon of the Kerr-Newman black hole, respectively.
The event horizon is located at $r=r_+$.

   Proceeding in a similar way as above, the action for a massless complex scalar field, in the near horizon limit, reduces to the following form \cite{Iso:2006ut,Umetsu:2009ra}:
\begin{eqnarray}
A &=& -\int d^4x \sqrt{-g} \Phi^*(\nabla_\mu + iA_\mu)(\nabla^\mu - iA^\mu)\Phi 
\nonumber
\\
&=& -\displaystyle{\sum_{l,n}}\int dt dr(r^2+a^2)\phi^{*}_{ln}\Big[\frac{r^2+a^2}{\Delta}\Big(\partial_t-iA_t\Big)^2
\nonumber
\\
&-& \partial_{r}\frac{\Delta}{r^2+a^2}\partial_{r}\Big]
\phi_{ln},
\label{app1.12}
\end{eqnarray}
where
\begin{eqnarray}
A_t=-eV(r)-n\Omega(r);\,\,\, V(r)= \frac{Qr}{r^2+a^2}, \Omega(r)=\frac{a}{r^2+a^2}.
\label{app1.13}
\end{eqnarray}
Here $e$ is the charge of the scalar field.
This shows that each partial wave mode of the fields can be described near the horizon as a ($1+1$) dimensional complex scalar field with two $U(1)$ gauge potentials $V(r)$, $\Omega(r)$ and the dilaton field $\psi=r^2+a^2$. It should be noted that the above action for each $l,n$ can also be obtained from the complex scalar field action in the background of the metric 
\begin{eqnarray}
ds^2 = F(r)dt^2 - \frac{dr^2}{F(r)}; \,\,\ F(r)=\frac{\Delta}{r^2+a^2}
\label{app1.14}
\end{eqnarray}
with the dilaton field $\psi=r^2+a^2$. 
%For detailed study on dimensional reduction see \cite{Umetsu:2009ra} where the result (\ref{app13}) has also been obtained.
Thus, the effective near horizon theory is two dimensional with a metric given by (\ref{app1.14}). Although here we have presented the dimensional reduction technique for the $4$ dimensional case, it can also be generalized to higher dimensional black holes. In that case one again gets a two dimensional ($t-r$) metric near the event horizon. For example see \cite{Iso:2006xj}. 
\end{subappendices}

%%%%%%%%%%%%%%%%%%%%%%%%%%%%%%%%%%%%%%%%%%%%%%%%%%%%
\chapter{\label{chap:spectroscopy}Quantum tunneling and black hole spectroscopy}
%%%%%%%%%%%%%%%%%%%%%%%%%%%%%%%%%%%%%%%%%%%%%%%%%%%%
Since the birth of Einstein's theory of gravitation, black holes have been one of the main
topics that attracted the attention and consumed a big part of the working time of the scientific community.
In particular, the computation of black hole entropy in the semi-classical and furthermore in the quantum regime has been a very
difficult and (in its full extent) unsolved problem that has created a lot of controversy. A closely related issue is the spectrum of this entropy as well as that of the horizon area. This will be our main concern.

     Bekenstein was the first to show
that there is a  lower bound (quantum) in the increase of the area of the black hole horizon when
a neutral (test) particle is absorbed \cite{Bekenstein:1973ur}
\begin{eqnarray}
(\Delta{A})_{min}=8\pi \l_{pl}^{2}
\label{6.01}
\end{eqnarray}
where we use gravitational units, i.e. $G=c=1$, and $\l_{pl}=(G \hbar /c^{3})^{1/2}$ is the Planck length.
Later on, Hod  considered the case of a charged particle assimilated by a Reissner-Nordstr\"om black hole and
derived a smaller bound for the increase of the black hole area \cite{Hod:1999nb}
\begin{eqnarray}
(\Delta{A})_{min}=4 \l_{pl}^{2}~.
\label{6.02}
\end{eqnarray}
At the same time, a new research direction was pursued; namely the derivation of
the area as well as the entropy spectrum of black holes  utilizing  the quasinormal modes of black holes \cite{Hod:1998vk}
\footnote{For some works on this direction see, for instance, \cite{Setare:2003bd} and references therein.}.
In this framework, the result obtained is of the form
\begin{eqnarray}
(\Delta{A})_{min}=4 \l_{pl}^{2} \ln k
\label{6.03}
\end{eqnarray}
where $k=3$. A similar expression was first put forward by
Bekenstein and Mukhanov \cite{muk} who employed the ``bit counting'' process.
However in that case $k$ is equal to $2$. Such a spectrum can also be derived in the context of quantum
geometrodynamics \cite{vaz}. Furthermore, using this result one can
find the corrections to entropy consistent with Gibbs' paradox \cite{kif}.

      Another significant attempt was to fix the Immirzi parameter in the framework of Loop Quantum Gravity  \cite{Dreyer:2002vy}
but it was unsuccessful \cite{Domagala:2004jt}. Furthermore, contrary to Hod's statement for a uniformly spaced
area spectrum of generic Kerr-Newman black holes, it was proven that the area spacing of Kerr black hole
is not equidistant \cite{Setare:2004uu}.
However, a new interpretation for the black hole quasinormal modes was proposed \cite{Maggiore:2007nq}
which rejuvenated the interest in this direction.
In this framework the area spectrum is evenly spaced and the area quantum for the Schwarschild
as well as for the Kerr black hole is given by (\ref{6.01}) \cite{Vagenas:2008yi}.
While this is in agreement with the old result of Bekenstein, it disagrees with (\ref{6.02}).

      In this chapter, we will use a modified version of the tunneling mechanism, discussed in chapter \ref{chap:anomaly}, to derive the entropy-area spectrum of a black hole. In this formalism, as explained earlier, a virtual pair of particles is produced just inside the black hole. One member of this pair is trapped inside the black hole while the other member can quantum mechanically tunnel through the horizon. This is ultimately observed at infinity, giving rise to the Hawking flux. Now the uncertainty in the energy of the emitted particle is calculated from a simple quantum mechanical point of view. Then exploiting information theory ({\it{entropy as lack of information}}) and the first law of thermodynamics, we infer that the entropy spectrum is evenly spaced for both {\it{Einstein's gravity}} as well as {\it{Einstein-Gauss-Bonnet gravity}}. Now, since in Einstein gravity, entropy is proportional to horizon area of black hole, the area spectrum is also evenly spaced and the spacing is shown to be exactly identical with one computed by Hod \cite{Hod:1999nb} who studied the assimilation of charged particle by a Reissner-Nordstr\"om black hole. On the contrary, in more general theories like Einstein-Gauss-Bonnet gravity, the entropy is not proportional to the area and therefore area spacing is not equidistant. This also agrees with recent conclusions \cite{Daw,Wei}.

    The organization of the chapter goes as follows. In section 7.1, we briefly  review the results of dimensional reduction presented earlier in Appendix \ref{appendixdim1} which will be used in this chapter. In section 7.2,
we compute the entropy and area spectrum of black hole solutions of both Einstein gravity and Einstein-Gauss-Bonnet gravity.
Finally, section 7.3 is devoted to a brief summary of our results and concluding remarks.
%%%%%%%%%%%%%%%%%%%%%%%%%%%%%%%%%%%%%%%%%%%%%%%%%%%%%%%%%%%%%%%%
\section{Near horizon modes}
     According to the no hair theorem,
{\it{collapse leads to a black hole endowed with mass, charge, angular momentum and no other free parameters}}.
The most general black hole in four dimensional Einstein theory is given by the Kerr-Newman metric (\ref{app1.10}).

      Now considering complex scalar fields in the Kerr-Newmann black hole background and then substituting the partial wave decomposition of the scalar field in terms of spherical harmonics it has been shown in Appendix \ref{appendixdim1} that near the horizon the action reduces to an effective 2-dimensional action (\ref{app1.12}) for free complex scalar field.
From (\ref{app1.12}) one can easily derive the equation of motion of the field $\phi_{lm}$ for the $l=0$ mode. We will denote this mode as $\phi$. This equation is given by the Klein-Gordon equation:
\begin{eqnarray}   
\Big[\frac{1}{F(r)}(\partial_t-iA_t)^2 - F(r)\partial^2_r - F'(r)\partial_r\Big]\phi=0 ~.
\label{6.19}
\end{eqnarray}
%Observe that this is just the Klein-Gordon equation for a free scalar field with $U(1)$ gauge field $A_t$ in the following 2-dimensional space-time metric
%\begin{eqnarray}
%ds^2=-F(r)dt^2+\frac{dr^2}{F(r)}~.
%\label{6.20}
%\end{eqnarray} 
%This shows that near the horizon the theory is dimensionally reduced to a 2-dimensional theory with the metric (\ref{6.20}) \cite{Iso:2006ut,Umetsu:2009ra}.

     Now proceeding in a similar way as presented in chapter \ref{chap:anomaly},
we obtain the relations between the modes defined inside and outside the black hole event horizon, which are given by (\ref{4.27}) and (\ref{4.28}).
In this case, the surface gravity $\kappa$ is defined by,
\begin{eqnarray}
\kappa = \frac{1}{2} \frac{d F(r)}{dr} \Big|_{r=r_{+}}
\label{6.21}
\end{eqnarray}
and the energy of the particle ($\omega$) as seen from an asymptotic observer is identified as,
\begin{eqnarray}
\omega=E-eV(r_+)-m\Omega(r_+).
\label{6.22}
\end{eqnarray}
Here $E$ is the conserved quantity corresponding to a timelike Killing vector ($1,0,0,0$). The other variables $V(r_+)$ and $\Omega(r_+)$ are the electric potential and the angular velocity calculated on the horizon.

   The same analysis also goes through for a D-dimensional spherically symmetric static black hole which is a solution for Einstein-Gauss-Bonnet theory \cite{Myers}:
\begin{eqnarray}
ds^2 = F(r)dt^2-\frac{dr^2}{F(r)}-r^2 d\Omega^2_{(D-2)}.
\label{6.22n1}
\end{eqnarray}
Here $F(r)$ is given by
\begin{eqnarray}
F(r)=1+\frac{r^2}{2\alpha}\Big[1-\Big(1+\frac{4\alpha\bar{\omega}}{r^{D-1}}\Big)^{\frac{1}{2}}\Big]
\label{6.23}
\end{eqnarray}
with
\begin{eqnarray}
\alpha&=&(D-3)(D-4)\alpha_{GB}
\label{6.24}
\\
\bar{\omega}&=&\frac{16\pi}{(D-2)\Sigma_{D-2}}M
\label{6.25}
\end{eqnarray}
where $\alpha_{GB}$, $\Sigma_{D-2}$ and $M$ are the coupling constant for the Gauss-Bonnet term in the action,
the volume of unit ($D-2$) sphere and the ADM mass, respectively.
Approaching in a similar manner for the dimensional reduction near the horizon, as discussed in Appendix \ref{appendixdim1} (also see \cite{RN1}) for arbitrary dimensional case), one can show that the physics can be effectively described by the 2-dimensional
form (\ref{app1.14}). 
%Here $F(r)$ is given by
%\begin{eqnarray}
%F(r)=1+\frac{r^2}{2\alpha}\Big[1-\Big(1+\frac{4\alpha\bar{\omega}}{r^{D-1}}\Big)^{\frac{1}{2}}\Big]
%\label{6.23}
%\end{eqnarray}
%with
%\begin{eqnarray}
%\alpha&=&(D-3)(D-4)\alpha_{GB}
%\label{6.24}
%\\
%\bar{\omega}&=&\frac{16\pi}{(D-2)\Sigma_{D-2}}M
%\label{6.25}
%\end{eqnarray}
%where $\alpha_{GB}$, $\Sigma_{D-2}$ and $M$ are the coupling constant for the Gauss-Bonnet term in the action,
%the volume of unit ($D-2$) sphere and the ADM mass, respectively. 
Therefore, in the Einstein-Gauss-Bonnet theory
one will obtain the same transformations, namely equations (\ref{4.27}) and (\ref{4.28}),
between the inside and outside modes.

    In the analysis to follow, using the aforementioned transformations,
i.e. equations (\ref{4.27}) and (\ref{4.28}), we will discuss about the spectroscopy
of the entropy and area of black holes.

\section{Entropy and area spectrum}
   In this section we will derive the spectrum for the entropy as well as the area of the black hole defined both in Einstein and Einstein-Gauss-Bonnet gravity. It has already been mentioned that the pair production occurs inside the horizon. The relevant modes are $\phi_{in}^{(L)}$ and $\phi_{in}^{(R)}$.
It has also been shown in chapter \ref{chap:anomaly} that the left mode is trapped inside the black hole while the right mode can tunnel through the horizon which is observed at asymptotic infinity.
Therefore, the average value of $\omega$ will be computed as
\begin{eqnarray}
<\omega> = \frac{\displaystyle{\int_0^\infty  \left(\phi^{(R)}_{in}\right)^{*}
\omega  \phi^{(R)}_{in} d\omega}}
{\displaystyle{\int_0^\infty  \left(\phi^{(R)}_{in}\right)^*
\phi^{(R)}_{in} d\omega}} ~.
\label{6.26}
\end{eqnarray}
It should be stressed that the above definition is unique since the pair production occurs inside the black hole and it is the right moving mode that eventually escapes (tunnels) through the horizon.

      To compute this expression it is important to recall that the observer is located outside the event horizon. Therefore it is essential to recast the ``$in$'' expressions into their corresponding ``$out$'' expressions using the map (\ref{4.27}) and then perform the integrations.
Consequently, using (\ref{4.27}) in the above we will obtain the average energy of the particle, as seen by the external observer. This is given by,
\begin{eqnarray}
<\omega>&=& \frac{\displaystyle{\int_0^\infty e^{-\frac{\pi\omega}{\hbar \kappa}} \left(\phi^{(R)}_{out}\right)^{*}
\omega e^{-\frac{\pi\omega}{\hbar \kappa}} \phi^{(R)}_{out} d\omega}}
{\displaystyle{\int_0^\infty e^{-\frac{\pi\omega}{\hbar \kappa}} \left(\phi^{(R)}_{out}\right)^*
e^{-\frac{\pi\omega}{\hbar \kappa}} \phi^{(R)}_{out} d\omega}}
\nonumber
\\
&=&\frac{\displaystyle{\int_0^\infty  \omega e^{-\beta\omega}d\omega}}
{\displaystyle{\int_0^\infty  e^{-\beta\omega}d\omega}}
\nonumber
\\
&=& \frac{\displaystyle{-\frac{\partial}{\partial\beta}\left(\int_0^\infty  e^{-\beta\omega}d\omega\right)}}
{\displaystyle{\int_0^\infty  e^{-\beta\omega}d\omega}}=\beta^{-1}
\label{6.27}
\end{eqnarray}
where $\beta$ is the inverse Hawking temperature
\begin{eqnarray}
\beta=\frac{2\pi}{\hbar \kappa}=\frac{1}{T_H}.
\end{eqnarray}
In a similar way one can compute the average squared energy of the particle detected by
the asymptotic observer
\begin{eqnarray}
<\omega^{2}>&=&
\frac{\displaystyle{\int_0^\infty e^{-\frac{\pi\omega}{\hbar \kappa}}\left(\phi^{(R)}_{out}\right)^{*}
\omega^2 e^{-\frac{\pi\omega}{\hbar \kappa}} \phi^{(R)}_{out} d\omega}}
{\displaystyle{\int_0^\infty e^{-\frac{\pi\omega}{\hbar \kappa}}\left(\phi^{(R)}_{out}\right)^{*}
e^{-\frac{\pi\omega}{\hbar \kappa}} \phi^{(R)}_{out}d\omega}}
=\frac{2}{\beta^{2}}~.
\label{6.28}
\end{eqnarray}
Now it is straightforward to evaluate the uncertainty, employing equations (\ref{6.27}) and (\ref{6.28}),
in the detected energy $\omega$
\begin{eqnarray}
\left(\Delta\omega \right)=\sqrt{<\!\!\omega^{2}\!\!>-<\!\!\omega\!\!>^2}\,=\, \beta^{-1} = T_H
\label{6.29}
\end{eqnarray}
which is nothing but the Hawking temperature $T_{H}$.
Hence the characteristic frequency of the outgoing mode is given by,
\begin{eqnarray}
\Delta f = \frac{\Delta\omega}{\hbar}=\frac{T_H}{\hbar}.
\label{6.30}
\end{eqnarray}

   Now the uncertainty (\ref{6.29}) in $\omega$ can be seen as the lack of information in energy of the black hole due to the particle emission. This is because $\omega$ is the effective energy defined in (\ref{6.22}).
Also, since in information theory the entropy is lack of information, 
then the first law of black hole mechanics can be exploited to connect these quantities,
\begin{eqnarray}
S_{bh}=\int \frac{\Delta \omega}{T_H}.
\label{6.31}
\end{eqnarray}
Substituting the value of $T_H$ from (\ref{6.30}) in the above we obtain
\begin{eqnarray}
S_{bh} = \frac{1}{\hbar}\int \frac{\Delta\omega}{\Delta f}.
\label{6.32}
\end{eqnarray}
Now according to the Bohr-Sommerfeld quantization rule
\begin{eqnarray}
\int \frac{\Delta\omega}{\Delta f} = n\hbar
\label{6.33}
\end{eqnarray}
where $n=1,2,3....$. Hence, combining (\ref{6.32}) and (\ref{6.33}), we can immediately infer that the entropy is quantized and the spectrum is given by
\begin{eqnarray}
S_{bh}=n.
\label{6.34}
\end{eqnarray}
This shows that the entropy of the black hole is quantized in units of the identity, $\Delta S_{bh} = (n+1) - n = 1$. Thus the corresponding
spectrum is equidistant for both {\it Einstein} as well as {\it Einstein-Gauss-Bonnet} theory.
Moreover, since the entropy of a black hole in {\it{Einstein}} theory is given by the Bekenstein-Hawking formula,
\begin{eqnarray}
S_{bh}=\frac{A}{4\l_{pl}^{2}}.
\label{6.35}
\end{eqnarray}
the area spectrum is evenly spaced and given by,
\begin{eqnarray}
 A_{n}=4\l_{pl}^{2}\, n\,
\label{6.37}
\end{eqnarray}
with  $n=1,2,3,\ldots $~.
Consequently, the area of the black hole horizon is also quantized with the area quantum given by,
\begin{eqnarray}
\Delta A=4\l_{pl}^{2}~.
\label{6.36}
\end{eqnarray}
%implying that the area spectrum is evenly spaced
%\begin{eqnarray}
% A_{n}=4\l_{pl}^{2}\, n\,
%\label{6.37}
%\end{eqnarray}
%with  $n=1,2,3,\ldots $~.

      A couple of comments are in order here. First, in Einstein gravity, the area quantum is universal in the sense that
it is independent of the black hole parameters. This universality was also derived in the
context of a new interpretation of quasinormal moles of black holes \cite{Maggiore:2007nq, Vagenas:2008yi}.
Second, the same value was also obtained earlier by Hod
by considering the Heisenberg uncertainty principle and Schwinger-type charge emission process \cite{Hod:1999nb}.

    On the contrary, in Einstein-Gauss-Bonnet theory, the black hole entropy is given by
\begin{eqnarray}
S_{bh}=\frac{A}{4}\Big[1+2\alpha\Big(\frac{D-2}{D-4}\Big)\Big(\frac{A}{\Sigma_{D-2}}\Big)^{-\frac{2}{D-2}}\Big]
\label{6.38}
\end{eqnarray}
which shows that entropy is not proportional to area. Therefore in this case the area spacing is not equidistant. The explicit form of the area spectrum is not be given here since (\ref{6.38}) does not have any analytic solution for $A$ in terms of $S_{bh}$. This is compatible with recent findings \cite{Daw,Wei}.

\section{Discussions}
We have calculated the entropy and area spectra of a black hole which is a solution of either Einstein or Einstein-Gauss-Bonnet (EGB) theory.
The computations were pursued in the framework of the tunneling method as reformulated in chapter \ref{chap:anomaly}.
In both cases entropy spectrum is equispaced and the quantum of spacing is identical.
Since in Einstein gravity, the entropy is proportional to the horizon area, the spectrum for the corresponding
area is also equally spaced.
The area quantum obtained here is equal to $4\l^{2}_{pl}$. This exactly reproduces the result of
Hod who studied the assimilation of a charged particle by a Reissner-Nordstr\"om black hole \cite{Hod:1999nb}.
In addition, the area quantum $4\l^{2}_{pl}$ is smaller than that
given by Bekenstein for neutral particles \cite{Bekenstein:1973ur} as well as the one computed in the
context of black hole quasinormal modes \cite{Maggiore:2007nq,Vagenas:2008yi}.

     Furthermore, for the computation of the area quantum obtained here, concepts from
statistical physics, quantum mechanics and black hole physics were combined in the following sense. First the uncertainty in energy of a emitted particle from the black hole horizon was calculated from the simple quantuam mechanical averaging process. Then exploiting the statistical information theory ({\it entropy is lack of information}) in the first law of black hole mechanics combined with Bohr-Sommerfeld quantization rule, the entropy/area quantization has been discussed.  Since this is done on the basis of the fundamental concepts of physics, it seems
that the result reached in our analysis is a better approximation (since a quantum theory of gravity which will give a definite
answer to the quantization of black hole entropy/area is still lacking).
Finally, the equality between our result and that of Hod for the area quantum may be due to the similarity
between the tunneling mechanism and the Schwinger mechanism (for a further discussion on this similarity see \cite{Paddy,Kim}).
On the other hand in Einstein - Gauss - Bonnet gravity, since  entropy is not proportional to area, the spectrum of area is not evenly spaced. 
This method is general enough to discuss entropy and area spectra for the black holes in other type of gravity theories like Ho\v{r}ava-Lifshitz gravity \cite{Majhi:2009xh}. Here also the entropy spectrum comes out to be evenly spaced while that of area is not. Hence, it may be legitimate to say that for gravity theories, in general, {\it{the notion of the quantum of entropy is more natural than the quantum of area}}.
However, one should mention that since our calculations are based on a semi-classical approximation,
the spacing obtained here is valid for large values of $n$ and for $s$-wave ($l=0$ mode).

%%%%%%%%%%%%%%%%%%%%%%%%%%%%%%%%%%%%%%%%%%%%%%
\chapter{\label{chap:statistical}Statistical origin of gravity}
%%%%%%%%%%%%%%%%%%%%%%%%%%%%%%%%%%%%%%%%%%%%%%%%%%%%%%%%%%%%%%%%%%
       There are numerous evidences \cite{Bekenstein:1973ur,Hawking:1974rv,Bardeen:1973gs} which show that gravity and thermodynamics are closely connected to each other. Recently, there has been a growing consensus \cite{Jacobson:1995ab,Kothawala:2007em,Verlinde:2010hp} that gravity need not be interpreted as a fundamental force, rather it is an emergent phenomenon just like thermodynamics and hydrodynamics. The fundamental role of gravity is replaced by thermodynamical interpretations leading to similar or equivalent results. Nevertheless, understanding the entropic or thermodynamic origin of gravity is far from complete since the arguments are more heuristic than concrete and depend upon specific ansatz or assumptions.

    In this chapter, using certain basic results derived in the earlier chapters (also see \cite{Banerjee:2008sn,Banerjee:2009pf}) in the context of tunneling mechanism, we are able to provide a statistical interpretation of gravity. The starting point is the standard definition of entropy given in statistical mechanics. We show that this entropy gets identified with the action for gravity. Consequently the Einstein equations obtained by a variational principle involving the action can be equivalently obtained by an extremisation of the entropy.

    Furthermore, for a black hole with stationary metric we derive the relation $S_{bh}=E/2T_H$, connecting the entropy ($S_{bh}$) with the Hawking temperature ($T_H$) and energy ($E$). We prove that this energy corresponds to Komar's expression \cite{Komar:1958wp,Wald:1984rg}. Using this fact we show that the relation $S_{bh}=E/2T_H$ is also compatible with the generalised Smarr formula \cite{Smarr:1972kt,Bardeen:1973gs,Gibbons:1976ue}. We mention that this relation was also obtained and discussed in \cite{Padmanabhan:2003pk,Padmanabhan:2009kr}.

\section{Partition function and the relation {\bf{$S_{bh} = \frac{E}{2T_H}$}}}
       We start with the partition function for the space-time with matter field \cite{Gibbons:1976ue},
\begin{eqnarray}
{\cal{Z}} = \int ~ D[g,\Phi] ~  e^{i I[g,\Phi]}
\label{8.01}
\end{eqnarray}
where $I[g,\Phi]$ is the action representing the whole system and $D[g,\Phi]$ is the measure of all field configurations ($g,\Phi$).  
Now consider small fluctuations in the metric ($g$) and the matter field ($\Phi$) in the following form:
\begin{eqnarray}
g=g_0 + {\tilde{g}};\,\,\,\,\ \Phi = \Phi_0 + {\tilde{\Phi}}
\label{8.02}
\end{eqnarray}
where $g_0$ and $\Phi_0$ are the stable background fields satisfying the periodicity conditions and which extremise the action. So they satisfy the classical field equations. Whereas ${\tilde{g}}$ and $\tilde{\Phi}$, the fluctuations around these classical values, are very very small. Expanding $I[g,\Phi]$ around ($g_0,\Phi_0$) we obtain
\begin{eqnarray}
I[g,\Phi] = I[g_0,\Phi_0] + I_2[\tilde{g}] + I_2[\tilde{\Phi}]+{\textrm{higher order terms}}.
\label{8.03}
\end{eqnarray}
The dominant contribution to the path integral (\ref{8.01}) comes from fields that are near the background fields ($g_0,\Phi_0$). Hence one can neglect all the higher order terms. The first term $I[g_0,\Phi_0]$ leads to the usual Einstein equations and gives rise to the standard area law \cite{Gibbons:1976ue}. On the other hand the second and third terms give the contributions of thermal gravitation and matter quanta respectively on the background contribution $I[g_0,\Phi_0]$. They lead to the (logarithmic) corrections to the usual area law \cite{Hawking:1976ja}. Here, since we want to confine ourself within the usual semi-classical regime, we shall neglect these quadratic terms for the subsequent analysis.
Therefore, keeping only the term $I[g_0,\Phi_0]$ in (\ref{8.03}) we obtain the partition function (\ref{8.01}) as \cite{Gibbons:1976ue},
\begin{eqnarray}
{\cal{Z}} \simeq e^{i I[g_0,\Phi_0]}.
\label{8.04}
\end{eqnarray}
Therefore, adopting the standard definition of entropy in statistical mechanics,
\begin{eqnarray}
S_{bh} = \ln{\cal{Z}} + \frac{E}{T_H}
\label{8.05}
\end{eqnarray}
and using (\ref{8.04}), the entropy of the gravitating system is given by {\footnote{In this chapter we have chosen units such that $k_B=G=\hbar=c=1$.}},
\begin{eqnarray}
S_{bh} = i I[g_0,\Phi_0] + \frac{E}{T_H}
\label{8.06}
\end{eqnarray}
where $E$ and $T_H$ are respectively the energy and temperature of the system.

      It may be pointed out that it is possible to interpret (\ref{8.04}) as defining the partition function of an emergent theory without specifying the detailed configuration of the gravitating system. The validity of such an interpretation is borne out by the subsequent analysis.

     In order to get an explicit expression for $E$, let us consider a specific system - a black hole.  Now thermodynamics of a black hole is universally governed by its properties near the event horizon. It is also well understood that near the event horizon the effective theory becomes two dimensional whose metric is given by the two dimensional ($t-r$)- sector of the original metric \cite{Carlip:1998wz,Robinson:2005pd}.
Correspondingly, the left ($L$) and right ($R$) moving (holomorphic) modes are obtained by solving the appropriate field equation using the geometrical (WKB) approximation. Furthermore, the modes inside and outside the horizon are related by the transformations (\ref{4.27}) and (\ref{4.28}) \cite{Banerjee:2008sn,Banerjee:2009pf}. 
Concentrating on the modes inside the horizon, the $L$ mode gets trapped while the $R$ mode tunnels through the horizon and is eventually observed at asymptotic infinity as Hawking radiation \cite{Banerjee:2008sn,Banerjee:2009pf} {\footnote{For a unified treatment of these issues, see \cite{Banerjee:2010rx}}}.
The average value of the energy, measured from outside, is given by (\ref{6.27}).
Therefore if we consider that the energy $E$ of the system is encoded near the horizon and the total number of pairs created is $n$ among which this energy is distributed, then we must have,
\begin{eqnarray}
E=nT_H
\label{8.07}
\end{eqnarray}
where only the $R$ mode of the pair is significant.

     Now to proceed further, it must be realised that the effective two dimensional curved metric can always be embedded in a flat space which has exactly two space-like coordinates. This is a consequence of a modification in the original GEMS (globally embedding in Minkowskian space) approach of \cite{Deser:1998xb} and has been elaborated by us in Chapter \ref{chap:GEMS}. Hence we may associate each $R$ mode with two degrees of freedom. Then the total number of degrees of freedom for $n$ number of $R$ modes is $N=2n$. Hence, 
from (\ref{8.07}), we obtain the energy of the system as
\begin{eqnarray}
E = \frac{1}{2}NT_H.
\label{8.09}
\end{eqnarray}
As a side remark, it may be noted that (\ref{8.09}) can be interpreted as a consequence of the usual law of equipartition of energy. For instance, if we consider that the energy $E$ is distributed equally over each degree of freedom, then 
(\ref{8.09}) implies that each degree of freedom should contain an energy equal to $T_H/2$, which is nothing but the {\it equipartition law of energy}. The fact that the energy is equally distributed among the degrees of freedom may be understood from the symmetry of two space-like coordinates ($z^1\longleftrightarrow z^2$) such that the metric is unchanged \cite{Banerjee:2010ma} (see chapter \ref{chap:GEMS}). In our subsequent analysis, however, we only require (\ref{8.09}) rather than its interpretation as the law of equipartition of energy.

       Now since there are $N$ number of degrees of freedom in which all the information is encoded, the entropy ($S_{bh}$) of the system must be proportional to $N$. Hence using (\ref{8.06}) we obtain
\begin{eqnarray}
N = N_0 S_{bh} = N_0 (i I[g_0,\Phi_0]  + \frac{E}{T_H}),
\label{8.10}
\end{eqnarray} 
where $N_0$ is a proportionality constant, which will be determined later.
Substituting the value of $N$ from (\ref{8.09}) in (\ref{8.10}) we obtain the expression for the energy of the system as
\begin{eqnarray}
E=\frac{N_0}{2 - N_0}iT_H I[g_0,\Phi_0].
\label{8.11}
\end{eqnarray}
This shows that in the absence of any fluctuations, the energy of a system is actually given by the classical action representing the system. In the following we shall use this expression to find the energy of a stationary black hole. Before that let us substitute the value of $I[g_0,\Phi_0]$ from (\ref{8.11}) in (\ref{8.06}). This immediately leads to a simple relation between the entropy, temperature and energy of the black hole:
\begin{eqnarray}
S_{bh} = \frac{2E}{N_0T_H}.
\label{8.12}
\end{eqnarray}
Now in order to fix the value of ``$N_0$'' we consider the simplest example, the Schwarzschild black hole for which the entropy, energy and temperature are given by,
\begin{eqnarray}
S_{bh} = \frac{A}{4} = 4\pi M^2, \,\,\  E = M,  \,\,\ T_H = \frac{1}{8\pi M},
\label{8.13}
\end{eqnarray}
where ``$M$'' is the mass of the black hole.
Substitution of these in (\ref{8.12}) leads to $N_0 = 4$.

     At this point we want to make a comment on the value of $N_0$. According to standard statistical mechanics one would have thought that $1/N_0 = \ln c$, where $c$ is an integer. Whereas to keep our analysis consistent with semi-classical area law, we obtained $c=e^{1/4}$, which is clearly not an integer. Indeed, any departure from this value of $N_0$ would invalidate the semi-classical area law and hence our analysis. Such a disparity is not peculiar to our approach and has also occurred elsewhere \cite{Maggiore:2007nq,Paddy1}. This may be due to the fact that our analysis is confined within the semi-classical regime, which is valid for large degrees of freedom. In this regime, it is not obvious that a semi-classical computation can reproduce $c$ to be an integer. Furthermore, the above value of $N_0$ is still valid even for very small number of degrees of freedom ($N$), where this semi-classical calculation is unjustified. This also happens in the semi-classical computation of the entropy spectrum of a black hole \cite{Maggiore:2007nq}. The entropy spectrum is found there to be $S_{bh}=2\pi N$ rather than $S_{bh}=N\ln c$ and this discrepancy is identified with the semi-classical approximation. A possible way to resolve such disagreement from standard statistical mechanics may be the full quantum theoretical computation of the number of microstates which is beyond the scope of the present chapter.

     Finally, putting back  $N_0=4$ in (\ref{8.12}) we obtain,
\begin{eqnarray}
S_{bh} = \frac{E}{2T_H}.
\label{8.14}
\end{eqnarray}
Such a relation was later obtained by us for higher dimensional Einstein gravity where $E$ is the Komar conserved quantity \cite{Banerjee:2010ye}.
Before discussing the significance and implications of this relation, we observe that substituting the value of $E$ from (\ref{8.14}) in (\ref{8.11}) with $N_0 = 4$, we obtain
\begin{eqnarray}
S_{bh} = -iI[g_0,\Phi_0].
\label{8.15}
\end{eqnarray}
Consequently, extremization of entropy leads to Einstein's equations.

\section{Identification of $E$ in Einstein's gravity}
    The relation (\ref{8.14}) is significant for various reasons which will become progressively clear. It is valid for all black hole solutions in Einstein gravity with appropriate identifications consistent with the area law. Here $S_{bh}$ and $T_H$ are easy to identify. These are, respectively, the entropy and Hawking temperature of the black hole. Since energy is one of the most diversely defined entities in general theory of relativity, special care is needed to identify $E$ in (\ref{8.14}). We now show that this $E$ corresponds to Komar's definition \cite{Komar:1958wp,Wald:1984rg}. Simplifying (\ref{8.11}) using $N_0 = 4$ and $T_H=\kappa/2\pi$, we obtain,
\begin{eqnarray}
E = -\frac{i\kappa I[g_0,\Phi_0]}{\pi}.
\label{8.16}
\end{eqnarray}
The classical action $I[g_0,\Phi_0]$ has already been calculated in \cite{Gibbons:1976ue}. The result is,
\begin{eqnarray}
I[g_0,\Phi_0] &=& 2i\pi\kappa^{-1}\Big[\frac{1}{16\pi}\int_{\Sigma}R\xi^ad\Sigma_a + \int_{\Sigma}(T_{ab} - \frac{1}{2}Tg_{ab})\xi^bd\Sigma^a 
\nonumber
\\
&-& \frac{1}{16\pi}\int_{\cal{H}}\epsilon_{abcd}\nabla^c\xi^d\Big],
\label{8.17}
\end{eqnarray}
where $\xi^a\partial/\partial x^a = \partial/\partial t$ is the time translation Killing vector and $\Sigma$ is the space-like hypersurface whose boundary is given by ${\cal{H}}$. Here $T_{ab}$ is the energy-momentum tensor of the matter field whose trace is given by $T$.  
Now for a stationary geometry, $\xi^a\nabla_aR=0$ \cite{Carrol}. Hence for a volume ${\cal{A}}$, we have
\begin{eqnarray}
0 = \int_{\cal{A}}\xi^a\nabla_aR d{\cal{A}} = \int_{\cal{A}}\Big[\nabla_a(\xi^a R) -(\nabla_a\xi^a)R\Big] d{\cal{A}} = \int_{\cal{A}}\nabla_a(\xi^a R)d{\cal{A}}
\label{8.18}
\end{eqnarray}
where in the last step the Killing equation $\nabla_a\xi_b+\nabla_b\xi_a=0$ has been used.
Finally, the Gauss theorem yields,
\begin{eqnarray}
\int_{\Sigma}\xi^a Rd{\Sigma_a} = 0.
\label{8.19}
\end{eqnarray}
Using this in (\ref{8.17}) we obtain,
\begin{eqnarray}
I[g_0,\Phi_0] = 2i\pi\kappa^{-1}\Big[\int_{\Sigma}(T_{ab} - \frac{1}{2}Tg_{ab})\xi^bd\Sigma^a - \frac{1}{16\pi}\int_{\cal{H}}\epsilon_{abcd}\nabla^c\xi^d\Big].
\label{8.20}
\end{eqnarray}
Substituting this in (\ref{8.16}) we obtain the expression for the energy of the gravitating system as
\begin{eqnarray}
E = 2\int_{\Sigma}(T_{ab} - \frac{1}{2}Tg_{ab})\xi^bd\Sigma^a - \frac{1}{8\pi}\int_{\cal{H}}\epsilon_{abcd}\nabla^c\xi^d
\label{8.21}
\end{eqnarray}
which is the Komar expression for energy \cite{Komar:1958wp,Wald:1984rg} corresponding to the time translation Killing vector. Similarly, if there is a rotational Killing vector, then there must be a Komar expression for rotational energy \cite{Carrol,Katz} and the total energy will be their sum.

    Incidentally, (\ref{8.14}) was obtained earlier in \cite{Padmanabhan:2003pk} for static space-time and its implications were discussed in \cite{Padmanabhan:2009kr}. However a specific `ansatz' for entropy compatible with the area law was taken and, more importantly, the Komar energy expression was explicitly used as an input in the derivation. Hence our analysis is different, since we do not invoke any ansatz for the entropy; neither is the Komar expression required at any stage. Rather we prove its occurence in the relation (\ref{8.14}).

     As an explicit check of (\ref{8.14}) for different black hole solutions, we consider a couple of examples. Take the Reissner-Nordstr$\ddot{\textrm{o}}$m (RN) black hole. In this case the entropy and temperature are given by,
\begin{eqnarray}
S_{bh} = \pi r_+^2, \,\,\,\ T_H = \frac{r_+ - r_-}{4\pi r_+^2}; \,\,\ r_{\pm} = M \pm \sqrt{M^2 - Q^2}
\label{8.22}
\end{eqnarray}
where ``$Q$'' is the charge of the black hole.
Substitution of these in (\ref{8.14}) yields,
\begin{eqnarray}
E = M - \frac{Q^2}{r_+},
\label{8.23}
\end{eqnarray}
which is the Komar energy of RN black hole \cite{Banerjee:2009tz}.

     Next we consider the Kerr black hole for which the entropy and temperature are respectively,
\begin{eqnarray}
S_{bh} &=& \pi (r_+^2 + a^2), \,\,\,\ T_H = \frac{r_+ - r_-}{4\pi (r_+^2 + a^2)};
\nonumber
\\
r_{\pm} &=& M \pm \sqrt{M^2 - a^2}, \,\,\,\ a = \frac{J}{M}.
\label{8.24}
\end{eqnarray}
Here ``$J$'' is the angular momentum of the black hole. Substituting (\ref{8.24}) in (\ref{8.14}) we obtain,
\begin{eqnarray}
E = M - 2J\Omega
\label{8.25}
\end{eqnarray}
which is the total Komar energy for Kerr black hole \cite{Dadich,Banerjee:2009tz}. Here $\Omega = \frac{a}{r_+^2 + a^2}$ is the angular velocity at the event horizon $r=r_+$.

   We thus find that, in all cases where $S_{bh}$, $E$, $T$ are known, they satisfy (\ref{8.14}) apart from the area law. In fact, it is possible to take (\ref{8.14}) as the defining relation for the Komar energy.
% in those examples where its direct calculation from (\ref{8.21}) is difficult.
Such an instance is provided by the Kerr-Newman black hole.
% Its Komar energy, as far as we aware, is not known in closed form. However
The entropy and temperature of Kerr-Newman black hole are given by,
\begin{eqnarray}
S_{bh} = \pi (r_+^2 + a^2); \,\,\,\ T_H =\frac{r_+ - r_-}{4\pi(r_+^2 + a^2)}
\label{8.26}
\end{eqnarray}
where
\begin{eqnarray}
r_{\pm} = M \pm \sqrt{M^2-Q^2-a^2}; \,\,\,\ a = \frac{J}{M}.
\label{8.27}
\end{eqnarray}
Now substituting (\ref{8.26}) in (\ref{8.14}) and then using (\ref{8.27}) we obtain the total Komar energy of Kerr-Newman black hole:
\begin{eqnarray}
E = \sqrt{M^2 - Q^2 - a^2} = M - \frac{Q^2}{r_+} - 2J\Omega\Big(1-\frac{Q^2}{2Mr_+}\Big) = M - QV - 2J\Omega,
\label{8.28}
\end{eqnarray}  
where $\Omega = \frac{a}{r_+^2 + a^2}$ is the angular velocity at the event horizon and $V = \frac{Q}{r_+} - \frac{QJ\Omega}{M r_+}$.
This value exactly matches with the direct evaluations of Komar expressions for energies 
%within the first order approximation 
\cite{Katz,Dadich,Banerjee:2009tz}. It is also reassuring to note that the definition of $M$ following from (\ref{8.14}) and (\ref{8.28}) reproduces the generalised Smarr formula \cite{Smarr:1972kt,Bardeen:1973gs,Gibbons:1976ue},
\begin{eqnarray}
\frac{M}{2} = \frac{\kappa A}{8\pi} + \frac{VQ}{2} + \Omega J.
\label{8.29}
\end{eqnarray}

\section{Discussions}
     In this chapter we have further clarified the possibility of considering gravity as an emergent phenomenon. Taking the standard definition of entropy from statistical mechanics we were able to show the equivalence of entropy with the action. Consequently, extremisation of the action leading to Einstein's equations is equivalent to the extremisation of the entropy. We derived the relation $S_{bh} = E/2T_H$ for stationary black holes with $S_{bh}$ and $T_H$ being the entropy and Hawking temperature. The nature of energy $E$ appearing in this relation was clarified. It was proved to be Komar's expression valid for stationary asymptotically flat space-time. An explicit check of $S_{bh}=E/2T_H$ was done for all black hole solutions of Einstein gravity. This relation was also seen to reproduce the generalised mass formula of Smarr \cite{Smarr:1972kt,Bardeen:1973gs,Gibbons:1976ue}. In this sense the Smarr formula can be interpreted as a thermodynamic relation further illuminating the emergent nature of gravity. As a final remark we feel that although our results were derived for Einstein gravity, the methods are general enough to include other possibilities like higher order theories.

%\include{chap-beyondsemiclassical}
%%%%%%%%%%%%%%%%%%%%%%%%%%%%%%%%%%%%%%%%%%%%%%%%%%%%%%%%%%%%%%%%%%%%%%
\chapter{\label{chap:conclusions} Conclusions}
%%%%%%%%%%%%%%%%%%%%%%%%%%%%%%%%%%%%%%%%%%%%%%%%%%%%%%%%%%%%%%%%%%%%%%
The motivation of this thesis was to study certain field theory aspects of
black holes, with particular emphasis on the Hawking effect, using various semi-classical techniques. We now summarize the results obtained in last seven chapters and briefly comment on future prospects.

In the second chapter, we gave a general framework of tunneling mechanism for a static, spherically symmetric black hole metric. Both Hamilton - Jacobi and radial null geodesic approaches were elaborated. The tunneling rate was found to be the Boltzmann factor. Then Hawking's expression for the temperature of a black hole - proportional to surface gravity - was derived. 
%Finally, using this explicit form of Hawking temperature for some particular black hole metrics were given.  

    In the third chapter, we provided an application of this general framework for null geodesic method. Back reaction as well as noncommutative effects in the space-time were incorporated. Here the main motivation was to find the modifications to the thermodynamic entities, such as temperature, entropy etc. 

   First the back reaction, which is just the effect of space-time fluctuations, was considered. It was shown in \cite{Lousto:1988sp} that even in the presence of this effect the metric remains in the static, spherically symmetric form, but with a modified surface gravity. So it was possible to use the method elaborated in the previous chapter. In this case, we showed the following results:
\begin{itemize}
\item  The temperature was modified and also the entropy received corrections. The leading order correction was found to be the logarithmic of area while the non-leading corrections are just the inverse powers of area. 
\item The coefficient of the logarithmic term was related to the trace anomaly of energy-momentum tensor.
\end{itemize} 
Both these results agreed with the earlier findings \cite{Lousto:1988sp,Fursaev:1994te} by other methods.   

     We also discussed the effect of noncommutativity in addition to the back reaction effect in the black hole space-time. Here again the corrections to the thermodynamic quantities were given. For consistency, we showed that in the proper limit the usual (commutative space-time) results were recovered.

    In the fourth chapter we discussed another method, the chiral anomaly method, to derive the fluxes of Hawking radiation. Here the chiral anomaly expressions were obtained from the non-chiral theory by using the trace anomaly and the chirality conditions. Then the Hawking flux was derived following the path prescribed in \cite{Banerjee:2007qs,Banerjee:2008az}. Another portion of this chapter was dedicated to show that the same chirality conditions were enough to find the Hawking temperature in the quantum tunneling method. Here the explicit form of modes created inside the black hole were obtained by solving the chirality condition. Then using the Kruskal coordinates  relations between the ``inside'' modes and ``outside'' modes were established. Finally, calculation of the respective probabilities yielded that the left moving mode was actually trapped inside the horizon while right moving mode can come out from the horizon with a finite probability. Thus this analysis manifested the crucial role of the chirality to give a unified description of both tunneling and anomaly approaches.

  In the fifth chapter, the Hawking emission spectrum from the event horizon was derived based on our reformulated tunneling mechanism introduced in the previous chapter. Using the density matrix technique the average number of emitted particle from the horizon was computed. The spectrum was exactly that of the black body with the Hawking temperature. Thereby we provided a complete description of the Hawking effect in the tunneling mechanism. The absence of any derivation of the spectrum was a glaring omission within the tunneling paradigm.

   In the next chapter, a unified description of Unruh and Hawking effects was discussed by introducing a new type of global embedding. Since the thermodynamic quantities of a black hole are determined by the horizon properties and near the horizon the effective theory is dominated by the two dimensional ($t-r$) metric, it is sufficient to consider the embedding of this two dimensional metric. Considering this fact, a new reduced global embedding of two dimensional curved space-times in higher dimensional flat ones was introduced to present a unified description of Hawking and Unruh effects. Our analysis simplified as well as generalised the conventional embedding approach.

    In chapter - \ref{chap:spectroscopy}, based on the modified tunneling mechanism, introduced in the previous chapters, we obtained the entropy spectrum of a  black hole. Our conclusions were following:
\begin{itemize}
\item In {\it{Einstein's gravity}}, both entropy and area spectrum are evenly spaced. 
\item On the other hand in more general theories (like {\it{Einstein-Gauss-Bonnet gravity}}), although the entropy spectrum is equispaced, the corresponding area spectrum is not. 
\end{itemize}
In this sense, it was legitimate to say that {\it quantization of entropy is more fundamental than that of area}.

     Finally, based on the above conceptions and findings, we explored an intriguing possibility that gravity can be thought as an emergent phenomenon. Starting from the definition of entropy, used in statistical mechanics, we showed that it was proportional to the gravity action. For a stationary black hole this entropy was expressed as $S_{bh} = E/ 2T_H$, where $T_H$ and $E$ were the Hawking temperature and the Komar energy respectively. This relation was also compatible with the generalised Smarr formula for mass.

    There are certain issues which are worthwhile for future study. 
\begin{itemize}
\item The inclusion of grey body effect within the tunneling approach would be an interesting exercise. The analysis given here did not include the grey body effect. Consequently, the flux obtained was compared with that associated with the perfect black body.  
\item Another important issue is the computation of black hole entropy by using the anomaly approach. There are strong reasons to believe that the black hole entropy, like  Hawking flux can be related to the diffeomorphism anomaly \cite{Solodukhin:1998tc,Carlip:2007za,Carlip:2004mn,Cvitan:2003vq,Banerjee:2008aa,Banerjee:2010zz,Chung:2010xz}. For example, in the analysis of \cite{Carlip:2007za,Carlip:2004mn} the counting of microstates was done by imposing the ``horizon constraints''. The algebra among these ``horizon constraints'' commutes only after modifying the generators for diffeomorphism symmetry. This modification in the generators give rise to the desired central charge, which ultimately leads to the Bekenstein-Hawking entropy. This is roughly similar to the diffeomorphism anomaly mechanism.
\item  So far, not much progress has been achieved in the understanding of the Unruh effect by the gravitational anomaly method. The main difficulty lies in the fact that the Unruh effect is basically related to flat space-time and the observer must be uniformly accelerated. So a naive use of the anomaly expressions is unjustified. In this thesis it was shown that the flat space embedding of the near horizon effective two dimensional ($t-r$) metric was enough for giving a unified description of Hawking and Unruh effects and it simplified as well as generalized earlier facts. The local Hawking temperature was exactly equivalent to the one detected by the Unruh observer. Again, in the gravitational (chiral) anomaly expressions the metric that contributed was the aforesaid effective metric.
 It may be possible to translate these expressions for the anomaly in the embedded space and establish a connection with the Unruh effect.
\item  The last point I want to mention is that in chapter-8, an emergent nature of gravity was illustrated from a statistical point of view. These discussions were confined to the four dimensional Einstein gravity without cosmological constant.
It would be fascinating to extend our discussion to higher dimensional Einstein gravity (with or without cosmological constant)
and more general gravity theories (e.g. Lovelock gravity). If this attempt is successful, then one will be able to give a unified form of the Smarr formula for all such theories. 
\end{itemize}
 It is thus clear that  the quantum tunneling mechanism,
  provided in this thesis, could illuminate the subject of thermodynamics of gravity, more precisely, the black hole.

\backmatter
%\include{bibliography}
%%%%%%%%%%%%%%%%%%%%%%%%%%%%%%%%%%%%%%%%%%%%%%%%%%%%%%%%%%%%%%%%%%%%%%%%%%%%%%%

%%%%%%%%%%%%%%%%%%%%%%%%%%%%%%%%%%%%%%%%%%%%%%%%%%%%%%%%%%%%%%%%%%

%\chapter*{}
\pagenumbering{roman}
\thispagestyle{empty}
\vspace*{8.5 cm}
\begin{center}
\large \uppercase{Reprints}
\end{center}

%%%%%%%%%%%%%%%%%%%%%%%%%%%%%%%%%%%%%%%%%%%%%%%%%%%%%%%%%%%%%%%%%%

%\include{part_reprints}

%%%%%%%%%%%%%%%%%%%%%%%%%%%%%%%%%%%%%%%%%%%%%%%%%%%%%%%%%%%%%%%%%%

\end{document}